\documentclass[aps,prc,amsfonts,preprintnumbers,superscriptaddress,showpacs,nofootinbib]{revtex4-1}
\usepackage{graphicx}
\usepackage{amsmath}
\usepackage{amssymb}
\usepackage{psfrag}
\usepackage{epsfig}
\usepackage{epsf}
\usepackage{float}
\usepackage{slashed}
\usepackage{longtable}

\usepackage[export]{adjustbox}
\usepackage{hyphenat}
\usepackage{booktabs}
\usepackage{siunitx}
\usepackage{physics}
\usepackage{dsfont} 
\usepackage{acronym}
\usepackage[greek,english]{babel}
\allowdisplaybreaks

\usepackage{color}

\usepackage{natbib}

\newcommand{\trfl}[1]{\operatorname{Tr}\left(#1\right)}

\newcommand{\hc}{\text{H.c.}}
\newcommand{\adj}{^{\dagger}}
\newcommand{\newvec}[1]{\textit{\textbf{#1}}}

\newcommand{\newvecgr}[1]{\textit{\textbf{\textgreek{#1}}}}
\newcommand{\newvecgrconst}[1]{\textbf{\textgreek{#1}}}

\newcommand{\diag}{\operatorname{diag}}

\newcommand{\identity}{\mathds{1}}
\newcommand{\efkt}{\operatorname{e}}
\newcommand{\ci}{\textup{i}}
\newcommand{\pimath}{\textrm{\textgreek{p}}}
\newcommand{\matrixgamma}{\textrm{\textgreek{g}}}
\newcommand{\tensorsigma}{\textrm{\textgreek{s\noboundary}}}
\newcommand{\levicivita}{\textsf{\textgreek{e}}}
\newcommand{\kronecker}{\text{\textgreek{d}}}
\newcommand{\minkowski}{\text{g}}
\renewcommand{\paulimatrix}{\textsf{\textgreek{t}}}
\newcommand{\paulimatrixvector}{\newvecgrconst{t}}
\newcommand{\covdc}{\operatorname{D}}
\newcommand{\eulergamma}{\textrm{\textgreek{g}}_\textrm{E}}
\newcommand{\lambar}{\bar{\lambda}}

\newcommand{\Cov}{\textrm{Cov}}

\newcommand{\pion}{\textit{\textgreek{p}}}
\newcommand{\pionfield}{\textit{\textgreek{p}}}
\newcommand{\pionfieldvector}{\newvecgr{p}}

\newcommand{\pionmass}{M_{\pion}}
\newcommand{\barepionmass}{M}

\newcommand{\pionZfactor}{Z_{\pion}}
\newcommand{\ompi}{\varOmega_{\pion}}

\newcommand{\phipi}{\varphi_{\pion}}
\newcommand{\epi}{E_{\pion}}
\newcommand{\pionmomentum}{\newvec{q}}
\newcommand{\Udag}{U^{\dagger}}
\newcommand{\delZpi}{\kronecker Z_{\pion}}
\newcommand{\delmpi}{\kronecker M}
\newcommand{\piondecayconstant}{F_{\pion}}
\newcommand{\bareFpi}{F}
\newcommand{\delFpi}{\kronecker F}

\newcommand{\photon}{\gamma}

\newcommand{\omgam}{\varOmega_{\photon '}}

\newcommand{\phigamma}{\varphi_{\photon '}}
\newcommand{\egi}{E_{\photon}}
\newcommand{\ego}{E_{\photon '}}
\newcommand{\photonmomentumout}{\newvec{k}'}

\newcommand{\nucleon}{N}

\newcommand{\nucleonmass}{m_{\nucleon}}
\newcommand{\barenucleonmass}{m}
\newcommand{\proton}{p}
\newcommand{\neutron}{n}

\newcommand{\eno}{E_{\nucleon '}}

\newcommand{\nucleonZfactor}{Z_{\nucleon}}
\newcommand{\delmn}{\kronecker \barenucleonmass}
\newcommand{\delZn}{\kronecker Z_{\nucleon}}
\newcommand{\nucleoncharge}{Q_{\nucleon}}

\newcommand{\deltapart}{\Delta}

\newcommand{\deltamass}{m_{\deltapart}}
\newcommand{\baredeltamass}{\mathring{m}_{\deltapart}}
\newcommand{\deltafield}{\varPsi}

\newcommand{\deltasplit}{\deltapart}
\newcommand{\deltamdm}{\mu_{\deltapart}}
\newcommand{\deltamdmind}[1]{\mu_{\deltapart^{#1}}}
\newcommand{\deltacharge}{Q_{\deltapart}}

\newcommand{\pindcoupling}{h_A}
\newcommand{\barehA}{h}
\newcommand{\axialcoupling}{g_A}
\newcommand{\baregA}{g}
\newcommand{\delgA}{\kronecker g}
\newcommand{\gOne}{g_1}

\newcommand{\lgen}{\mathcal{L}}

\newcommand{\lpipi}{\lgen_{\pion\pion}}
\newcommand{\lpin}{\lgen_{\pion\nucleon}}
\newcommand{\lpid}{\lgen_{\pion\deltapart}}
\newcommand{\lpind}{\lgen_{\pion\nucleon\deltapart}}

\newcommand{\leff}{\lgen^{\text{eff}}}

\newcommand{\chiralbreaking}{\varLambda_{b}}
\newcommand{\matel}{\mathcal{M}}
\newcommand{\unpol}{|\matel |^2}
\newcommand{\ws}{\sqrt{s}}

\newcommand{\obs}{\mathcal{O}}
\newcommand{\dobs}{\kronecker\mathcal{O}}

\begin{document}

\title{Radiative pion photoproduction in covariant chiral perturbation
  theory}

\author{J.~R\ij neveen}
\email[]{Email: jan.rijneveen@rub.de}
\affiliation{Ruhr-Universit\"at Bochum, Fakult\"at f\"ur Physik und
        Astronomie, Institut f\"ur Theoretische Physik II,  D-44780
        Bochum, Germany}

\author{N.~R\ij neveen}
\email[]{Email: nora.rijneveen@rub.de}
\affiliation{Ruhr-Universit\"at Bochum, Fakult\"at f\"ur Physik und
        Astronomie, Institut f\"ur Theoretische Physik II,  D-44780
        Bochum, Germany}

\author{H.~Krebs}
\email[]{Email: hermann.krebs@rub.de}
\affiliation{Ruhr-Universit\"at Bochum, Fakult\"at f\"ur Physik und
        Astronomie, Institut f\"ur Theoretische Physik II,  D-44780
        Bochum, Germany}

\author{A.~M.~Gasparyan}
\email[]{Email: ashot.gasparyan@rub.de}
\affiliation{Ruhr-Universit\"at Bochum, Fakult\"at f\"ur Physik und
        Astronomie, Institut f\"ur Theoretische Physik II,  D-44780
        Bochum, Germany}
\affiliation{NRC “Kurchatov Institute” - ITEP, B. Cheremushkinskaya 25, 117218 Moscow, Russia}
      
\author{E.~Epelbaum}
\email[]{Email: evgeny.epelbaum@rub.de}
\affiliation{Ruhr-Universit\"at Bochum, Fakult\"at f\"ur Physik und
        Astronomie, Institut f\"ur Theoretische Physik II,  D-44780
        Bochum, Germany}

\date{\today}

\begin{abstract}
We present a calculation of radiative pion photoproduction in the framework 
of covariant chiral perturbation theory with explicit $\deltapart(1232)$ degrees of freedom.
The analysis is performed employing the small scale expansion scheme
adjusted for the $\deltapart$ region. Depending on the channel, we include contributions
up to next-to-next-to-leading order.
We fit the available experimental data for the reaction $ \photon\proton\to\photon\proton\pion^0 $
and extract the  value of the $\deltapart^+$ magnetic moment.
Errors from the truncation of the small scale expansion are estimated using the Bayesian approach.
We compare our results both with the previous studies within the $\delta$-expansion scheme and
with the $\deltapart$-less theory.
We also give predictions for radiative charged-pion photoproduction.
\end{abstract}


\maketitle

\section{Introduction}
The properties of the $\deltapart(1232)$ resonance are important for understanding the
low-energy dynamics of the strong interaction owing to its nature as the lowest nucleon excitation. 
One of the main characteristics of the $\deltapart$ isobar are its electromagnetic moments, 
which are rather poorly known at present.
Since the $\deltapart$ isobar is unstable and its lifetime is very
short, a direct experimental determination of the electromagnetic moments 
is impossible. One, therefore, has to rely only on an analysis of scattering experiments where 
the $\deltapart$ resonance can be produced. In particular, the magnetic dipole
moment ({MDM}) of the $\deltapart^+$ particle 
can be accessed through the measurement of the radiative photoproduction of neutral pions
in the $\deltapart$-resonance region.
An accurate determination of the {MDM} of the $\deltapart^+$ from experiment is particularly important 
as it would allow to test predictions based on a variety of
theoretical approaches such as quark models, Dyson-Schwinger equations, hadron-string duality, QCD sum rules,
large-$N_c$ constraints, chiral perturbation theory calculations
(covariant and heavy-baryon),
etc., see Refs.~\cite{an:2006zf,Ledwig:2008es,Schlumpf:1993rm,Wagner:2000ii,Ramalho:2010xj,Kim:2019gka,Nicmorus:2010sd,Hashimoto:2008zw,
Azizi:2008tx,Lee:1997jk,Luty:1994ub,FloresMendieta:2009rq,Ahuatzin:2010ef,Butler:1993ej,Geng:2009ys,Li:2016ezv}.
The spread of the theoretical predictions for the $\deltapart^+$ {MDM}
is quite large, namely 
 $\deltamdmind{+} \simeq (1.7-3.5) \mu_{\nucleon}$\footnote{Throughout this paper, $\deltamdmind{}$ denotes the 
real part of the corresponding magnetic moment, see Sec.~\ref{sec:DeltaFF}.},
where $
 \mu_{\nucleon}=e/(2\nucleonmass) $ is the nucleon magneton.
Lattice-QCD calculations of  the $\deltapart^+$ {MDM} are also
available, see
e.g.~Refs.~\cite{Alexandrou:2007we,Alexandrou:2008bn,Lee:2005ds,Aubin:2008qp,Boinepalli:2009sq},
but not fully conclusive yet.
These studies report $\deltamdmind{+}$-values in the range $\deltamdmind{+} \simeq (1.0-2.4) \mu_{\nucleon}$.

The current Particle Data Group ({PDG}) value of the $\deltapart^+$ {MDM} is
$\deltamdmind{+} = 
\left(2.7^{+1.0}_{-1.3}(\textrm{stat.})\pm 1.5(\textrm{syst.})\pm 3(\textrm{theor.}) \right)\mu_{\nucleon}\,$.
It is based on the analysis of the radiative neutral-pion photoproduction observables measured 
in the TAPS/A2 experiment at MAMI \cite{Kotulla:2002cg}.
The analysis was performed employing a phenomenological model of
Ref.~\cite{Drechsel:2001qu}, see also
Refs.~\cite{Machavariani:1999fr,Drechsel:2000um} for earlier studies
along this line. 
In a subsequent experiment at MAMI \cite{Schumann:2010js},
a considerable improvement on statistics was achieved.
The extraction of the $\deltapart^+$ {MDM} was performed using the more advanced 
unitarized dynamical model of Ref.~\cite{Chiang:2004pw}
resulting in $\deltamdmind{+} =\left(2.89^{+0.30}_{-0.34}(\textrm{stat.})\pm 0.30(\textrm{syst.}) \right)\mu_{\nucleon}$,
although the data were not described by the model very well.

The main disadvantage of using phenomenological models for the analysis of the experimental data is 
impossibility to obtain a reliable estimate of theoretical uncertainties.
A more systematic way to analyze radiative neutral-pion photoproduction
is achieved in the effective field theory ({EFT}) framework, which
allows one to account for theoretical uncertainties. 
The low-energy effective field theory of the standard model is chiral
perturbation theory ({$\chi$PT}). It is based on the effective
chiral Lagrangian constructed from the pion, nucleon and, for the case
at hand, also $\deltapart$ fields in the presence of external sources.
The Lagrangian is organized as a series of terms with increasing number of derivatives and powers of the quark masses 
(proportional to the pion mass squared). The scattering amplitude can
then be written as a systematic expansion 
in terms of a small parameter (pion mass, external momenta, $\deltapart$-nucleon mass difference),
see Sec.~\ref{sec:renormalization} for the details. 

In the single-nucleon sector, calculations are performed both using
the heavy-baryon \cite{Jenkins:1990jv, Bernard:1992qa, Bernard:1995dp}  
and covariant \cite{Becher:1999he,Fuchs:2003qc} formulations of 
{$\chi$PT}.
In the energy regime considered in the present work (the $\deltapart$ region),
the covariant approach appears to be the natural choice because the initial momentum  
may be too high for the nucleon to be treated non-relativistically.
The analysis of radiative pion photoproduction in the
$\deltapart$ region, i.e.~close to the $\deltapart$ pole, makes it inevitable to include
explicit $\deltapart$ degrees of freedom.
In the following, we will explicitly demonstrate this feature by also 
showing predictions within the $\deltapart$-less approach
and studying the convergence pattern in such a scheme.

The first analysis of radiative $\pion^0$-photoproduction within covariant $\deltapart$-full {$\chi$PT}
was done in Refs.~\cite{Pascalutsa:2004je,Pascalutsa:2007wb} within the so-called 
$\delta$-expansion scheme \cite{Pascalutsa:2002pi}.
The $\delta$-counting treats the $\deltapart$-nucleon mass difference $\deltasplit$
as being a lower order quantity than the pion mass $\pionmass$ ($\deltasplit \sim\delta$, $\pionmass\sim\delta^2$).
The $\deltapart$-width is regarded to be of order $M_\pion^3$, and, therefore, all $\deltapart$-pole graphs
are assumed to be dominant in the $\deltapart$ region.
The analysis of the old experimental data from Ref.~\cite{Kotulla:2002cg} based on this scheme
resulted in the values of the $\deltapart^+$ {MDM} in the range of 
$\deltamdmind{+} =\left(1-3 \right)\mu_{\nucleon}$  \cite{Pascalutsa:2007wb}.
However, when applied to the combined set of data including the newer higher-statistics experiment at MAMI,
the extracted value of the {MDM} turned out to be 
$\deltamdmind{+} =\left(3.77^{+0.14}_{-0.15}(\textrm{stat.})\pm 0.65(\textrm{syst.}) \right)\mu_{\nucleon}$\cite{Schumann:2010js}.
This value lies outside the range of the theoretical predictions
mentioned above and might be an indication of 
a slow convergence of such a scheme.

In this work, we analyze radiative pion photoproduction utilizing the
covariant $\deltapart$-full {$\chi$PT} framework within the so-called
small scale expansion (SSE) scheme \cite{Hemmert:1997ye},
where one treats the $\deltapart$-nucleon mass difference as $\deltasplit \sim O(M_\pion)$.
We also take into account that certain tree-level diagrams involving
$\deltapart$ are enhanced in the vicinity of the $\deltapart$ pole.
We include the leading pion-nucleon loop graphs, which provide a sizable background (with respect to the $\deltapart$ poles) contribution.
For the treatment of the meson-baryon loops, we adopt the extended
on-mass-shell ({EOMS}) renormalization scheme \cite{Fuchs:2003qc}.
The details of our approach and its differences from the scheme of Ref.~\cite{Pascalutsa:2007wb} are discussed
in subsequent sections.

The only free parameter of our calculation is the $\deltapart$ {MDM}
since other low-energy constants are 
taken from analyses of other reactions.
We also analyze pion photoproduction, a
subprocess of the reaction under consideration, in order to
determine various $\gamma N \Delta$ low-energy constants
in full consistency with our treatment of radiative
pion photoproduction.
We also take into account the theoretical uncertainty related to the truncation of the 
small scale expansion within a Bayesian approach, thus providing a
reliable extraction of the $\deltapart$ {MDM} from the data. 

In addition to the $ \photon\proton\to\photon\proton\pion^0 $
reaction, we also study radiative photoproduction of charged pions
$ \photon\proton\to\photon\neutron\pion^+ $, which provides access to
the {MDM} of the $\deltapart^0$ resonance.
Since no experimental data are available for this reaction, we give our predictions for
various observables using the isospin symmetry and the value of the $\deltapart^{++}$ {MDM}
extracted previously from the reaction $ \pion^+ p\to \pion^+\proton\photon $  \cite{Bosshard:1991zp,LopezCastro:2000cv},
and discuss their sensitivity to the $\deltapart^+$ and $\deltapart^0$ {MDM}.

Our paper is organized as follows.
In Sec.~\ref{sec:kinematics} we introduce the notation and define
various kinematical quantities and discuss the observables used in our analysis.
In Sec.~\ref{sec:Lagrangian} we provide the effective Lagrangian relevant for our calculation. 
Construction of the reaction amplitude, power counting and renormalization are discussed in  
Secs.~\ref{sec:PowerCounting},~\ref{sec:renormalization}.
Next, Sec.\ref{sec:photoproduction} is devoted to the extraction of the low-energy constants
from pion photoproduction while  the numerical results of our study
are presented and discussed in 
Sec.~\ref{sec:results}. The main results of our work are summarized in Sec.~\ref{sec:summary}.

\section{Kinematics, reaction amplitude and observables for radiative
  pion photoproduction} \label{sec:kinematics}
Radiative pion photoproduction is a reaction with a photon  
and a nucleon in the initial state and a photon, a nucleon and a pion in the final state:
\begin{align}
\photon(\lambda,k) + \nucleon_i(s,p) \; \to \; \photon(\lambda',k') +
  \nucleon_j '(s',p') + \pion_a(q),
\end{align}
where $ p $ ($ p' $),  $ s $ ($ s' $), $i$ ($j$) are the momentum, helicity and the isospin index of
the incoming (outgoing) nucleon;
$ k $ ($ k' $) and  $ \lambda $ ($ \lambda' $) are the momentum and helicity of the incoming (outgoing) photon
while $ q $ and $ a $ are the momentum and the isospin index of the outgoing pion.

For each set of the isospin indices $a$, $i$, $j$,
the radiative-pion-photoproduction matrix element 
\begin{align}
\matel_{a;ji}=\bar{u}^{(s')}_{j}(p') \, 
\varepsilon^{(\lambda)}_{\mu}(k) \, \varepsilon^{*(\lambda
  ')}_{\nu}(k') \, \matel_{a;ji}^{\mu\nu}
\, u^{(s)}_{i}(p)\,, 
\end{align}
where $u^{(s)}_{i}(p)$ ($\bar{u}^{(s')}_{j}(p')$) stands for the initial (final) nucleon spinor and
$\varepsilon^{(\lambda)}_{\mu}(k)$ ($\varepsilon^{*(\lambda ')}_{\nu}(k')$) for the initial (final) photon
polarization vector,
can be parameterized in terms of $16$ scalar invariant amplitudes
$A_{l;mn}$, see   e.g., \cite{Bardeen:1969aw}, by
introducing the projector
$P^{\mu\nu}=\minkowski^{\mu\nu}-\frac{k'^{\mu}k^{\nu}}{k\cdot k'}$ via
\begin{align}
\matel^{\mu\nu} &=
P^{\mu\alpha}P^{\nu\beta}\sum_{l=1}^4\sum_{m,n=1}^2 Q_{m;\alpha}Q_{n;\beta}X_l\; A_{l;mn}\,,\nonumber\\
Q_{1;\alpha}&=p_\alpha\,, \quad Q_{2;\alpha}=p'_\alpha\,, \quad  X_1=\matrixgamma^5 \,, \quad  X_2=\slashed k\matrixgamma^5 \,, \quad  X_3=\slashed k'\matrixgamma^5 \,, \quad  X_4=[ \slashed k,\slashed k' ]\matrixgamma^5 \,.
\label{eq:inv_amplitudes_A}
\end{align}
This ensures that the amplitude is explicitly transverse: $ k_{\mu}\matel^{\mu\nu}=k'_{\nu}\matel^{\mu\nu}=0 $.
The invariant amplitudes $A_{l;mn} \equiv A_{l;mn}(s,s_1,s_2,t_1,t_2)$ are functions of 
the scalar Mandelstam variables
\begin{align}
s=(p+k)^2\,, \quad s_1=(p'+k')^2\,, \quad  s_2=(p'+q)^2\,, \quad t_1&=(k-k')^2\,,  \quad t_2=(k-q)^2.
\end{align}
The number of invariant amplitudes coincides with the number of helicity amplitudes.
Note that parity conservation does not lead to a reduction of the number of independent amplitudes
because a parity violating structure can be made parity conserving by multiplying it with 
the pseudoscalar quantity $ \levicivita^{\mu\nu\rho\sigma} p_\mu p'_\nu k_\rho k'_\sigma$,
which cannot be expressed unambiguously as a function of the Mandelstam variables introduced above.
The amplitudes $A_{l;mn}$ are not free of kinematical singularities or constraints, 
which can be fixed by finding an appropriate linear transformation.
This might be relevant for an analysis based on dispersion relations
but is not crucial for our purely perturbative calculation.
Therefore, we stick to the above-mentioned basis.

In isospin space, each $A_{l;mn}$ can be decomposed into a linear
combination of four structures\footnote{Here, we suppress the indices
  $l$, $m$, $n$ but show explicitly  the isospin indices.}
\begin{align}
A_{a}=B^1 \, \kronecker_{3a}\, \identity +B^2 \, \kronecker_{3a}\,
  \paulimatrix_3 + B^3 \, \paulimatrix_{a}+B^4\,
  \ci\levicivita_{3ab}\, \paulimatrix_b \,,
\end{align}
where $\paulimatrix_a$ are the Pauli matrices.
There are four possible channels of the radiative pion photoproduction
reaction, namely
\begin{align}
&\text{1.}\quad\photon\proton \to\photon\proton\pion^0 \nonumber\\
&\text{2.}\quad\photon\proton \to\photon\neutron\pion^+ \nonumber\\
&\text{3.}\quad\photon\neutron \to\photon\neutron\pion^0 \nonumber\\
&\text{4.}\quad\photon\neutron \to\photon\proton\pion^- \,.
\end{align}
The amplitude in the isospin basis can be transformed into the  particle basis
by means of the following relations
\begin{align}
&\text{1.}\quad \zeta\adj_{\proton}A_{3}\zeta_{\proton}= B^1+B^2+B^3 \nonumber\\
&\text{2.}\quad \frac{1}{\sqrt{2}} \zeta\adj_{\neutron}(A_{1}-\ci A_{2}) \zeta_{\proton}= \sqrt{2}(B^3-B^4) \nonumber\\
&\text{3.}\quad \zeta\adj_{\neutron}A_{3}\zeta_{\neutron}= B^1-B^2-B^3 \nonumber\\
&\text{4.}\quad \frac{1}{\sqrt{2}}\zeta\adj_{\proton}(A_{1}+\ci A_{2})\zeta_{\neutron}= \sqrt{2}(B^3+B^4)\,,
\end{align}
with $ \zeta_p=(1,0)^T $, $ \zeta_n=(0,1)^T $.

Our
calculations are carried out in the center of mass ({CM}) frame with $k=(\egi, \newvec{k})$, 
$k'=(\ego, \photonmomentumout)$, $q=(\epi, \pionmomentum)$,
where $\egi=|\newvec{k}|$, $\ego=|\photonmomentumout|$, $\epi=\sqrt{M_\pion^2+\pionmomentum^2}$.
We choose our coordinate system such that 
\begin{equation}
\newvec{k}=%
\begin{pmatrix} 0 \\ 0 \\ \egi \end{pmatrix},\quad%
\photonmomentumout=\ego%
\begin{pmatrix} \sin\vartheta_{\photon'}\cos\phigamma \\ \sin\vartheta_{\photon'}\sin\phigamma \\ \cos\vartheta_{\photon'} \end{pmatrix}\quad\text{and}\quad%
\pionmomentum=|\newvec{q}|%
\begin{pmatrix}\sin\vartheta_{\pion}\cos\phipi \\\sin\vartheta_{\pion}\sin\phipi\\\cos\vartheta_{\pion}\end{pmatrix}. \label{eq:cmframdef}
\end{equation}
The differential cross section for a reaction $1+2\to 3+4+5$ is given by (the notation is obvious)
\begin{align}
 \dd \sigma = \frac{1}{2\sqrt{\lambda(s,m_1^2,m_2^2)}}(2\pimath)^4\,\kronecker^{(4)}\left(\sum_{j=3}^5 p_{j}-p_{1}-p_{2}\right)
\unpol\prod_{j=3}^5\frac{\dd^3p_{j}}{(2\pimath)^3 2E_j}\,,
 \end{align}
with $\lambda(x,y,z)=x^2+y^2+z^2-2xy-2xz-2yz$. For radiative
pion photoproduction, this equation turns into
 \begin{align}
 \dd \sigma =\frac{1}{s-\nucleonmass^2}\frac{1}{16(2\pimath)^5 \eno} \, \kronecker(\eno+\ego+\epi-\ws) \unpol
\ego|\pionmomentum|\dd \ego \dd \omgam \dd \epi \dd {\ompi}\,,
\label{eq:dsigma}
 \end{align}
 where $\eno=\sqrt{\nucleonmass^2+(\pionmomentum+\photonmomentumout)^{2}}$
is the energy of the outgoing nucleon and $\omgam$ ($\ompi$) is the solid angle corresponding to 
the emitted photon (pion).

In this work we focus on the polarized and unpolarized differential cross sections
$\dv{\sigma}{\omgam}$ and $\dv{\sigma}{\ompi}$, $\dv{\sigma}{\ego}$,
which are readily obtained from Eq.~(\ref{eq:dsigma}) by
integrating over remaining variables.
For the unpolarized cross section, one sums in Eq.~(\ref{eq:dsigma}) over the polarizations of the outgoing particles
and averages over the polarizations of the incoming particles.
We also consider several further polarization-dependent observables.
The linear photon polarization asymmetry for a fixed direction of the outgoing pion is given by
\begin{align}
\varSigma^{\pion}=\frac{(\dd\sigma_{\perp}/\dd\ego\dd\ompi) - (\dd\sigma_{\parallel}/\dd\ego\dd\ompi)}{(\dd\sigma_{\perp}/\dd\ego\dd\ompi) 
+ (\dd\sigma_{\parallel}/\dd\ego\dd\ompi)},
\end{align}
where $ \sigma_{\perp} $ ($ \sigma_{\parallel} $) are the cross
sections for the initial photon polarizations 
that are perpendicular (parallel) to the plane spanned by the incoming photon and outgoing pion momenta.
The superscript $\pion$ signifies the choice $\phipi=0$.
The circular photon polarization asymmetry for a fixed direction of the outgoing pion is defined according to Ref. \cite{Pascalutsa:2007wb} as 
\begin{align}
\varSigma^{\pion}_{\textrm{circ}}=\frac{2}{\pimath}\frac{\int\dd\omgam 2\sin\phigamma\left(\dd\sigma_{+} - \dd\sigma_{-}\right)}
{\int\dd\omgam \left(\dd\sigma_{+} + \dd\sigma_{-}\right)}\,,
\end{align}
where we introduce the shorthand notation for the polarized cross
sections according to
\begin{align}
\dd\sigma_{\pm}=\frac{\dd\sigma_{\pm}}{\dd\ego\dd\omgam\dd\ompi}\,.
\end{align}
The subscript $\pm$ stands for the incoming photon helicity $\lambda=\pm 1$.
Analogously, we define the circular photon polarization asymmetry for a fixed direction of the outgoing photon 
\begin{align}
\varSigma^{\photon}_{\textrm{circ}}=\frac{2}{\pimath}\frac{\int\dd\ompi 2\sin\phipi\left(\dd\sigma_{+} - \dd\sigma_{-}\right)}
{\int\dd\ompi \left(\dd\sigma_{+} + \dd\sigma_{-}\right)}\,,
\end{align}
where the superscript $\gamma$ indicates the choice $ \phigamma=0 $.

Whenever the integration over the energy of the outgoing photon $ \ego $ is performed, 
the lower limit (infrared cutoff) is set to $ \ego^-=\SI{30}{\MeV} $ in accordance with the
experimental methodology of Refs.~\cite{Schumann:2010js,Kotulla:2002cg}.

We also analyze the ratio of the differential cross sections for
radiative and ordinary pion photoproduction
weighted with the bremsstrahlung factor 
introduced in Ref.~\cite{Chiang:2004pw}. 
For the neutral channel, it is defined as
\begin{align}
R=\frac{1}{\sigma_{\pion^{\hspace{.05em}0}}}\ego\dv{\sigma}{\ego},
\end{align}
with
\begin{align}
\sigma_{\pion^{\hspace{.05em}0}}=\frac{e^2}{2\pimath^2}\int\dd\ompi W(v)\left(\dv{\sigma}{\ompi}\right)^{\photon\proton \to \proton\pion^0}\,,\quad
W(v)=-1+\frac{v^2+1}{2v}\ln\left(\frac{v+1}{v-1}\right)\,,\quad
v=\sqrt{1-4\nucleonmass^2/(p'-p)^2} 
\end{align}
whereas for the charged channel, it is given by
\begin{align}
R=\frac{1}{\sigma_{\pion^+}}\ego\dv{\sigma}{\ego},
\end{align}
with
\begin{align}
\sigma_{\pion^+}=\frac{e^2}{2\pimath^2}\int\dd\ompi W(v')\left(\dv{\sigma}{\ompi}\right)^{\photon\proton \to \neutron\pion^+}\,,\quad
v'=\sqrt{1+4\nucleonmass\pionmass/((\nucleonmass-\pionmass)^2-(q-p)^2)}\,.
\end{align}
The soft-photon theorem ensures that $R\overset{\ego \to
  0}{\longrightarrow}1$  \cite{Chiang:2004pw}.

\section{Effective Lagrangian} \label{sec:Lagrangian}
Chiral perturbation theory is based on the effective Lagrangian consistent with symmetries of  {QCD}. 
It contains an infinite set of terms with increasing number of derivatives and powers of quark masses.
The terms of the effective Lagrangian relevant for the calculation of the radiative-pion-photoproduction amplitude
at the order we are working are given by
\begin{align}
\leff=\sum_{i=1}^2\lpipi^{(2i)}+\lgen^{(4)}_{\textrm{WZW}}+\sum_{j=1}^3\lpin^{(j)}+\sum_{k=1}^2\lpid^{(k)}+\sum_{l=1}^3\lpind^{(l)}
 \label{eq:effectiveL}\,,
\end{align}
where the superscripts denote the order of the corresponding term.
The building blocks of the Lagrangian are the pion matrix field entering via $U=u^2$,
\begin{align}
U=1+\frac{\ci}{\bareFpi}\paulimatrixvector\cdot\pionfieldvector-
\frac{1}{2\bareFpi^2}\pionfieldvector^2-
\alpha\frac{\ci}{\bareFpi^3}\pionfieldvector^2\paulimatrixvector\cdot\pionfieldvector+
\left(\alpha-\frac{1}{8}\right)\frac{1}{\bareFpi^4}\pionfieldvector^4+\order{\pionfield^5}\,,
\end{align}
where $\bareFpi$ is the pion decay constant in the chiral limit
and $\alpha$ is an arbitrary parameter that does not affect observables;
the nucleon isodoublet field $\nucleon$;
the $\deltapart$ isospin-$3/2$ Rarita-Schwinger-spinor field $\deltafield^{\mu}_{i}$, satisfying $\paulimatrix_i\deltafield^{\mu}_{i}=0$;
the vector source $ v_{\mu}=-e\nucleoncharge A_{\mu}=-e\frac{\identity+\paulimatrix_3}{2}A_{\mu} $, 
with the electric charge $ e\approx\num{0.303} $ and the electromagnetic field $ A_{\mu} $;
and the axial source $a_{\mu}$.

The covariant derivatives are defined as
\begin{align}
\nabla_{\mu}U&=\partial_{\mu}U-\ci r_{\mu}U+\ci U l_{\mu}\,,\nonumber\\
\covdc_{\mu}\nucleon&=\left(\partial_{\mu}+\varGamma_{\mu}\right)\nucleon\,, \nonumber\\
\covdc^{\mu}_{ij}\deltafield^{\nu}_{j}&=\left(\partial^{\mu}+\varGamma^{\mu}\right)\deltafield^{\nu}_{i}
-\ci\levicivita_{ijk}\trfl{\paulimatrix_k\varGamma^{\mu}}\deltafield^{\nu}_{j}
                                        \,,
\end{align}
with 
\begin{equation}                                     
 l_{\mu}=v_{\mu}-a_{\mu}\,,\quad  r_{\mu}=v_{\mu}+a_{\mu}\,,\quad
\varGamma_{\mu} = \frac{1}{2}\left[u^{\dagger}(\partial_{\mu}-\ci r_{\mu})u+u(\partial_{\mu}-\ci l_{\mu})u^{\dagger}\right]\,.
\end{equation}
We also introduce the quantities
\begin{equation}
u_{\mu}=\ci\left[u^{\dagger}(\partial_{\mu}-\ci r_{\mu})u-u(\partial_{\mu}-\ci l_{\mu})u^{\dagger}\right]\,,\quad
w^{\mu}_i=\frac{1}{2}\trfl{\paulimatrix_i u^{\mu}}\,,\quad
w^{\mu\nu}_i=\frac{1}{2}\trfl{\paulimatrix_i
  \left[\covdc^{\mu},u^{\nu}\right]}\,, \quad
\chi_{\pm} = u\adj\chi u\adj \pm u\chi\adj u\,,
  \end{equation}
where $\chi=\diag(M^2,M^2)$ and $\barepionmass$ is the pion mass to leading order in quark masses.
The field strength tensors are given  by
\begin{eqnarray}
F_{L}^{\mu\nu} &=&\partial^{\mu}
                   l^{\nu}-\partial^{\nu}l^{\mu}-\ci\comm{l^{\mu}}{l^{\nu}}\,,
                   \nonumber \\
F_{R}^{\mu\nu} &=&\partial^{\mu} r^{\nu}-\partial^{\nu}r^{\mu}-\ci\comm{r^{\mu}}{r^{\nu}}\,, \nonumber \\
F_{\mu\nu}^{\pm} &=& u F_{L,\mu\nu}u\adj\pm u\adj F_{R,\mu\nu}u\,,\nonumber \\
F^{\pm}_{i,\mu\nu}&=&\trfl{\paulimatrix_i F^{\pm}_{\mu\nu}} \,.
\end{eqnarray}

The pionic part of the effective Lagrangian reads \cite{Gasser:1983yg}
\begin{eqnarray}
\lpipi^{(2)}&=&\frac{\bareFpi^2}{4}\trfl{(\nabla_{\mu}U)^{\dagger} \nabla^{\mu}U}+\frac{\bareFpi^2}{4}\trfl{\chi_+}\,,\nonumber\\
\lpipi^{(4)} &=&\frac{l_3}{16}\trfl{\chi_+}^2 +\frac{l_4}{16}\left(2\trfl{\nabla_{\mu}U(\nabla^{\mu}U)^{\dagger}}\trfl{\chi_{+}}+\trfl{2((\chi \Udag)^2 + (U\chi^{\dagger})^2)-4\chi^{\dagger}\chi-\chi_-^2}\right)\nonumber\\
&+&
                                                                                                                                                                                                                                \frac{l_5}{2}\left(2\trfl{F_{R,\mu\nu}UF_{L}^{\mu\nu}\Udag}-\trfl{F_{L,\mu\nu}F_{L}^{\mu\nu}+F_{R,\mu\nu}F_{R}^{\mu\nu}}\right) \nonumber\\
                                                                                                                                                                                                                                &+&\frac{\ci l_6}{2}\trfl{F_{R,\mu\nu}\nabla^{\mu}U(\nabla^{\nu}U)^{\dagger} + F_{L,\mu\nu}(\nabla^{\mu}U)^{\dagger}\nabla^{\nu}U}+\dotsi.
 \label{eq:lpipi}
\end{eqnarray}
The term from the Wess-Zumino-Witten ({WZW}) Lagrangian relevant for our calculation has the form\footnote{We use the convention $ \levicivita^{0123}=1 $.} \cite{Wess:1971yu,Witten:1983tw}
\begin{align}
\lgen_{\textrm{WZW}}^{(4)}=\frac{e^2}{32\pimath^2\bareFpi}\levicivita^{\kappa\lambda\mu\nu}F_{\kappa\lambda}F_{\mu\nu}\pionfield_3\,,
\end{align}
where $ F_{\mu\nu} =\partial_{\mu}A_{\nu}-\partial_{\nu}A_{\mu}$.

The leading-order pion-nucleon Lagrangian reads
\begin{align}
\lpin^{(1)} = \bar{\nucleon}\left(\ci\slashed \covdc -\barenucleonmass +\frac{\baregA}{2}\slashed u \matrixgamma^5\right)\nucleon\,,
\end{align}
where $ \barenucleonmass$ is the bare nucleon mass and $ \baregA $ is
the bare axial coupling constant of the nucleon.
The second- and third-order pion-nucleon Lagrangian depends on the low-energy constants $c_i$ and $d_i$\footnote{In
 the definitions of the constants, the physical nucleon mass $\nucleonmass$ is used.}:
\begin{eqnarray}
\lpin^{(2)}&=&\bar{\nucleon}\left\{c_1\trfl{\chi_+}+\tensorsigma^{\mu\nu}\left[\frac{c_6}{8\nucleonmass}
F^+_{\mu\nu}+\frac{c_7}{8\nucleonmass}\trfl{F^+_{\mu\nu}}\right] \right\}\nucleon+\dotsi\,,\nonumber\\
\lpin^{(3)}&=&\bar{\nucleon}\bigg[-\ci\frac{d_8}{2\nucleonmass}\levicivita^{\mu\nu\alpha\beta}\trfl{\tilde{F}^+_{\mu\nu}u_{\alpha}}\covdc_{\beta}
-\ci\frac{d_9}{2\nucleonmass}\levicivita^{\mu\nu\alpha\beta}\trfl{F^+_{\mu\nu}}u_{\alpha}\covdc_{\beta}
-\ci\frac{d_{20}}{8\nucleonmass^2}\matrixgamma^{\mu}\matrixgamma^5\left[\tilde{F}^+_{\mu\nu},u_{\lambda}\right]\covdc^{\lambda\nu}\bigg]\nucleon
+\hc \nonumber\\
&+&\bar{\nucleon}\bigg[\frac{d_{16}}{2}\matrixgamma^{\mu}\matrixgamma^5\trfl{\chi_+}u_{\mu}+
                   \frac{d_{18}}{2}\ci\matrixgamma^{\mu}\matrixgamma^5\left[\covdc_{\mu},\chi_-\right]
                   +\frac{d_{21}}{2}\ci\matrixgamma^{\mu}\matrixgamma^5\left[\tilde{F}^+_{\mu\nu},u^{\nu}\right]+
\frac{d_{22}}{2}\matrixgamma^{\mu}\matrixgamma^5\left[\covdc^{\nu},F^-_{\mu\nu}\right]\bigg]\nucleon+\dotsi\,,
\end{eqnarray}
with $ \tensorsigma^{\mu\nu}= \frac{\ci}{2}\comm{\matrixgamma^{\mu}}{\matrixgamma^{\nu}} $ 
and $ \tilde{F}_{\mu\nu}^{\pm}=F_{\mu\nu}^{\pm}-\frac{1}{2}\trfl{F_{\mu\nu}^{\pm}} $, see Ref.~\cite{Fettes:2000gb}
for the full list of terms.

The relevant terms quadratic in the $\deltapart$ field are given by \cite{Hemmert:1997ye,Hemmert:1997wz}
\begin{align}
\lpid^{(1)} &=-\bar{\deltafield}_{i,\mu}\Big[ (\ci \slashed \covdc_{ij}-
\baredeltamass\kronecker_{ij})\minkowski^{\mu\nu}-\ci(\matrixgamma^{\mu}\covdc^{\nu}_{ij}+
\matrixgamma^{\nu}\covdc^{\mu}_{ij})+\ci\matrixgamma^{\mu}\slashed \covdc_{ij}\matrixgamma^{\nu}+
\baredeltamass\matrixgamma^{\mu}\matrixgamma^{\nu}\kronecker_{ij} + \frac{g_1}{2}\minkowski^{\mu\nu}\slashed u\matrixgamma^5\kronecker_{ij}\Big]\deltafield_{j,\nu}\,,\nonumber\\ 
\lpid^{(2)} &=
-\ci c_1^{\deltapart} \bar{\deltafield}_{i,\mu}  \trfl{\chi_+}\tensorsigma^{\mu\nu} 
\deltafield_{i,\nu}  +
\frac{c_6^{\deltapart}}{8\Re(\deltamass)} \bar{\deltafield}_{i,\mu}
 F^+_{\kappa\lambda} \tensorsigma^{\kappa\lambda} \deltafield_{i}^\mu + 
\frac{c_7^{\deltapart}}{8 \Re(\deltamass)} \bar{\deltafield}_{i,\mu} 
\trfl{F^+_{\kappa\lambda}} \tensorsigma^{\kappa\lambda} \deltafield_{i}^\mu+\dotsi\,,
\end{align}
where $\baredeltamass$ stands for the bare $\deltapart$ mass and
$\deltamass$ for the physical $\deltapart$ mass (complex pole mass).
The terms in $\lpid^{(2)}$ are modified as compared to Ref.~\cite{Hemmert:1997wz} 
in analogy with the pion-nucleon Lagrangian. However, they are equivalent up to 
a $\deltapart$-field redefinition.

We also need the following terms from the  $\deltapart$-to-nucleon transition Lagrangian \cite{Hemmert:1997wz,Zoeller:2016z}
\begin{eqnarray}
\lpind^{(1)}&=&\barehA\left(\bar{\deltafield}_{i,\mu}w^{\mu}_i \nucleon
+\bar{\nucleon}w^{\mu}_i \deltafield_{i,\mu} \right), \nonumber \\
\lpind^{(2)} &=& \ci\frac{b_1}{2}\bar{\deltafield}_{i}^\mu
 F^{+}_{i,\mu\kappa}\matrixgamma^{\kappa}\matrixgamma^5\nucleon   
+ \ci b_3 \bar{\deltafield}_{i}^\mu   w_{i,\mu\kappa} \matrixgamma^{\kappa} \nucleon
- \frac{b_6}{\nucleonmass} \bar{\deltafield}_{i}^\mu  w_{i,\mu\kappa} \covdc^{\kappa} \nucleon 
  +\hc +\dotsi \,, \nonumber \\
\lpind^{(3)} &=& \frac{h_1}{\nucleonmass} \bar{\deltafield}_{i}^\mu 
 F^+_{i,\mu\kappa} \matrixgamma^5\covdc^{\kappa} \nucleon - \ci\frac{h_{15}}{2}  \bar{\deltafield}_{i}^\mu 
 \trfl{\comm{\covdc_{\kappa}}{F^+_{\mu\lambda}}\paulimatrix^{i}} \tensorsigma^{\kappa\lambda}\matrixgamma^5\nucleon
+ \ci\frac{h_{16}}{2\nucleonmass}  \bar{\deltafield}_{i}^\mu
 \trfl{\comm{\covdc_{\kappa}}{F^+_{\mu\lambda}}\paulimatrix^{i}}
                 \matrixgamma^{\lambda}\matrixgamma^5\covdc^{\kappa}\nucleon
                 \nonumber \\
  &+&\hc+ \dotsi\;.
\end{eqnarray}
Note that all redundant off-shell parameters in $\lpind$ and $\lpid$ 
are set to zero as they have no observable effects, see Refs.~\cite{Tang:1996sq,Krebs:2009bf}.

The renormalization of the low-energy constants (LECs) appearing in the effective Lagrangian 
as well as the relations between the bare and renormalized constants are discussed in Sec.~\ref{sec:renormalization}.

\section{Power counting} \label{sec:PowerCounting}
\subsection{Small scale expansion}\label{sec:SSE}
In chiral perturbation theory, the perturbative expansion of the amplitude in small parameters is organized according
to a certain power counting. We start with considering the power
counting in the pion-nucleon threshold region, i.e.~with the external
particle momenta being $|\vec q \, |\sim M_\pion$, and then discuss its modification in the $\deltapart$ region.
In this work, we employ the so-called small scale expansion scheme ($ \epsilon $ scheme), which treats the 
$\deltapart$-nucleon mass difference $ \deltasplit=\deltamass-\nucleonmass\approx\SI{300}{\MeV} $ as being of order $O(M_\pion)$.
Therefore, the expansion is performed in the parameter
\begin{align}
\epsilon\in\left\{\frac{q}{\varLambda_b},\frac{\pionmass}{\varLambda_b},\frac{\deltasplit}{\varLambda_b}\right\},\quad \varLambda_b\in\left\{M_\rho,4\pimath\piondecayconstant,\nucleonmass\right\},
\end{align}
where the mass of the $\rho$-meson  $M_\rho$, the nucleon mass
$\nucleonmass$ and the scale $4\pimath\piondecayconstant$  emerging from pion loops 
are regarded as hard scales.
The small scale expansion was introduced for heavy-baryon {$\chi$PT} in Ref.~\cite{Hemmert:1997ye}.
In the covariant formulation of {$\chi$PT} it was applied e.g.~to pion-nucleon scattering \cite{Yao:2016vbz}
and to nucleon Compton scattering \cite{Bernard:2012hb}.

Since both the $\deltapart$ and the nucleon propagators count as $O(\epsilon^{-1})$, 
the order $D$ of any Feynman diagram can be computed according to the formula \cite{Weinberg:1991um}
\begin{align}
D=1+2L+\sum_{n}(2n-2)V_{2n}^M+\sum_d(d-1)V_d^B, \label{eq:PC2}
\end{align}
where  $ L $ is the number of loops, $ V^M_{2n} $ is the number of purely mesonic vertices of order $ 2n $ and $ V^B_d $ is the number of vertices 
involving baryons of order $d$. 
We label purely pion-nucleon contributions as $ q^D $ and those containing $\deltapart$ lines as $ \epsilon^D $.
All Feynman diagrams relevant for the present study are collected in Appendix~\ref{sec:Feynman-diagrams}
for radiative pion photoproduction and in Appendix~\ref{sec:Feynman-diagrams_photoproduction} for ordinary pion photoproduction. 
The latter reaction is used to determine several LECs serving as input for our calculation.

The leading tree-level contributions appear at order $O(q)$. However, for radiative neutral-pion photoproduction as well as for 
ordinary neutral-pion photoproduction, such diagrams are suppressed as $1/m_N$ 
(in particular, because the diagrams involving photon-pion couplings  vanish).

Tree-level graphs involving $\deltapart$ lines start to contribute at order $\epsilon^2$ for both reactions. 
However, for radiative pion photoproduction in the neutral channel, they are also $1/m_N$-suppressed and
thus shifted to order $\epsilon^3$. 

For the process $\gamma N\to\gamma N\pion$, diagrams with the emitted photon coupled to the outgoing or the incoming nucleon, 
are enhanced in case of ultrasoft photons ($\ego\ll\pionmass$).
The $\gamma N\to\gamma N\pion$ amplitude in this regime is determined by the photoproduction amplitude \cite{Chiang:2004pw}
according to the soft-photon theorem \cite{Low:1958sn}.
This theorem is satisfied automatically in our scheme since the amplitude satisfies gauge invariance. That is why we do not modify the power counting for 
this small part of the phase space.

Loop diagrams first appear at order $O(q^3)$ for the purely nucleonic
graphs and $\epsilon^3$ diagrams involving $\deltapart$ lines.
Notice that not all $\epsilon^3$ diagrams are taken into account in our study as will be discussed in detail in the next subsections.

In what follows, we denote by $q^i$ ($\epsilon^i$) the order $i$ of a
diagram that follows directly from  Eq.~\ref{eq:PC2} without taking
into account possible additional enhancements or suppressions in
specific kinematical regions.

\subsection{Power counting in the $\deltapart$ region for radiative pion photoproduction} \label{sec:PowerCountingInTheDeltaRegion}
The main goal of our study is to investigate the electromagnetic properties of the $\deltapart$ resonance.
Therefore, the energy region of interest is the $\deltapart$ region, i.e.~$\sqrt{s}\approx \deltamass$.
In this energy regime, there are two sources of enhancement (suppression) for some of the contributions.
The first one is due to a numerically large value of the initial particle momenta
($\egi\sim\Delta$) as compared to the threshold
kinematics and, especially, to the final particle momenta. This makes
the $1/\nucleonmass$ suppression of the $\epsilon^2$ tree-level
diagrams of Fig.~\ref{fig:deltadiagslo} 
to be of order $\Delta / \nucleonmass \sim1/3$
in contrast to the ``genuine'' $\epsilon^3$ diagrams of Fig.~\ref{fig:deltaTreeE3},
which are suppressed by a factor $\ego / \varLambda_b$ coming from the second order vertices 
from $\lpin^{(2)}$, $\lpind^{(2)} $ and $\lpid^{(2)}$ containing $F_{\mu\nu}$.
On the other hand, for the same reason, some of the nucleon
propagators (at least in the tree diagrams) are less enhanced as
compared to the threshold region, 
i.e. they are of order $1/\Delta$ rather than $1/\pionmass$ (even
though they are not distinguished in the $\epsilon$ counting). 

The second and main source of enhancement are the $s$-channel
$\deltapart$ propagators in the $1\deltapart$-reducible graphs. 
Those propagators that have a pole in the $s$-variable are enhanced by
a factor\footnote{We employ
  the complex-mass scheme, see Sec.~\ref{sec:renormalization} for details.}
 \begin{align}
\gamma=\frac{\Delta}{|\Im(\deltamass)|}\,,
 \end{align}
as compared to the threshold region (the maximal enhancement is obtained for the energy $\sqrt{s}=\Re(\deltamass)$).
Formally, the imaginary part of the $\deltapart$ pole-mass (or the
$\deltapart$ half-width), being a one-loop effect, is
of order $O(q^3)$ and, naively, we must regard $\gamma\sim\epsilon^{-2}$
and promote many contributions from orders (in the threshold power counting) $\epsilon^3$, $\epsilon^4$ and even higher.
However, numerically ($\gamma\approx 6$), this estimate is not justified
and $\gamma$ is rather of order $O(\epsilon^{-1})$ due to a large value of the $\pion N\Delta$ coupling constant.
The enhancement with respect to the nucleon propagator (counted as $O(q^{-1})$) is even smaller: $\frac{\pionmass}{|\Im(\deltamass)|}\approx 3$.
Therefore, we simply keep such factors of $\gamma$ in the following
analysis when promoting the formally higher-order diagrams
mentioned above.
There are also $\deltapart$ propagators with a pole in the $s_2$-variable, i.e.~those which couple to the $\pion N$-system in the final state.
Such propagators are enhanced by a factor 
\begin{align}
\tilde\gamma\approx\frac{\Delta}{|\sqrt{s}-\ego-\deltamass|}\,. 
\end{align}
The enhancement is maximal ($\tilde\gamma\approx 6$) in the part of the phase space where $\ego\approx \sqrt{s}-\Re(\deltamass)$. 
For the energy closest to the $\deltapart$ pole $\sqrt{s}=\Re(\deltamass)$,
this enhancement affects only the phase-space region of very soft emitted photons.

In our study, we concentrate predominantly on the neutral channel, i.e.~on the reaction $\gamma p\to\gamma p\pion^0$.
In this case, motivated by the above-mentioned modifications in the $\deltapart$ region,
we attribute various contributions to leading and next-to-leading orders,
to which we assign effective orders $\epsilon^2_\text{eff}$ and $\epsilon^3_\text{eff}$, respectively,
according to the following power counting rules: 

\begin{itemize}
\item Leading order ($\epsilon^2_\text{eff}$):
\begin{itemize}
 \item[--] Nucleonic order-$q^1$ tree-level diagrams in Fig.~\ref{fig:basictrees}, which are $1/\nucleonmass$ suppressed for the neutral-pion channel.
 \item[--] Tree-level diagrams with $\deltapart$ lines of order
   $O(\epsilon^2)$ shown in Fig.~\ref{fig:deltadiagslo}. 
While they are also suppressed by a factor $\sim
\Delta /\nucleonmass$,
certain diagrams within this set appear to be enhanced
by the factor of $\gamma$ such as graphs (b),~(c)\footnote{We mention
  only diagrams that yield non-vanishing contributions for the neutral channel.} and/or by
a factor of $\tilde\gamma$ such as diagrams (a),~(b).
Therefore, taking into account a numerically rather weak $1/m_N$-suppression and a sizable enhancement due to 
the factors of $\gamma$ and $\tilde\gamma$, we expect  these
diagrams to be, at least, not less important than the above-mentioned
$q^1$ ones. 
In fact, numerically, they do provide the dominant contribution due to
a large  value of 
the $\gamma N \Delta$-coupling $\bar{b}_1\approx 6$
$\nucleonmass^{-1}$.
\end{itemize} 
Note that we consider all subsets of diagrams containing the same vertices (and therefore the same combinations of LECs)
together even if only some of them are subject of a certain enhancement (suppression)
in order to ensure that gauge and chiral symmetries are not violated.
\item
  Next-to-leading order ($\epsilon^3_\text{eff}$):
\begin{itemize}
 \item[--] Nucleonic order-$q^2$ tree-level diagrams in 
   Fig.~\ref{fig:qTo2Diagrams}, which are $1/\nucleonmass$
   suppressed.
 \item[--] Nucleonic order-$q^3$ tree-level diagrams in
   Figs.~\ref{fig:qTo3Diagrams} and \ref{fig:highertrees}.
 \item[--] Pion-nucleon loop diagrams of order $O(q^3)$ in Figs.~\ref{fig:Aloops}-\ref{fig:Ex2loops}.
 \item[--] Tree-level order-$\epsilon^3$ diagrams with $\deltapart$
   lines, including the diagrams in Fig.~\ref{fig:deltaTreeE3} (a)-(i)
   proportional to 
the $\deltapart$ magnetic moment vertex.
This set contains the $s$-channel pole diagrams,
Fig.~\ref{fig:deltaTreeE3} (a), (b), (e), (h), and is enhanced by a
factor of $\gamma$. 
However, in contrast to the second set of the leading-order diagrams,
these enhanced diagrams are proportional to $\ego/\Lambda_b$
due to insertions of vertices from $\lpin^{(2)}$, $\lpind^{(2)} $ and $\lpid^{(2)}$ containing $F_{\mu\nu}$
(and not to $\egi/\nucleonmass$), which leads to a numerically 
smaller total enhancement factor. 
For the same reason,
there is no additional $\tilde\gamma$-enhancement for the diagram
depicted in Fig.~\ref{fig:deltaTreeE3} (h)
close to the $\deltapart$ pole ($\sqrt{s}=\deltamass$).
A moderate enhancement due to a factor $\tilde\gamma$ takes place only at the energies above the $\deltapart$ pole, 
which we also include in the analysis (see Sec.~\ref{sec:results}).
From dimensional arguments, this set of diagrams falls somewhere
between the leading-order terms and other next-to-leading order 
terms. Numerically, they turn out to yield rather small contributions comparable with other next-to-leading order terms. 
Nevertheless, we checked that promoting them to leading order has very
little effect on our results
including the truncation error estimation.
 \item[--] Tree-level order-$\epsilon^3$ diagrams with $\deltapart$ lines
   containing the subleading $\gamma N\Delta$ vertex from
   $\lpind^{(3)}$, see Fig.~\ref{fig:deltaTreeE3} (j)-(s).
This set of diagrams is completely analogous to the second set of the
leading-order diagrams, the only difference being the replacement of  
the second-order $\gamma N\Delta$ vertex by the third-order one 
leading to an extra factor of $\egi/\nucleonmass\sim
\Delta/\nucleonmass$. 
The same kinds of suppression and enhancement as in the case of
the corresponding leading-order diagrams apply also here. 
 \item Loop corrections of order $O(\epsilon^3)$ to the $\deltapart$ pole graphs. 
They include corrections to the $\gamma N\Delta$ vertex in the previous set of diagrams (Fig.~\ref{fig:DeltaLoops1}),
which are subject to the same types of enhancement and suppression.
The loop diagrams shown in Figs.~\ref{fig:DeltaLoops2} form a gauge-invariant set 
together with the ones from Fig.~\ref{fig:DeltaLoops1} and are also taken into account.
Given the rather narrow energy domain we consider, the real parts of the loops in Fig.~\ref{fig:DeltaLoops1}
merely renormalize the $\gamma N\Delta$ LECs $b_1$ and $h_1$. We have
explicitly verified this feature numerically by switching them on and off.
Thus, only the imaginary parts of such loops yield non-trivial contributions.
This is also the reason why we do not include the analogous but technically more complicated 
loop diagrams with $\deltapart$ lines inside the loops which generate
no imaginary parts in the considered energy region.\footnote{The same
  argument applies to the $\deltapart$ loop corrections to
  $\deltamass$, $g_A$, $\bar{c}_6$, $\bar{c}_7$.} 
 \item[--] Tree-level diagrams of order $\epsilon^3$ with a
   $\deltapart$ line and one insertion of the Wess-Zumino-Witten anomalous
$\pion^0\gamma\gamma$ vertex are taken into account as they are enhanced by
a factor of $\tilde\gamma$ (but only for energies above  
the exact $\deltapart$ region since the amplitude is proportional to
$\ego$), see Fig.~\ref{fig:deltaTreeE3} (t)-(u).
Moreover, for forward angles of the emitted photon, we have $t\approx 0$ and the pion propagator is maximally enhanced.
The numerical effect of such diagrams turns out to be insignificant, which can justify 
neglecting further $\epsilon^3$ diagrams that are not subject to the $\gamma$-enhancement.
In particular, other $\epsilon^3$ loop diagrams that are not enhanced
by a factor of $\gamma$ are not included in our analysis. 
\end{itemize}
As already pointed out above, we do not take into account order-$\epsilon^3$
diagrams with  $\deltapart$ lines inside loops.
Calculating such diagrams requires evaluating high-rank-tensor loop
integrals (up-to 5-point functions), which is computationally extremely demanding.
Our expectation that such contributions are not important at the order
we are working is based, apart from the arguments given above, 
on the relative numerical insignificance of the pion-nucleon order-$q^3$ loop
diagrams, see Sec.~\ref{sec:results} for details, and the additional suppression 
of the $\deltapart$ propagators (inside loops) compared to the nucleon propagators.
\end{itemize}

\begin{figure}[tb]
\includegraphics[width=0.25\textwidth]{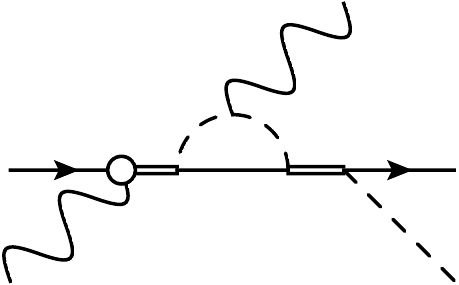}
\caption{An example of a loop diagram included in Ref.~\cite{Pascalutsa:2007wb} and
neglected in the present work. All vertices are from the leading order
Lagrangians $\lpipi^{(2)}$, $\lpin^{(1)}$ and $\lpind^{(1)}$,
except the open circle that denotes the subleading vertex from $\lpind^{(2)}$.}
\label{fig:LoopExample}
\end{figure}

For radiative charged-pion photoproduction we apply the power counting analogous to the neutral channel.
The main difference is the absence of the $1/\nucleonmass$ suppression for the sets of 
order-$O(q^1)$, $O(q^2)$, $O(\epsilon^2)$, $O(\epsilon^3)$ diagrams
due to the possibility for photons to couple directly to pions.  
Therefore, we obtain the following sets of leading ($\epsilon_\text{eff}$), next-to-leading ($\epsilon^2_\text{eff}$), 
and next-to-next-to-leading ($\epsilon^3_\text{eff}$) order diagrams
based on the enhancement arguments discussed above.
\begin{itemize}
  \item
Leading order ($\epsilon_\text{eff}$):
\begin{itemize}
 \item[--] Nucleonic order-$q^1$ tree-level diagrams, see Fig.~\ref{fig:basictrees}.
 \item[--] Tree-level diagrams with $\deltapart$ lines of order
   $O(\epsilon^2)$ enhanced by a factor of $\gamma$, see Fig.~\ref{fig:deltadiagslo}. 
\end{itemize}
\item
Next-to-leading order ($\epsilon^2_\text{eff}$):
\begin{itemize}
 \item[--] Nucleonic $q^2$-tree-level diagrams, see Fig.~\ref{fig:qTo2Diagrams}.
 \item[--] Tree-level $\epsilon^3$ diagrams with $\deltapart$ lines
   involving one insertion of the subleading $\gamma N\Delta$ vertex from $\lpind^{(3)}$ 
 enhanced by a factor of $\gamma$, see Fig.~\ref{fig:deltaTreeE3} (j)-(s).
 \item Loop corrections of order $O(\epsilon^3)$ to the
   $\deltapart$-pole graphs, see Figs.~\ref{fig:DeltaLoops1}-\ref{fig:DeltaLoops3}. 
\end{itemize}
\item
Next-to-next-to-leading order ($\epsilon^3_\text{eff}$):
\begin{itemize}
 \item[--] Nucleonic order-$q^3$ tree-level diagrams, see Figs.~\ref{fig:qTo3Diagrams},~\ref{fig:highertrees}.
 \item[--] Pion-nucleon loop diagrams of order $O(q^3)$, see Figs.~\ref{fig:Aloops}-\ref{fig:Ex2loops}.
 \item[--] Tree-level order-$\epsilon^3$ diagrams with $\deltapart$ lines, including the diagrams proportional to 
the $\deltapart$ magnetic moment vertex, see
Fig.~\ref{fig:deltaTreeE3} (a)-(i).
Note that for the reaction $ \photon\proton\to\photon\neutron\pion^+
$, the diagram in Fig.~\ref{fig:deltaTreeE3} (i) is proportional to the 
$\Delta^0$ magnetic moment. However, this particular diagram contains
no enhancement factors like $\gamma$ and,
therefore, the sensitivity of the results to $\deltamdmind{0}$ is expected to be very weak. 
 \item[--] Tree-level diagrams of order $\epsilon^3$ with the
   $\deltapart$ line and the Wess-Zumino-Witten anomalous
$\pion^0\gamma\gamma$ vertex, see Fig.~\ref{fig:deltaTreeE3} (t)-(u).
\end{itemize}
\end{itemize}
Finally, we would like to underline the key differences between our scheme and the
$\delta$-counting approach of Ref.~\cite{Pascalutsa:2007wb}, which we use for comparison.
In the $\delta$-counting scheme, the single and multiple $\deltapart$-pole graphs 
are stronger promoted as the $\deltapart$ half-width is regarded 
as being of order $O(q^3)\sim O(\delta^3)$.
Therefore, one includes 
the pion-nucleon
loop corrections to the $\deltapart$-pole graphs 
(including the $\deltapart$-magnetic-moment contribution) such as the one shown in Fig.\ref{fig:LoopExample},
which appear at higher order in our scheme (numerically, we found them
indeed being small).
On the other hand, at next-to-leading order in the $\delta$-scheme considered in Ref.~\cite{Pascalutsa:2007wb},
the purely nucleonic order-$q^3$ loop and tree-level graphs (Figs.~\ref{fig:qTo3Diagrams}-\ref{fig:Ex2loops}) are not included (being of higher order).
However, our analysis shows that these contributions are important in order to achieve 
a better agreement with experimental data.
The authors of Ref.~\cite{Pascalutsa:2007wb} also perform the expansion of the amplitude in energy of 
the emitted photon $\ego$, which converges slowly apart from the
region of very soft photons.

\subsection{Power counting for pion photoproduction} \label{sec:PowerCountingInTheDeltaRegionPhotoproduction}
In order to determine several LECs needed as an input for radiative
pion photoproduction, 
we consider ordinary pion photoproduction from the threshold region to 
the lower $\deltapart$ energy region, see
Sec.~\ref{sec:photoproduction} for details.
We apply the power counting scheme consistent with the one used for
radiative pion photoproduction and
described in Sec.~\ref{sec:PowerCountingInTheDeltaRegion}.
In particular, we take into account all tree-level and loop contributions up to order $q^3$.
We also include the order-$\epsilon^2$ and $\epsilon^3$ tree-level
diagrams enhanced in the vicinity of the $\deltapart$ pole 
with the leading and subleading $\gamma N \Delta$-vertices as well as the loop corrections to them, 
see Appendix~\ref{sec:Feynman-diagrams_photoproduction} for the
whole set of considered diagrams.
Analogously to radiative pion photoproduction, we neglect diagrams
involving loops with $\deltapart$ lines inside.

Since we analyze simultaneously both the threshold region and the
$\deltapart$ region, we choose to 
assign to each diagram an order that follows from the standard
threshold $\epsilon$-counting, see Eq.~(\ref{eq:PC2}),
when estimating the truncation uncertainty.
One could, in principle, introduce a different power counting for different energy regions in order to take into account
the enhancement of the $\deltapart$-pole graphs.
However, this would lead to unnecessary complications without
significantly affecting the results.
We have verified this explicitly by promoting the $\deltapart$-pole graphs one order lower
for the multipoles coupled to the $\deltapart$ in the $s$-channel.

Note further that the fits to the photoproduction multipoles are
performed in the isospin
basis, which corresponds to a linear combination of the neutral and charged
channels. 
We, therefore, do not introduce any special treatment for the neutral channels,
where certain $1/\nucleonmass$-suppressions appear.

\section{Renormalization} \label{sec:renormalization}
In this section we describe the
renormalization of the low-energy constants 
and the relations between the bare parameters of the effective
Lagrangian in Eq.~(\ref{eq:effectiveL}), their renormalized values
and physical quantities.

The loop integrals appearing in our calculation contain ultraviolet divergencies, which we 
handle by means of dimensional regularization. Divergent parts of the integrals are cancelled by the counter terms 
entering the bare parameters of the Lagrangian so that the resulting amplitude is
expressed in terms of the finite renormalized LECs, physical masses and coupling constants.
Due to the presence of an extra hard scale  corresponding to the nucleon or $\deltapart$ mass, 
baryonic loops also generate power-counting-violating terms, i.e.~terms of a lower order 
as compared to the estimation based on dimensional power counting in Eq.~(\ref{eq:PC2}) \cite{Gasser:1987rb}.
These terms are local and can be absorbed by a redefinition of the LECs of the effective Lagrangian at lower orders.
Such a procedure is realized in a systematic way in the extended on-mass-shell scheme (EOMS) \cite{Fuchs:2003qc}.
However, for radiative (ordinary) pion photoproduction, there are
no contact interactions at order lower than $q^5$ ($q^3$)  
except for the ones fixed by symmetries.
Therefore, no power-counting-breaking terms appear at the order we are working, and the EOMS scheme 
is essentially equivalent to the $\widetilde{\mathrm MS}$
\cite{Fuchs:2003qc,Gasser:1983yg} scheme for the case at hand.

For the masses and wave functions as well as for the
$\pion NN$, $\pion N\Delta$, $\gamma N N$, $\gamma N \Delta$, 
$\gamma \Delta \Delta$ coupling constants we impose the on-shell renormalization conditions
as this is more appropriate for calculating physical on-shell amplitudes.
Notice that in the course of renormalization, when calculating the matrix elements of subprocesses
with two, three and four external lines, the same sets of diagrams as in the corresponding subgraphs
in the radiative-pion-photoproduction diagrams are taken into account.

In the subsections below, we provide the renormalization conditions for all relevant LECs.
The explicit expressions for the counter terms are given in Appendix~\ref{sec:counterrerms}.
\subsection{Pion mass, field and decay constant}
For the pion field, the renormalization conditions 
\begin{align}
\varSigma_{\pion}(\pionmass^2)=0,\quad
\varSigma_{\pion}'(\pionmass^2)=0\qq{and}\bra{0}A_i^{\mu}(0)\ket{\pion_j(q)}=\ci q^{\mu}\kronecker_{ij}\piondecayconstant\,,
\end{align}
with $\varSigma_{\pion}(q^2)$ and $ A_i^{\mu} $ being the pion
self-energy and the axial current, respectively, 
relate the constants $\barepionmass$, $ \pionZfactor $, $\bareFpi$, $ l_3 $, $ l_4 $  
to the physical
quantities $ \pionmass $, $ \piondecayconstant $ and the $Z$-factor $ \pionZfactor $. 
The fourth-order LECs $ l_3 $ and $ l_4 $ do not explicitly enter the amplitude of the considered processes at the order
we are working after renormalization.
\subsection{Nucleon mass and field}
For the nucleon, analogous renormalization conditions
\begin{align}
\varSigma_{\nucleon}(\nucleonmass)=0\qq{and}
\varSigma_{\nucleon}'(\nucleonmass)=0,
\end{align}
with $\varSigma_{\nucleon}(\slashed p)$ being the nucleon self-energy, 
fix the constants $ \barenucleonmass $ and $ \nucleonZfactor $. 
In turn, an explicit dependence of the radiative (ordinary) pion-photoproduction amplitude on $ c_1 $ disappears after renormalization.
\subsection{$\deltapart$-resonance mass and field}
We implement the complex-mass scheme \cite{Denner:1999gp,Denner:2006ic} for the $\deltapart$ resonance and take into account its width explicitly.
Within this scheme, loop corrections to the self-energy of the $\deltapart$ turn out to contribute beyond
the order we are working. This holds also in the $\deltapart$ region.
The renormalization condition at the $\deltapart$ pole ($\deltamass=\Re(\deltamass)-\ci|\Im(\deltamass)|$) reads 
\begin{align}
\varSigma_{\deltapart}\left(\deltamass\right)=0 \quad\textrm{and}\quad
\varSigma_{\deltapart}'\left(\deltamass\right)=0.
\end{align}
As in the case of the nucleon, the constant $ c_1^{\deltapart} $
does not appear explicitly in our calculation after renormalization.
%

\subsection{The $\boldmath \pion NN$ coupling constant}
Since the nucleon axial coupling constant $g_A$ enters our calculation only through the $\pion NN$ 
vertex, it is natural to use the renormalization condition for the
pseudoscalar coupling $g_{\pion\nucleon\nucleon}$
defined as the $\pion NN$ vertex function for all three particles being on mass shell:
\begin{align}
\bar{u}(p')\Gamma_i(p,p',q=p'-p)u(p)= g_{\pion\nucleon\nucleon}\paulimatrix_i 
\bar{u}(p')\matrixgamma^5 u(p)\,, \ q^2=\pionmass^2 \label{eq:gAphys}.
\end{align}
This condition relates the constants $\baregA$, $d_{18}$, $d_{16}$ to the physical quantity $g_{\pion\nucleon\nucleon}$.
We, therefore, follow the common procedure in the single- and
few-nucleon sectors of chiral EFT, see e.g.~\cite{Siemens:2016hdi,Siemens:2016jwj,Epelbaum:2014efa,Epelbaum:2014sza,Reinert:2017usi}, and introduce the effective 
axial coupling $ \axialcoupling $ defined via the Goldberger-Treiman relation \cite{Goldberger:1958tr}
\begin{align}
\axialcoupling=\frac{\piondecayconstant}{\nucleonmass}g_{\pion\nucleon\nucleon}\,,
\end{align}
which differs from the physical nucleon axial coupling, defined as
the matrix element of the axial current, by higher order contributions
that give rise to the Goldberger-Treiman discrepancy.
This ensures that the amplitudes relevant for the present work do not
explicitly depend on
the constants $ d_{16} $ and $ d_{18} $ anymore.
\subsection{Electromagnetic form factors of the nucleon}\label{sec:nucleon_FF}
For the renormalization of the $ \photon\nucleon\nucleon $ vertex, we consider the matrix element 
of the electromagnetic current $ J^{\mu} $ between the 1-nucleon states:
\begin{align}
\bra{N(p')}J^{\mu}(0)\ket{N(p)}=\bar{u}(p')\left(\matrixgamma^{\mu} F
_1(Q^2) + \frac{\ci\tensorsigma^{\mu\nu}k_{\nu}}{2\nucleonmass} F
_2(Q^2)\right)u(p)\,,
\end{align}
where $ Q^2=-(p-p')^2 $ and the functions $F_1(Q^2)$ and $F_2(Q^2)$ are the Dirac and Pauli form factors
of the nucleon, respectively. 
The renormalized constants $\bar{c}_6$ and $\bar{c}_7$ are related to the nucleon magnetic moment
and defined by the relations
\begin{align}
 \bar{c}_6=F_2^p(0)-F_2^n(0)\,,\
\bar{c}_7=F_2^n(0)\,,
\end{align}
where the superscript $p$($n$) stands for the proton (neutron).

\subsection{The $\boldmath \pion N\Delta$ coupling
constant}
In a complete analogy with  $ \axialcoupling $, we define the effective 
axial nucleon-to-$\deltapart$ transition coupling constant $ \pindcoupling $ 
through the corresponding $g_{\pion \nucleon\Delta}$  coupling
$g_{\pion \nucleon\Delta}\equiv g_{\pion \nucleon\Delta}(\pionmass^2)$
as
\begin{align}
\pindcoupling=\piondecayconstant\Re(g_{\pion\nucleon\Delta})\,,
\label{eq:piNDelta_coupling}
\end{align}
where the form factor $g_{\pion \nucleon\Delta}(q^2)$ is defined in
terms of the $\pion \nucleon\Delta$ vertex function \cite{Ellis:1997kc}:
\begin{align}
\bar{u}(p')\Gamma_{ij}^\mu(p,p',q=p'-p)u^{\deltapart}_{j;\mu}(p)= 
g_{\pion\nucleon\nucleon}(q^2) q^{\mu}\bar{u}(p') u^{\deltapart}_{i;\mu}(p). 
\end{align}
Here, the momentum of the $\deltapart$ resonance is taken at the pole:
$p^2=\deltamass^2$ \cite{Gegelia:2009py}.
Note that taking the real part in the definition in Eq.~(\ref{eq:piNDelta_coupling})
is not necessary at the order we are working since there are no  
loop corrections to $\pindcoupling$.
Eq.~(\ref{eq:piNDelta_coupling}) relates the bare constants $h$, $b_3$,  $b_6$ with $\pindcoupling$
and allows one to get rid of the redundant constants $b_3$ and  $b_6$
by a redefinition of $h_A$. There remains a residual contribution of $b_3$ and  $b_6$ to 
the (radiative) pion-photoproduction amplitude coming from the non-pole
parts of tree graphs involving  $\deltapart$ lines,
which can, up to terms of a higher order, be absorbed by the shifts in $d_i$'s given in Sec.~\ref{sec:other_LECs}.
Therefore, one can safely set $b_3=0$ and $b_6=0$.

\subsection{The $\boldmath \pion \Delta\Delta$
  coupling constant}
In our calculation, the $\pion\deltapart\deltapart$ vertex appears only at its leading order.
Therefore, the $\pion\deltapart\deltapart$ coupling $g_1$ does not get renormalized.

\subsection{Electromagnetic $N\deltapart$ transition form factors}
The electromagnetic $N\deltapart$ transition matrix elements can be parameterized using three form factors 
$ G_i $ e.g.~via \cite{Pascalutsa:2005vq}:
\begin{equation}
\mel{\deltapart(p')}{J^{\mu}(0)}{\nucleon(p)}
=-\sqrt{\frac{2}{3}}\bar{u}^{\deltapart}_{\nu}(p')
\Big(\frac{\matrixgamma^{\mu}k^{\nu}-\slashed k \minkowski^{\mu\nu}}{2}G_1(Q^2) +\frac{k\cdot p'\minkowski^{\mu\nu} - k^{\nu}p'^{\mu}}{\nucleonmass}G_2(Q^2) +\frac{k^{\mu}k^{\nu} - k^2\minkowski^{\mu\nu}}{\nucleonmass}G_3(Q^2) \Big)\ci\matrixgamma^5u(p)\,, \label{eq:EMTFFbasis}
\end{equation}
where $ Q^2=-(p-p')^2 $ and the isospin indices are suppressed, see also Ref.~\cite{Agadjanov:2014kha}
for a discussion of subtleties related with matrix elements of unstable particles.
We impose the renormalization conditions
\begin{align}
 \bar{b}_1= \Re[G_1(0)] \,,\quad \bar{h}_1=\Re[G_2(0)] \,.\label{eq:renormalization_b1_h1}
\end{align}
The $\deltapart$ momentum is taken at the pole: $(p')^2=\deltamass^2$.
Notice that the first contributions of contact terms to $ G_3 $ appear at order $\epsilon^4$.
Eq.~(\ref{eq:renormalization_b1_h1}) relates the bare constants $b_1$, $h_1$,  $h_{15}$, $h_{16}$ with $\bar{b}_1$
and $\bar{h}_1$ and allows one to get rid of the redundant constants $h_{15}$ and $h_{16}$.
The residual contributions of the LECs $h_{15}$ and $h_{16}$ to 
(radiative) pion-photoproduction amplitude coming from the non-pole
parts of tree graphs involving $\deltapart$ lines 
can, up to terms of a higher order, be absorbed by the shifts in $d_i$'s given in Sec.~\ref{sec:other_LECs}.
Therefore, we set $h_{15}=0$ and $h_{16}=0$.

Notice that in the literature, one also finds another convention for
the $\gamma N \Delta$ terms in the effective Lagrangian in terms of the couplings
$g_M$ and $g_E$, see
e.g.~Refs.~\cite{Pascalutsa:2007wb,Blin:2014rpa}. For the sake of
completeness, we give the relation between them and $\bar{b}_1$ and
$\bar{h}_1$ 
obtained from the on-shell matching:
\begin{align}
\bar{b}_1=3 \frac{\deltamass}{\nucleonmass(\nucleonmass+\deltamass)}g_M\,, \quad 
\bar{h}_1=\frac{3}{2}\frac{1}{\nucleonmass+\deltamass}(g_E+g_M).\label{eq:gM_gE}
\end{align}
\subsection{Electromagnetic form factors of the $\deltapart$ resonance}\label{sec:DeltaFF}
The matrix element of the electromagnetic current $ J^{\mu} $ between
the $\deltapart$ states
can be written in terms of the four form factors $ F_i^* $ as \cite{Nozawa:1990gt}\footnote{Notice that in this subsection we use the notation $\deltamass\equiv \Re(\deltamass)$.}
\begin{align}
\bra{\deltapart(p')}J^{\mu}(0)\ket{\deltapart(p)} = &-\bar{u}_{\alpha}(p') 
\Big\{F_1^*(Q^2)\minkowski^{\alpha\beta}\matrixgamma^{\mu} + \frac{\ci}{2\deltamass}
\Big[F_2^*(Q^2)\minkowski^{\alpha\beta} + F_4^*(Q^2)\frac{k^{\alpha}k^{\beta}}{4\deltamass^2}\Big]
\tensorsigma^{\mu\nu}k_{\nu} \nonumber \\
&+\frac{F_3^*(Q^2)}{4\deltamass^2}\Big[k^{\alpha}k^{\beta}\matrixgamma^{\mu} 
- \frac{1}{2}\slashed k(\minkowski^{\alpha\mu}k^{\beta} + \minkowski^{\beta\mu}k^{\alpha})\Big]\Big\}u_{\beta}(p)\,, \label{eq:DEMFFPara}
\end{align}
where $ Q^2=-(p-p')^2 $.
The form factors $F_1^*$ and $F_2^*$ for zero momentum transfer are given by 
the electric charge and the dipole magnetic moment of the $\deltapart$:
\begin{align}
F_1^*(0) = \deltacharge \equiv\frac{\identity+3\paulimatrix_3}{2}\,,\quad
\deltamdm = \frac{e}{2\deltamass}\left\{\deltacharge + \Re\left[F_2^*(0)\right]\right\}\,.
\end{align}
The imposed renormalization condition on $\bar{c}_6^{\deltapart}$ and $\bar{c}_7^{\deltapart}$
has the form:
\begin{align}
\deltamdm = \frac{e}{2\deltamass}\left(\deltacharge -\frac{1+\paulimatrix_3}{2} \bar{c}_6^{\deltapart} - \bar{c}_7^{\deltapart}\right).
\end{align} 
Contributions of contact terms to $ F_3^* $ and $ F_4^* $ start at higher orders.
Notice that at the order we are working, there are no loop corrections
to the $\deltapart$ electromagnetic form factor and no counter terms
for $c_6^{\deltapart}$ and $c_7^{\deltapart}$, and, therefore, 
$\bar{c}_6^{\deltapart}=c_6^{\deltapart}$ and $\bar{c}_7^{\deltapart}=c_7^{\deltapart}$.
For the same reason, the imaginary part of the $\deltapart$ magnetic moment is equal to zero
in our calculation.

In the particle basis, the {MDM} for each $\deltapart$ state reads
\begin{eqnarray}
\deltamdmind{++} &=& \frac{e}{2\deltamass}\left(2 -
                     \bar{c}_6^{\deltapart} -
                     \bar{c}_7^{\deltapart}\right), \nonumber \\
 \deltamdmind{+} &=& \frac{e}{2\deltamass}\left( 1 - \frac{2}{3}\bar{c}_6^{\deltapart} - \bar{c}_7^{\deltapart}\right),\nonumber\\
\deltamdmind{0} &=&  \frac{e}{2\deltamass}\left( - \frac{1}{3}\bar{c}_6^{\deltapart} - \bar{c}_7^{\deltapart}\right), \nonumber\\
\deltamdmind{-} &=& \frac{e}{2\deltamass}\left( -1 - \bar{c}_7^{\deltapart}\right)\,. \label{eq:MDMLEC}
\end{eqnarray}  

\subsection{Other LECs}\label{sec:other_LECs}
The renormalized constants $\bar{l}_5$ and $\bar{l}_6$ from the pionic Lagrangian $\lpipi^{(4)} $
are related to the corresponding bare quantities through \cite{Gasser:1983yg}
\begin{align}
l_i = \beta_{l_i}\frac{\bar l_i}{32\pimath^2} -
  \beta_{l_i}\frac{A_0(\pionmass^2)}{2\pionmass^2}\,,\quad \mbox{with} \quad
\beta_{l_{5}}=-\frac{1}{6}\,,\quad \beta_{l_{6}}=-\frac{1}{3}\,.
\label{eq:LECshift_l_i}
\end{align}
The pion tadpole function in $d\approx 4 $ dimensions is equal to (see Eq.~(\ref{eq:loop_integrals}))
\begin{equation}
A_0(\pionmass^2)
=-2\pionmass^2\left(\lambar+\frac{1}{32\pimath^2}\ln\left(\frac{\pionmass^2}{\mu^2}\right)\right)\,,
\end{equation}
with the divergent quantity $\lambar$ given by
\begin{equation}
\lambar=\frac{1}{16\pimath^2}\left(\frac{1}{d-4}+
\frac{1}{2}(\eulergamma-\ln(4\pimath)-1)\right).
\end{equation}
Here, $ \eulergamma $ is the Euler-Mascheroni constant and $ \mu $ is the renormalization scale. 
Notice that only the renormalization-scale-independent linear combination 
$ \tilde{l}_{5;6} \equiv \bar{l}_5-\bar{l}_6\propto (2l_5-l_6)$ appears in our calculation.

The renormalized constants $\bar{d}_8$, $\bar{d}_9$, $\bar{d}_{20}$, $\bar{d}_{21}$, $\bar{d}_{22}$
appearing in the photoproduction contact terms
are related to the bare quantities as follows: 
\begin{align}
d_i = \bar{d}_i+\delta_{d_i} - \frac{\beta_{d_i}}{\piondecayconstant^2} \frac{A_0(\pionmass^2)}{2\pionmass^2}\,, \label{eq:LECshift}
\end{align}
with the $\beta$ functions:
\begin{align}
\beta_{d_8}    &= \frac{\axialcoupling\pindcoupling^2}{18}\,,\quad
\beta_{d_9}    = 0\,,\quad
\beta_{d_{20}} = -\frac{2\axialcoupling\pindcoupling^2}{9}\,,\quad
\beta_{d_{21}} = \frac{\axialcoupling\pindcoupling^2}{9}\,,\quad
\beta_{d_{22}} = 0\,,\label{eq:beta_functions}
\end{align}
and the shifts due to the absorption of the constants $b_3$, $b_6$, $h_{15}$, $h_{16}$ by
the redefinition of $h_A$, $b_1$, $h_1$:
\begin{equation}
\delta_{d_8} = \delta_{d_{20}} = -\delta_{d_{21}} = \frac{-b_1(b_3+b_6) +
2\pindcoupling(h_{15}+h_{16})}{9}\,,\quad  \delta_{d_9} =
\delta_{d_{22}} = 0\,.
\end{equation}
We have calculated the $\beta$ functions in Eq.~(\ref{eq:beta_functions})
using the power counting described in Sec.~\ref{sec:PowerCountingInTheDeltaRegionPhotoproduction}.
In general, there are other divergent contributions proportional to $\axialcoupling\pindcoupling^2$,
such as $\pion\deltapart$ loops, that are neglected in our scheme.

Notice that in the $\deltapart$-less case, all the $\beta$-functions are equal to zero.
We further emphasize that the LECs $\bar{d}_{21}$ and $\bar{d}_{22}$
always  appear in the linear combination
$\bar{d}_{21;22} \equiv\bar{d}_{21}-\bar{d}_{22}/2$ in 
our calculation.

As was already mentioned, we use the $\widetilde{\mathrm MS}$
renormalization scheme throughout our work, 
i.e.~we set $ \lambar=0 $.
We have checked that the residual renormalization scale dependence of
the amplitude is of a higher order
than we are working. In the numerical calculations,
the renormalization scale is set to $\mu=\nucleonmass$.

\section{Determination of the LECs from pion photoproduction}
 \label{sec:photoproduction}
 We now focus on the determination of the  low-energy constants
$\bar{b}_1$, $\bar{h}_1$, $\bar{d}_8$, $\bar{d}_9$, $\bar{d}_{20}$ and
$\bar{d}_{21;22}$, that serve as input parameters for the study of
radiative pion photoproduction, from  the analysis of ordinary pion
photoproduction. 
The most important LECs we need to determine are $\bar{b}_1$ and
$\bar{h}_1$, which control the leading and subleading
$\gamma N\Delta$ couplings. They indeed are found to have the largest
impact on the radiative-pion-photoproduction amplitude. 

There have been several studies of pion photoproduction within covariant {$\chi$PT},
both in the $\deltapart$-less \cite{Hilt:2013uf,Hilt:2013fda,Bernard:2005dj} 
and the $\deltapart$-full approach
\cite{Blin:2014rpa,Blin:2016itn,Navarro:2019iqj,Navarro:2020zqn}, see also
Ref.~\cite{Bernard:1992nc} for a pioneering calculation in relativistic $\chi$PT and
\cite{Bernard:1992qa,Bernard:1993bq,Bernard:1994gm,Bernard:1996ti,Bernard:1996bi,Fearing:2000uy,Bernard:2001gz}
for related studies in the heavy-baryon approach.
However, we cannot rely on the values of LECs from these studies as we have to
treat pion photoproduction consistently with the scheme
that we implement for radiative pion photoproduction. For this 
reason, we have to perform our own analysis. 

We consider both $\deltapart$-less and $\deltapart$-full approaches to provide the input parameters for
the corresponding versions of the radiative-pion-photoproduction amplitude.
For the $\deltapart$-full analysis, we take into account, apart from
the low-energy region, also a part of 
the $\deltapart$ region: $\SI{1150}{\mega\electronvolt}\le\sqrt{s}\le\SI{1250}{\mega\electronvolt}$.
At energies $\sqrt{s}>\SI{1250}{\mega\electronvolt}$,
one can hardly apply {$\chi$PT} due to the strong non-perturbative dynamics in the pion-nucleon system.
We exclude energies very close to the $\pion\nucleon$ threshold ($\sqrt{s}\le\SI{1150}{\mega\electronvolt}$) from the analysis
to avoid possible threshold artifacts due to the constant $\deltapart$ width in our approach as 
a consequence of using the complex-mass scheme.
In the $\deltapart$-less approach, we exclude the $\deltapart$ region
completely and consider the energies
$\SI{1090}{\mega\electronvolt}\le\sqrt{s}\le\SI{1200}{\mega\electronvolt}$.
We work in the isospin-symmetric limit and, therefore,  exclude energies 
$\sqrt{s}\le\SI{1090}{\mega\electronvolt}$ from the analysis in order to minimize the impact of the pion mass difference.

Ideally, one would have to fit the whole set of available photoproduction observables in the considered energy region, 
a task which requires a considerable effort and deserves a separate study.
In this work, we follow a more pragmatic approach and fit the photoproduction multipoles
in the isospin basis
taken from empirical partial wave analyses.
The definition of the photoproduction multipoles and their relation to the invariant amplitudes
can be found e.g.~in the original paper by Chew et al.~\cite{Chew:1957tf}.
It is sufficient to consider only the real parts of the multipoles
because the imaginary parts are not independent and constrained by 
unitarity as follows from Watson's theorem \cite{Watson:1954uc}.

\begin{figure}[tb]
\includegraphics[width=\textwidth]{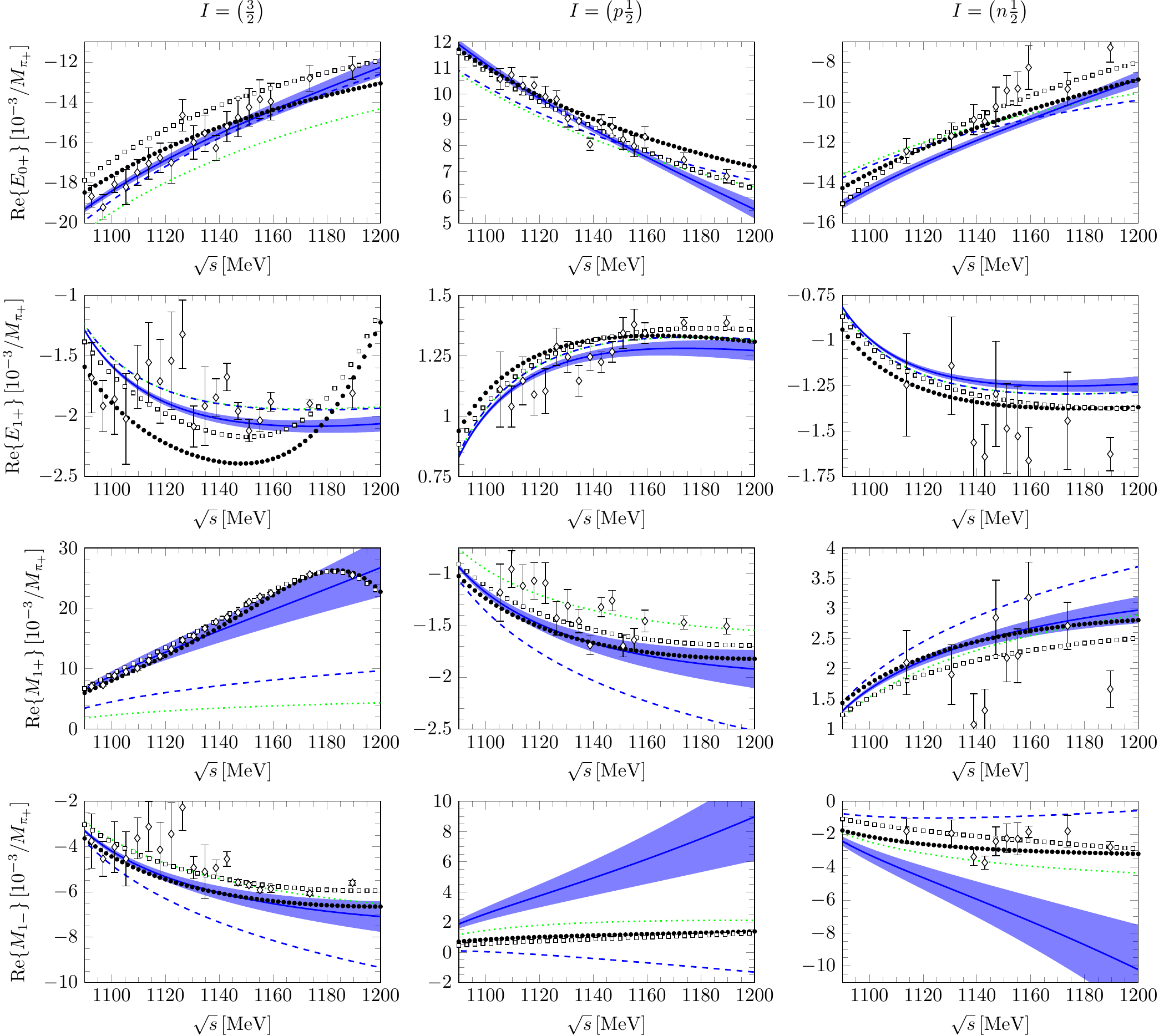}
\caption{$\deltapart$-less fits to the real parts of the $s$- and $p$-wave photoproduction multipoles. 
The solid, dashed and dotted lines denote the $q^3$, $q^2$
and $q^1$ calculations, respectively.
The bands indicate the estimated truncation errors at order
$q^3$.
The filled circles show the results of the MAID partial wave analysis from Ref.~\cite{Drechsel:2007if},
while the squares (diamonds) are the results of the energy
dependent (independent) SAID
analysis from 
 Ref.~\cite{Workman:2012jf}. }
 \label{fig:plotPiPhPdeltaless}
\end{figure}

In Fig.~\ref{fig:plotPiPhPdeltaless}, the real parts of the $s$- and $p$-wave photoproduction multipoles 
for the isospin $I=3/2$ channel and for the proton and neutron isospin $I=1/2$ channel
from the MAID analysis \cite{Drechsel:2007if} and from the energy-dependent and
energy-independent SAID analyses \cite{Workman:2012jf} are shown.
One can see that the dominant contributions come from the $s$-wave ($E_{0+}$) multipoles 
and $M^{3/2}_{1+}$-multipole, which corresponds to the magnetic excitation of the $\deltapart$ isobar in the $s$-channel.
Among the $s$-wave multipoles, the best agreement between various partial wave analyses is observed in the $I=3/2$ channel.
For both electric and magnetic $\deltapart$ multipoles ($E^{3/2}_{1+}$ and $M^{3/2}_{1+}$) the agreement
between the MAID and the SAID analyses is also very good, especially, if we consider the energy-dependent version of the SAID analysis.
Taking into account these observations and the fact that we are mostly concerned about the $\gamma N\Delta$ couplings,
we choose to fit first only the $I=3/2$ multipoles.
For the  $\deltapart$-full fit, we consider four $s$- and $p$-wave $I=3/2$ multipoles:
$E^{3/2}_{0+}$, $M^{3/2}_{1+}$, $M^{3/2}_{1-}$ $E^{3/2}_{1+}$, which
are most sensitive to the LECs $d_i$
and the $\deltapart$-pole contributions.
In the  $\deltapart$-less case, we exclude the $E^{3/2}_{1+}$ multipole from the fit
because it receives no contributions from the LECs that we adjust (in
the absence of the $\deltapart$-pole graphs), apart from 
the residual $1/m_N$-effects.
The constant $\bar{d}_9$ does not contribute to any of the $I=3/2$ multipoles
and must be determined subsequently  from a separate fit to $I=1/2$ multipoles as will be described below.

For the fit, we use the MAID partial wave analysis which, however,
does not provides uncertainties.
Therefore, we follow a common practice, 
see e.g.~the analysis of pion-nucleon elastic scattering in Ref.~\cite{Fettes:1998ud},
and assign the same relative error of \SI{5}{\percent}  
for all data points using energy steps of \SI{2}{\mega\electronvolt}.
We have varied the value of the relative error in the range
\SI{1}{\percent}-\SI{15}{\percent} and found 
that its choice has almost no impact on the result of the fit and very
little impact on the value of the $\chi^2$ since 
the resulting uncertainty appears to be dominated by the truncation
error within the small scale (chiral) expansion.
Our approach to estimating the  truncation errors is discussed in Sec.~\ref{sec:truncation_errors}.
We combine the truncation uncertainty with the ``experimental'' errors
and minimize the objective $\chi_{3/2}^2$ function 
\begin{align}
 \chi_{3/2}^2=\sum_i\left(\frac{\obs_i^{\textrm{exp}}-\obs_i^{(3)}}{\dobs_i}\right)^2\,,\quad 
\dobs =\sqrt{(0.05\; \obs_i^{\textrm{exp}})^2+(\dobs^{(3)})^2}\,,\label{eq:chi2_32}
\end{align}
to obtain the
central values of the fit parameters. 
Here, the summation runs over all fitted $I=3/2$ multipoles and energy points $\obs_i$.
$ \obs_i^{\textrm{exp}} $ refers to the empirical value of the corresponding multipole from 
the MAID analysis, while $\obs_i^{(3)}$ is its theoretical value calculated at order $q^3$ ($\epsilon^3$)
for the $\deltapart$-less ($\deltapart$-full) case.
The truncation errors at order $q^3$ ($\epsilon^3$) are denoted as $\dobs^{(3)}$.
The uncertainties of the parameters 
are extracted from the covariance matrix, which is approximated
by the inverse of the Hessian matrix:
\begin{equation}
\Cov(y_i,y_j)=H_{ij}^{-1}\,,\quad \mbox{with} \quad
 H_{ij}=\frac{1}{2} \left.\frac{\partial^2\chi_\text{3/2}^2}{\partial y_i\partial y_j}\right|_{\newvec{y} = \bar{\!\newvec{y}}}\,, \quad
\newvec{y}=(\bar{b}_1, \bar{h}_1, \bar{d}_8, \bar{d}_{20}, \bar{d}_{21;22})\,,
\end{equation}
and $ \bar{\!\newvec{y}}$ denoting the vector of the best fit parameters.

\begin{figure}[htbp]
\includegraphics[width=\textwidth]{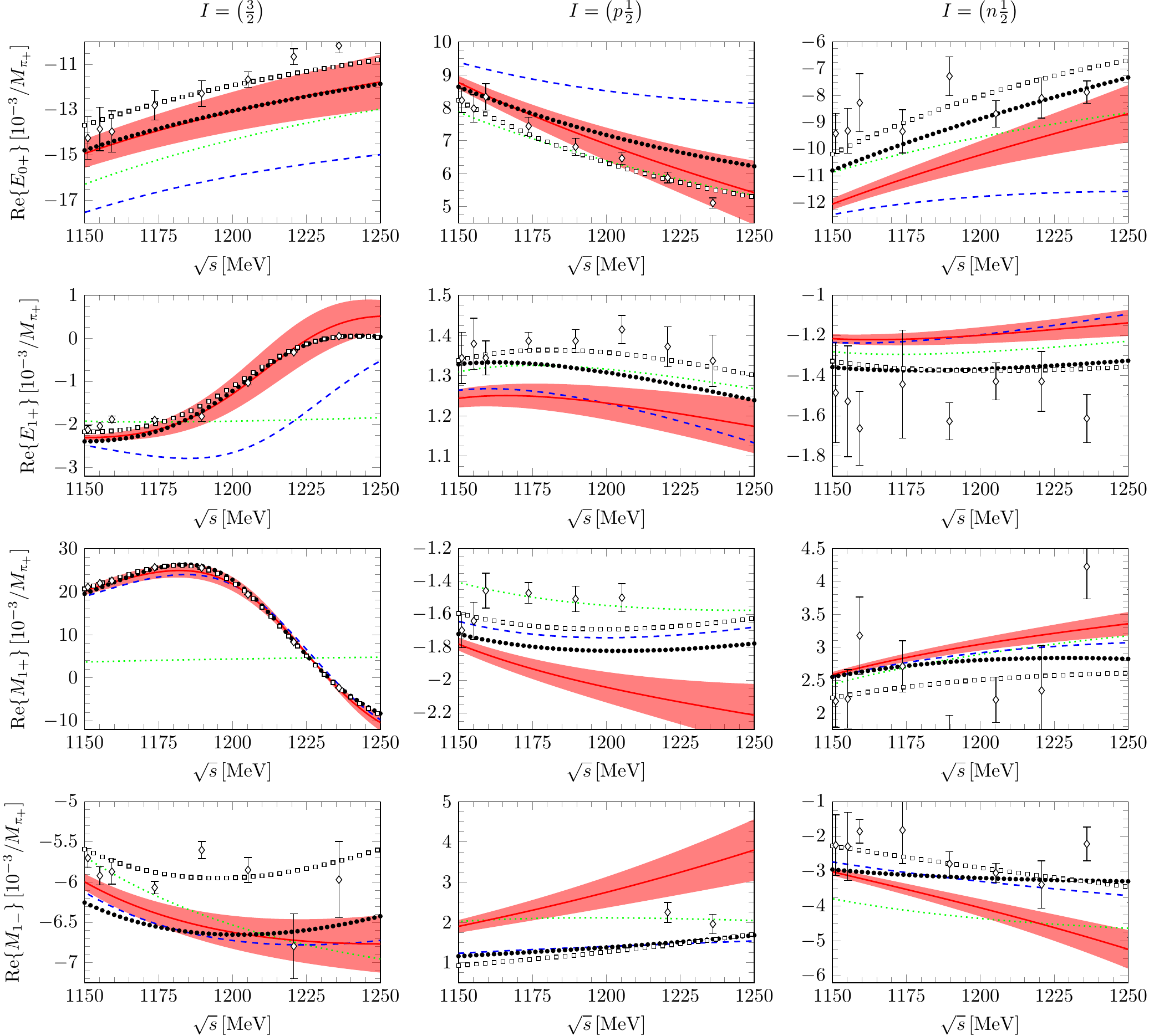}
\caption{$\deltapart$-full fits to the real parts of the $s$- and $p$-wave photoproduction multipoles. 
The solid, dashed and dotted lines denote the $\epsilon^3$, $\epsilon^2$
and $\epsilon^1$ calculations, respectively.
The bands indicate the estimated truncation errors at order
$\epsilon^3$.
The filled circles show the results of the MAID partial wave analysis from Ref.~\cite{Drechsel:2007if},
while the squares (diamonds) are the results of the energy
dependent (independent) SAID
analysis from 
 Ref.~\cite{Workman:2012jf}. } \label{fig:plotPiPhPdeltaful}
\end{figure}

To determine the LEC $\bar{d}_9$ we proceed as follows. 
We assume that the constants
$\bar{b}_1$, $\bar{h}_1$, $\bar{d}_8$, $\bar{d}_{20}$, $\bar{d}_{21;22}=\bar{d}_{21}-\bar{d}_{22}/2$
are relatively well constrained from the fit to the $I=3/2$ multipoles and can be used 
as input for a determination of $\bar{d}_9$ from $I=1/2$ multipoles.
We have verified this assumption by checking that the uncertainties of the LECs determined
from the $I=3/2$ fit have little impact on the results of the $I=1/2$ fit.
The constant $\bar{d}_9$ contributes to four multipoles (excluding the
$1/m_N$ effects), namely
$_pM^{1/2}_{1+}$, $_nM^{1/2}_{1+}$, $_pM^{1/2}_{1-}$, $_nM^{1/2}_{1-}$.
Unfortunately, reasonable convergence is not yet reached in the case of 
$_pM^{1/2}_{1-}$ and $_nM^{1/2}_{1-}$, as can be seen from Figs.~\ref{fig:plotPiPhPdeltaless},~\ref{fig:plotPiPhPdeltaful}.
The situation remains the same
independent of what value of $\bar{d}_9$ one adopts.
Moreover, the fit values of $\bar{d}_9$ do not (or very little) depend on whether these two multipoles
are included into the $\chi^2$ or not (of course, this affects the value of the $\chi^2$ itself).
We, therefore, retained only the $_pM^{1/2}_{1+}$ and $_nM^{1/2}_{1+}$ multipoles in the final fit.
The $\chi_{1/2}^2$ is defined analogously to the case of  $I=3/2$ fit (see Eq.~(\ref{eq:chi2_32})),
and the uncertainty of $\bar{d}_9$ is given by 
$\displaystyle{\kronecker \bar{d}_9=
\bigg[\frac{1}{2}\frac{\partial^2\chi_\text{1/2}^2}{(\partial \bar{d}_9)^{2}}\bigg]^{-1/2}}$, 
where the derivative is taken at the minimum.

Special attention should be paid to the choice of the renormalized $\deltapart$ mass $\deltamass$, which determines
our complex mass scheme. Although the scattering amplitude has a pole at $s=\deltamass^2$, it does not
necessarily mean that  $\deltamass$ must coincide with the physical $\deltapart$ pole mass
because our theory is not meant to  be applicable in a vicinity of the complex $\deltapart$ pole.
Rather, $\deltamass$ must be chosen in such way as to obtain an efficient scheme from the convergence point of view.
In fact, the photoproduction amplitude at order $\epsilon^2$ is quite sensitive to $\deltamass$
since the real part of the $s$-channel $\deltapart$-pole diagram vanishes for the magnetic 
$\deltapart$ multipole $M^{3/2}_{1+}$ at $\sqrt{s}=\deltamass$, and there are no other free parameters to compensate for the shift in the position of
the resonance. As a result, an inappropriate choice of $\deltamass$
would lead to large discrepancies 
with experimental data at order $\epsilon^2$,
which, in turn, would have a large impact on the estimated truncation errors making the fit less stable.
Therefore, we decided to fit $\deltamass$ (along with the coupling
constant $b_1$) to the $M^{3/2}_{1+}$ multipole at order $\epsilon^2$.
We obtain $\deltamass=\num{1219.3} - \num{53.7}\,\ci $ MeV, which is rather close to the 
PDG value of the pole mass $\num{1210} - \num{50}\,\ci$~\cite{Tanabashi:2018oca}.
The same value is then used in our order-$\epsilon^3$ calculations.
The value of the LEC $b_1$ obtained from the above-mentioned $\epsilon^2$-fit, $b_1=5.7m_N^{-1}$,
is almost the same as the one extracted from the $\epsilon^3$-fit ($b_1=5.4m_N^{-1}$),
which is a nice indication of the stability of the scheme.
As a consequence, the truncation errors do not depend on which of the two values of $b_1$
is chosen for the $\epsilon^2$ amplitude.

\begin{table}[t]
\caption{Low-energy constants obtained from a $\deltapart$-less fit to the photoproduction multipoles 
using the partial-wave analysis of Ref.~\cite{Drechsel:2007if} in
units of GeV$^{-2}$.}
\label{tab:fitresPiPhPdeltaless}
\begin{tabular*}{\textwidth}{@{\extracolsep{\fill}}ccccc} 
  \hline
  &&&& \\[-7pt]
& {$\bar{d}_8$} & {$\bar{d}_9$} &
                                                                      {$\bar{d}_{20}$} & {$\bar{d}_{21;22}$} \\[2pt]
  \hline
  &&&& \\[-7pt] 
order-$q^3$ fit value: & \num{-4.8\pm 0.2} &	\num{ 0.01 \pm 0.01}	&	\num{-8.4\pm 0.3}	&	\num{ 9.2\pm 0.3} \\[2pt]
\hline
\end{tabular*}
\end{table}

\begin{table}[t]
\caption{Low-energy constants obtained from a $\deltapart$-full fit to the photoproduction multipoles 
using the partial-wave analysis of Ref.~\cite{Drechsel:2007if}. The
LECs $d_i$ are given in units of GeV$^{-2}$ while $\bar{b}_1$ and
$\bar{h}_1$ are in units of $\nucleonmass^{-1}$}
\label{tab:fitresPiPhPdeltaful}
\begin{tabular*}{\textwidth}{@{\extracolsep{\fill}}ccccccc} 
 \hline
  &&&&&& \\[-7pt]
& {$\bar{d}_8$} & {$\bar{d}_9$} &
                                                                      {$\bar{d}_{20}$} & {$\bar{d}_{21;22}$} & {$\bar{b}_1$} & {$\bar{h}_1$} \\[2pt]
 \hline
  &&&&&& \\[-7pt]   
order-$\epsilon^3$ fit value: &\num{-0.19\pm 0.03} & \num{0.03\pm 0.02} & \num{-3.0\pm 0.1} & \num{2.7\pm 0.1} & \num{5.4\pm 0.1} & \num{1.0\pm 0.1}
  \\[2pt]
  \hline
\end{tabular*}
\end{table}

The values of the low-energy constants obtained from the $\deltapart$-less ($\deltapart$-full)
fit are collected in Table~\ref{tab:fitresPiPhPdeltaless} (Table~\ref{tab:fitresPiPhPdeltaful}).
The reduced $\chi^2/n_{\textrm{dof}} $ ($n_{\textrm{dof}}$ stands for the number of degrees of freedom)
for the $I=3/2$ ($I=1/2$) fit is equal to $0.4$ ($0.3$)
in the $\deltapart$-less case and to $0.2$ ($2.5$) in the $\deltapart$-full case, which we find satisfactory
given our somewhat simplistic approach to the statistical errors.
Small (large) values of the reduced $\chi^2$ may indicate the overestimation (underestimation)
of the truncation errors,
in particular, by using an inappropriate value of the breakdown scale $\chiralbreaking$.
Nevertheless, we prefer to follow the procedure consistent with 
radiative pion photoproduction
and adopt the value $\chiralbreaking= \SI{700}{\MeV}$, see
Secs.~\ref{sec:truncation_errors},~\ref{sec:results_neutral_channel}.
We emphasize again that the extracted values of $\bar{b}_1$ and $\bar{h}_1$ cannot be directly compared 
with other values from the literature (e.g., in Refs~\cite{Blin:2016itn,Pascalutsa:2005vq,Pascalutsa:2002pi}
$\bar{b}_1$ varies in the range $(2.6-4.9)$ $\nucleonmass^{-1}$
and $\bar{h}_1$ varies in the range $(-2.2-4.2)$ $\nucleonmass^{-1}$ if one translates them using Eq.~(\ref{eq:gM_gE}))
because they are calculated within different schemes.

The fit results for the multipoles at order $q^3$ ($\epsilon^3$) in the $\deltapart$-less ($\deltapart$-full) case are presented
in Fig.~\ref{fig:plotPiPhPdeltaless} (Fig.~\ref{fig:plotPiPhPdeltaful})
with the bands indicating the truncation errors.
Also shown are the results at order $q^1$ and $q^2$ ($\epsilon^1$ and $\epsilon^2$)
to demonstrate the convergence rate in various channels.
As expected, including explicit $\deltapart$ degrees of freedom significantly improves the
convergence for the $M^{3/2}_{1+}$ multipole.

\begin{figure}[htbp]
\includegraphics[width=\textwidth]{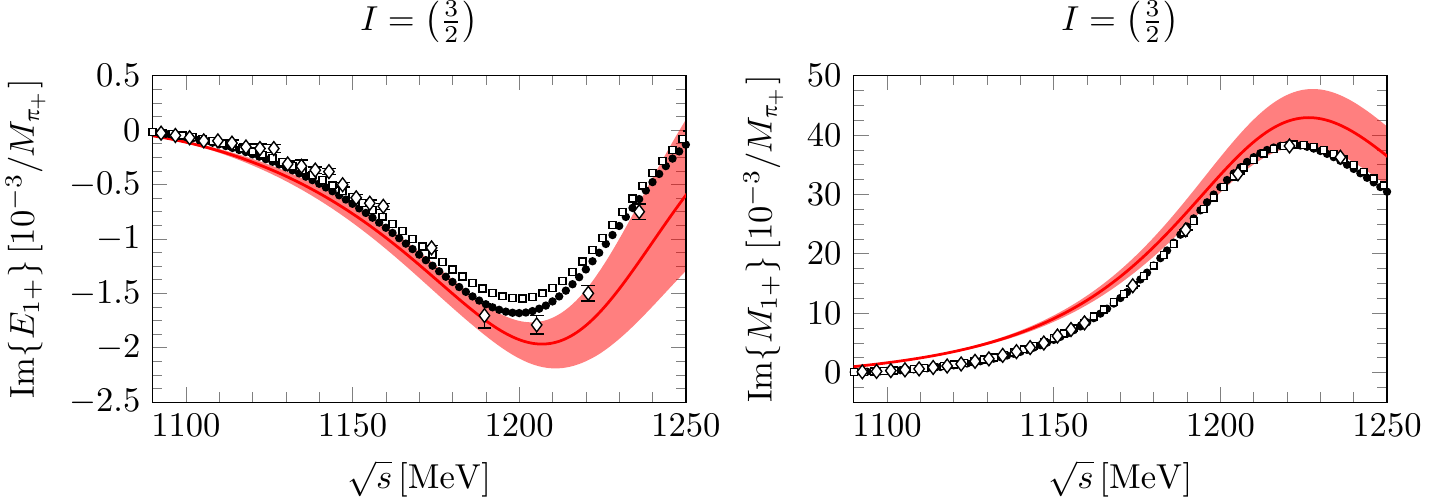}
\caption{$\deltapart$-full results at order $\epsilon^3$ for
the imaginary parts of the $E^{3/2}_{1+}$ and $M^{3/2}_{1+}$ multipoles.
The notation is as in Fig.~\ref{fig:plotPiPhPdeltaful}. } \label{fig:plotPiPhPimaginaryParts}
\end{figure}

In order to see that our choice of the renormalized $\deltapart$ mass $\deltamass$ is consistent
with the $\deltapart$-resonance contribution to the $\pi N$ elastic channel, 
we plotted the imaginary parts of the $E^{3/2}_{1+}$ and $M^{3/2}_{1+}$ multipoles 
(see Fig.~\ref{fig:plotPiPhPimaginaryParts}), since the phase of the photoproduction amplitude is
determined by the elastic $\pi N$ phase shifts.
Indeed, the agreement with the results of the partial wave analyses for these channels is 
reasonable.

It is instructive to analyze the difference between the $\deltapart$-full and 
$\deltapart$-full {LECs} ($d_i$'s) from the point of view of the $\deltapart$-resonance saturation
in pion photoproduction, see Refs.~\cite{Bernard:1996gq,Krebs:2018jkc}
for a similar discussion of the $\pion N$ {LEC}.
If we consider the heavy-baryon limit for the $\epsilon^2$ $\deltapart$-pole diagrams
and, in addition, go to the limit of $ \deltasplit \equiv \deltamass-\nucleonmass \to \infty$,
then their effect proportional to $1/\deltasplit$ will be given by the
following shifts in the photoproduction {LEC}:
\begin{equation}
\bar{d}_8(\Delta) =- \bar{d}_{21;22}(\Delta) = -\frac{h_Ab_1}{9\deltasplit}\simeq-\SI{3.3}{\per\GeV\squared}\,, \quad 
\bar{d}_9(\Delta)=\bar{d}_{20}(\Delta)  =0\,. \label{eq:deltasat}
\end{equation}
The actual differences in the $\bar{d}_8$ and $\bar{d}_{21;22}$ obtained from the fits are:
\begin{align}
\bar{d}_8^{\slashed\deltapart}-\bar{d}_8^{\deltapart}\simeq\SI{-4.6}{\per\GeV\squared}\,, \quad
\bar{d}_{21;22}^{\slashed\deltapart}-\bar{d}_{21;22}^{\deltapart}\simeq\SI{6.5}{\per\GeV\squared}\,,
\end{align}
which are, indeed, to a large extent saturated by the shifts from Eq.~(\ref{eq:deltasat}).
Moreover, in the $\Delta$-full scheme, the $d_i$'s appear to be
smaller in absolute value and more natural.

Last but not least, we emphasize that the considered LECs also
contribute to the longest-range two-nucleon electromagnetic current
\cite{Kolling:2011mt,Krebs:2019aka,Pastore:2009is} and are thus of
considerable interest for calculations in the few-nucleon
sector. These studies are, however, carried out in the heavy-baryon
approach.
The determination of the LECs in the heavy-baryon convention
and the extension to the fourth chiral order will be presented in a
separate publication.

\section{Results and discussion} \label{sec:results} 
In this section, we present the numerical results of our calculation.
The results are obtained using our own code written in \emph{Mathematica} \cite{mathematica12.0}, FORM \cite{Kuipers:2012rf}
and Fortran. For the numeric evaluation of loop integrals, 
the \emph{Mathematica}  packages \emph{Package}-{\textbf X} \cite{Patel:2016fam} and  LoopTools \cite{Hahn:1998yk}
have been used.

\subsection{Low-energy constants}
The radiative-pion-photoproduction amplitude, at the order we are working, depends 
only on the free parameters related to the dipole
magnetic moment of the $\deltapart$ resonance, i.e.~on
$\deltamdmind{+}$ for the neutral channel and $\deltamdmind{+}$ and $\deltamdmind{0}$ for the charged channel.
There are no free parameters in the $\deltapart$-less case.
The quantity $\deltamdmind{+}$ is related to the linear combination 
$ \tilde{c}_{67}^{\deltapart} =\bar{c}_6^{\deltapart} +
3\bar{c}_7^{\deltapart}/2$, see Sec.~\ref{sec:DeltaFF} for details.
The numerical values of all remaining {LEC}
from the effective Lagrangian in Eq.~(\ref{eq:effectiveL}), which appear in the radiative-pion-photoproduction amplitude
after the renormalization procedure described in Sec.~\ref{sec:renormalization},
are taken from other sources. 
The values of the particle masses and the coupling constants
from the leading-order effective Lagrangian are collected in Table~\ref{tab:LO_constants}.
\begin{table}[t]
\caption{Particle masses (in MeV) and leading-order coupling constants used in this work.
Unless specified, the values are taken from
PDG~\cite{Tanabashi:2018oca}. The $\deltapart$ mass is determined in Sec.~\ref{sec:photoproduction}
from the fit to pion photoproduction.
} \label{tab:LO_constants}
\begin{tabular*}{\textwidth}{@{\extracolsep{\fill}}cccccccc}
 \hline
  &&&&&&& \\[-7pt]
$ \pionmass$ & $ \nucleonmass $ & $ \deltamass  $ & $ e $ & $
                                                            \piondecayconstant
                                                            \; [\si{\MeV}] $ & $\axialcoupling$ & $\pindcoupling$ & $g_1$\\[2pt]
  \hline
  &&&&&&& \\[-7pt] 
\num{138.03} & \num{938.27} & \num{1219.3} - \num{53.7}\,\ci & \num{0.303}
        & \num{92.1} & \num{1.289} \cite{Baru:2010xn} & \num{1.43}
                                                       \cite{Bernard:2012hb,Yao:2016vbz}&
                                                                                          \num{-1.21} \cite{Yao:2016vbz}\\[2pt]
\hline
\end{tabular*}
\end{table}
For the $\deltapart$, the pole mass is used (following the complex-mass scheme) 
and for the $\pindcoupling$ coupling, the value extracted from the $\deltapart$ width is adopted \cite{Bernard:2012hb}.
This value practically coincides with the one extracted from the $\epsilon^3$ analysis of the $\pion N$ scattering
in Ref.~\cite{Yao:2016vbz}, from which we also take the value of the $\pion\Delta\Delta$ coupling $g_1$.

Below, we list other {LECs} from higher-order terms in the Lagrangian. 
The purely mesonic {LECs} $ l_5 $ and $ l_6 $ appear, in our calculation
of the charged-pion-photoproduction amplitude,
only as a linear combination $ \tilde{l}_{5;6} = \bar{l}_5-\bar{l}_6$. 
This quantity can be extracted from the decay $\pion^{+} \to e^{+} \nu \gamma$, see, e.g.,~\cite{Bijnens:2014lea,Bijnens:1996wm}:
$\tilde{l}_{5;6} = \num{-3.0}$.
The {LECs} $ \bar{c}_6 $ and $ \bar{c}_7 $ are fixed by the magnetic
moment of the nucleon,  
see Sec.~\ref{sec:nucleon_FF} and PDG~\cite{Tanabashi:2018oca}:
$\bar{c}_6=\num{3.706}$ and $\bar{c}_7=\num{-1.913}$.

As described in Sec.~\ref{sec:photoproduction}, we extracted the constants 
$\bar{b}_1$, $\bar{h}_1$, $\bar{d}_8$, $\bar{d}_9$, $\bar{d}_{20}$ and $\bar{d}_{21;22}$
from the fit to the pion-photoproduction multipoles.
Note also that the neutral-pion-photoproduction amplitude depends only on the linear combination 
of $\bar{d}_8$ and $\bar{d}_9$: $\bar{d}_{89}=\bar{d}_8+\bar{d}_9$.

\subsection{Fitting procedure} \label{sec:fitting}
In order to determine the $ \deltapart^+ $ {MDM}, 
we fit the radiative neutral-pion-photoproduction observables by minimizing the $\chi^2$
\begin{align}
\chi^2=\sum_i\left(\frac{\obs_i^{\textrm{exp}}-\obs_i^{(n)}}{\dobs_i}\right)^2\,,
\end{align}
where the summation runs over all available observables ($\dd\sigma/\dd\ego$)
and all kinematical data points. Here,  $ \obs_i^{\textrm{exp}} $ is the 
experimental value of a relevant observable
at a chosen kinematical data point  and $ \obs_i^{(n)} $ is
the corresponding theoretical value calculated at order $ n $
(in the case of radiative neutral-pion photoproduction $n=2$ or $3$). 
The uncertainty $\dobs$ (we omit the index $ i $ in what follows) originates from two independent sources:
the experimental error  $ \dobs^{\textrm{exp}}$,
and the error related with the truncation of the small scale expansion at order $ n $, $ \dobs^{(n)} $, 
see Sec:~\ref{sec:truncation_errors}.
Therefore, we add them quadratically:
\begin{align}
\label{eq:uncertainty}
\dobs =\sqrt{(\dobs^{\textrm{exp}})^2+(\dobs^{(n)})^2}\,.
\end{align}
Apart from the statistical error of $\deltamdmind{+}$
extracted from the fit, there are errors originating from the uncertainties of the
input parameters. In most cases, they are rather small and have no significant impact on the result, which
we have verified explicitly. The only exceptions are the
uncertainties of the LECs determined from pion photoproduction.
In particular, the radiative-pion-photoproduction amplitude is rather
sensitive to the leading $\photon N\Delta$ coupling $\bar{b}_1$ and, to a lesser extent, to the
subleading $\photon N\Delta$ coupling $\bar{h}_1$. Ideally, one should perform a combined fit
to observables of both reactions $\photon N\to \pion N\photon$ and $\photon N\to \pion N$ to extract 
the whole set of parameters.
However, we follow here a simpler and more pragmatic approach and adopt the reasonable assumption that
$\bar{b}_1$, $\bar{h}_1$ as well as $\bar{d}_8$, $\bar{d}_9$,
$\bar{d}_{20}$ and $\bar{d}_{21;22}$ 
and their uncertainties can be determined from pion photoproduction with a good
accuracy without additional information from radiative pion photoproduction.
This is motivated by the fact that the $\deltapart$ couples directly to the $\photon N$ and $\pion N$ systems,
and only very weakly to the $\pion N \photon$ system.
Therefore, we fit $\deltamdmind{+}$ (or $\tilde c_{67}^{\deltapart}$)
to the radiative-pion-photoproduction data
with $\bar{b}_1$, $\bar{h}_1$, $\bar{d}_8$, $\bar{d}_9$, $\bar{d}_{20}$, $\bar{d}_{21;22}$
as input parameters.
The condition of minimal $\chi^2$ defines indirectly the function 
$\deltamdmind{+}(\bar{b}_1, \bar{h}_1, \bar{d}_8, \bar{d}_9, \bar{d}_{20}, \bar{d}_{21;22})$,
and the errors of the parameters determined from pion photoproduction
are propagated through this function.
This is essentially equivalent to the following procedure, which we implement:
after finding the best value of $\deltamdmind{+}$, we combine the $\chi^2$ for the reaction 
$\gamma p\to\gamma p\pion^0$ with $\chi_{3/2}^2$ and $\chi_{1/2}^2$ from the fits to
the photoproduction multipoles with isospins $I=3/2$ and $I=1/2$, respectively (see Sec.~\ref{sec:photoproduction}),
to define the total $\chi_\text{tot}^2$: 
\begin{equation}
 \chi_\text{tot}^2(\newvec{z})
 =\chi^2(\newvec{z})+\chi_{3/2}^2(\newvec{y})+\chi_{1/2}^2(\newvec{y},\bar{d}_9)\quad
 \mbox{with} \quad 
\newvec{y}=(\bar{b}_1, \bar{h}_1, \bar{d}_8, \bar{d}_{20},
\bar{d}_{21;22})\,,\quad 
\newvec{z}=(\tilde c_{67}^{\deltapart},\newvec{y},\bar{d}_9)\,.
\end{equation}
As has been explained above, we assume that $\chi_\text{tot}^2$ takes its minimal value at $\newvec{z} = \bar{\!\newvec{z}}$,
where $\bar{\!\newvec{y}}$ is determined from the photoproduction $I=3/2$ fit, 
the central value for $\bar{d}_9$ from the photoproduction $I=1/2$ fit,
and the central value for $\tilde c_{67}^{\deltapart}$ from the radiative-pion-photoproduction fit.
In the vicinity of the minimum, we approximate the $\chi^2$ by the Taylor expansion up to quadratic terms:
\begin{align}
\chi_\text{tot}^2 &\approx \chi_\text{tot}^2(\bar{\newvec{z}})+H_{ij}(z_i-\bar z_i)(z_j-\bar z_j)\,,\quad
 H_{ij}=\frac{1}{2} \left.\frac{\partial^2\chi_\text{tot}^2}{\partial z_i\partial z_j}\right|_{\newvec{z} = \bar{\!\newvec{z}}}\,,
\end{align}
where we explicitly assume that linear terms $\sim
\partial\chi_\text{tot}^2 /\partial z_i$ can be neglected,
i.e.~there are no additional shifts in the central values of the parameters already
determined from the photoproduction fit, as has been discussed above.
The errors of the input parameters are propagated to $ \tilde c_{67}^{\deltapart} $ through the mixed derivatives
$\displaystyle{\frac{\partial^2\chi_\text{tot}^2}{\partial \tilde c_{67}^{\deltapart}\partial y_i}}$,
$\displaystyle{\frac{\partial^2\chi_\text{tot}^2}{\partial \tilde c_{67}^{\deltapart}\partial\bar{d}_9}}$.
Finally, the error of $\tilde c_{67}^{\deltapart}$ is given by the diagonal element of the covariance matrix:
\begin{align}
 \kronecker \tilde c_{67}^{\deltapart} = \Big[\Cov(c_{67}^{\deltapart},c_{67}^{\deltapart})\Big]^{\frac{1}{2}}\,, \quad \Cov(z_i,z_j)=H_{ij}^{-1}\,.
\end{align}

\subsection{Truncation errors} \label{sec:truncation_errors}
The truncation errors for all considered processes (radiative neutral- and charged-pion photoproduction
and ordinary pion photoproduction)
are calculated utilizing the Bayesian model considered in Refs.~\cite{Epelbaum:2019zqc,Epelbaum:2019kcf}
based on the ideas developed in Refs.~\cite{Furnstahl:2015rha,Melendez:2017phj}.
 
An analyzed observable $ \obs $ is represented as an expansion with dimensionless
coefficients $c_i$:
\begin{equation}
\obs = \obs^{(1)}+ \Delta \obs^{(2)} + \Delta \obs^{(3)} + \dotsi 
= \obs_{\textrm{ref}} \, \Big(c_1Q+ c_2Q^2+c_3Q^3+ \dotsi \Big),
\end{equation}
where $\Delta \obs^{(i)} = \obs^{(i)}- \obs^{(i-1)} $ and the superscript $i$
denotes the order in the small scale expansion. 
The expansion parameter $Q$ and the reference value $\obs_{\textrm{ref}}$ are chosen to be
\begin{align}
Q=\frac{\egi}{\chiralbreaking}\,,\quad
\obs_{\textrm{ref}}= \max\left(\frac{|\obs^{(1)}|}{Q},\, \frac{|\Delta
  \obs^{(2)}|}{Q^2},\, \frac{|\Delta \obs^{(3)}|}{Q^3}\right).
\label{eq:reference_value}
\end{align}
In order to estimate the truncation error at order $k$, $\kronecker \obs^{(k)}\equiv \sum_{i > k} \Delta \obs^{(i)}$,
it is assumed that all coefficients $c_i$ are distributed according to the
Gaussian prior  $\mathrm{pr} (c_i | \bar c )$:
\begin{align}
  \label{prior}
 \mathrm{pr} (c_i | \bar c ) = \frac{1}{\sqrt{2 \pimath} \bar c} \, e^{-c_i^2/(2 \bar c^2 )}\,,
\end{align} 
except $c_m=1$, which defines the overall scale, where $m$ is the
number of a maximal argument in the $\max$ function in Eq.~(\ref{eq:reference_value}).
In turn, the parameter  $\bar c$  is assumed to obey a log-uniform probability distribution
\begin{align}
\mathrm{pr} ( \bar c ) = \frac{1}{\ln ( \bar c_> / \bar c_< )} \, 
\frac{1}{\bar c} \, \theta (\bar c - \bar c_< ) \, \theta (\bar c_> - \bar c )\,.
\end{align}
The cutoffs $\bar c_<$ and $\bar c_>$ reflect the constraints imposed by the naturalness assumption.
Following Refs.~\cite{Epelbaum:2019zqc,Epelbaum:2019kcf}, we set $\bar c_<=\num{0.5}$, $\bar c_>=\num{10}$.
After performing marginalization over $h$ chiral orders $k+1, \ldots, k+h$ 
assumed to dominate the truncation error,
the resulting posterior probability distribution for  
the dimensionless quantity
\begin{align}
\deltapart_k=\sum_{n=k+1}^{\infty}c_nQ^n\approx \sum_{n=k+1}^{k+h}c_nQ^n,
\end{align} 
given the knowledge of  $\{c_{i  \le k} \}$ is given by 
\begin{align}
\text{pr}^{C}_h(\deltapart| \{c_{i\leq k}\}) &= \frac{1}{\sqrt{\pimath \bar q^2c_k^2}}\left(\frac{c_k^2}{c_k^2+\deltapart^2/\bar q^2}\right)
\frac{\Gamma\left(\frac{k}{2},\frac{1}{2\bar c^2_{>}}\left(c_k^2+\frac{\deltapart^2}{\bar q^2}\right)\right)-
\Gamma\left(\frac{k}{2},\frac{1}{2\bar c^2_{<}}\left(c_k^2+\frac{\deltapart^2}{\bar q^2}\right)\right)}
{\Gamma\left(\frac{k-1}{2},\frac{c_k^2}{2\bar c^2_{>}}\right)-\Gamma\left(\frac{k-1}{2},\frac{c_k^2}{2\bar c^2_{<}}\right)}\,, 
\nonumber\\
\Gamma (s,x)&=\int\limits_{x}^{\infty}\dd t\, t^{s-1}\efkt^{-t}
\label{eq:pdf},
\end{align}
where $\bar q^2=\sum_{i=k+1}^{k+h}Q^{2i}$, $ c_k^2=\sum_{i\in A}c_i^2 $, 
$A=\{n\in\mathbb N_0 | n\leq  k \wedge n\neq 1 \wedge n\neq m\}$. 
The $1\sigma$ truncation error $\kronecker \obs^{(k)}$ is defined in such a way that
the integral from the probability distribution $\text{pr}^{C}_h(\deltapart$ over the
region $|\deltapart|<\kronecker \obs^{(k)}/\obs_{\text{ref}}$ 
is equal to the confidence level $ 0.68 $. Following Refs.~\cite{Epelbaum:2019zqc,Epelbaum:2019kcf,Melendez:2017phj}, we
choose $h=10$. The breakdown scale is assumed to be $\chiralbreaking =
700$~MeV. It is chosen to be somewhat larger than the value used 
in Refs.~\cite{Epelbaum:2014efa,Epelbaum:2014sza,Furnstahl:2015rha} ($\chiralbreaking =600$~MeV)
because we explicitly include the $\deltapart$ degrees of freedom, and the $\deltapart$ pole should not
affect the convergence rate of the chiral expansion.

Notice that for radiative neutral-pion photoproduction, our analysis includes only 
two different orders (LO and NLO) in the small scale expansion.
This makes our probabilistic Bayesian approach to the uncertainty estimation not
quite
reliable. In order to  increase the reliability
of the estimated truncation errors, one has to calculate higher-order contributions to the amplitude explicitly.

\subsection{Radiative neutral-pion photoproduction}\label{sec:results_neutral_channel}
We start the discussion of the results with the reaction $ \photon\proton \to \photon\proton\pion^0 $. 
This channel is of particular interest since it is sensitive to 
the value of the dipole magnetic moment of the $ \deltapart^+$ particle,
and there is sufficient amount of experimental data for analysis. 

\begin{figure}[tb]
\includegraphics[width=\textwidth]{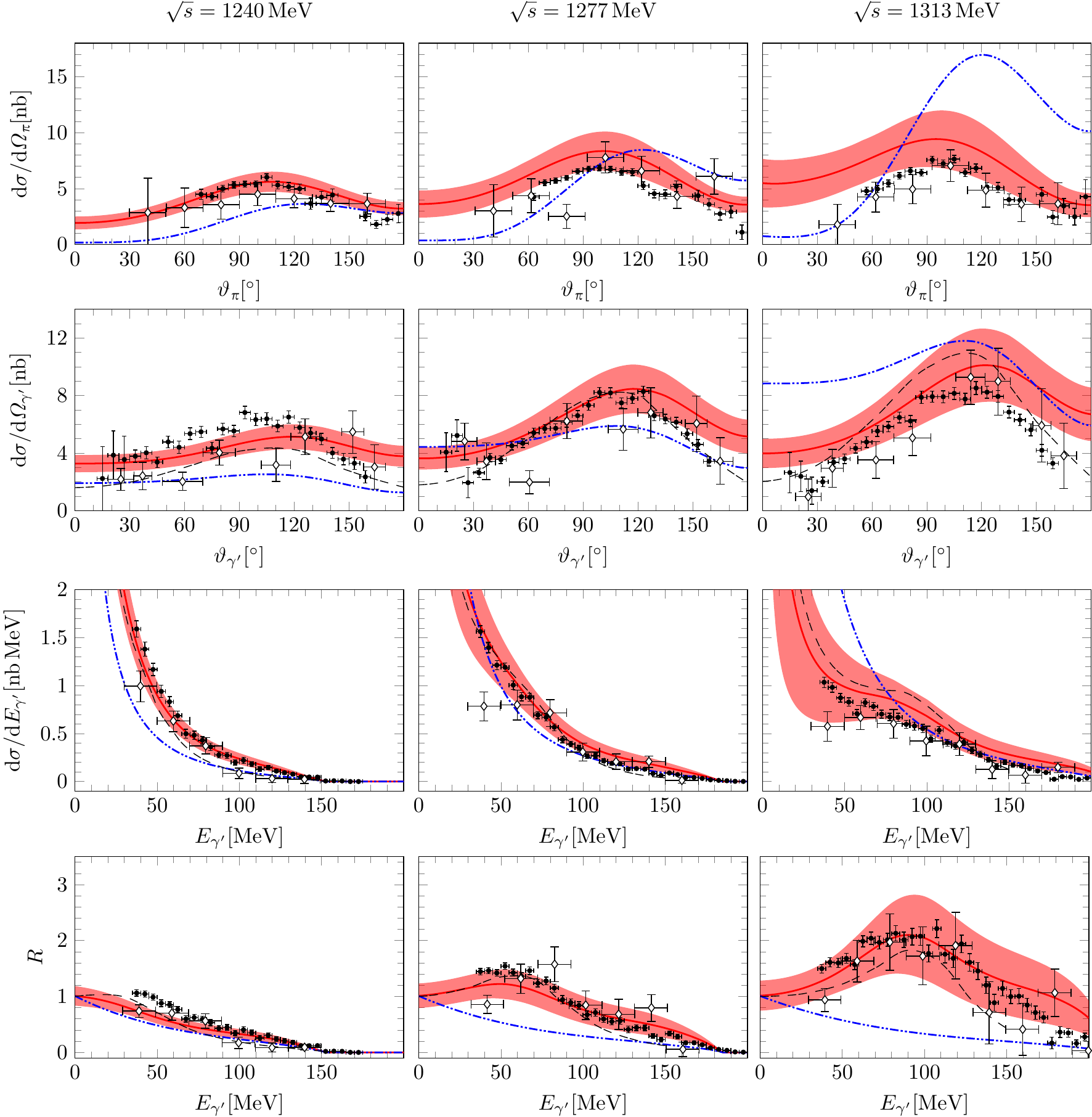}
\caption{Differential cross sections and the ratio $ R $ (defined in text) for the reaction 
$ \photon\proton \to \photon\proton\pion^0 $. 
The dashed-double-dotted lines correspond to the $\deltapart$-less
order-$ q^3 $ calculation. The solid lines denote  
the NLO $\deltapart$-full fit with the bands indicating the truncation errors. 
The double-dashed lines stand for the results of
Ref.~\cite{Pascalutsa:2007wb}. 
The data are from \cite{Schumann:2010js} (filled circles) and \cite{Kotulla:2002cg} (diamonds).} \label{fig:plots}
\end{figure}

Within the $\deltapart$-full approach at order $\epsilon_\text{eff}^3$
(NLO), 
we fit the available experimental data
for three observables: $\dv{\sigma}{\omgam}$, $\dv{\sigma}{\ompi}$ and
$\dv{\sigma}{\ego}$,  see Sec.~\ref{sec:kinematics} for definitions,
and three incident energies $\sqrt{s}=$\SIlist{1240;1277;1313}{\MeV}.
The results of the fit are shown in Fig.~\ref{fig:plots} by the solid lines
with the bands indicating the truncation errors
corresponding to $68\%$ degree-of-belief intervals.
As one can see, the results of the fit are in good agreement with the data within the error bars.
The fit quality is given by $\chi^2/n_{\textrm{dof}}=1.00 $, which
indicates, in particular, that the assumed value for the breakdown
scale $\chiralbreaking = 700$~MeV is reasonable.
For comparison, we have also considered the $\deltapart$-less approach.
Although one should not expect convergence of $\deltapart$-less {$\chi$PT}
in the considered energy region where the $\deltapart$-pole contributions are 
most prominent, we have performed the corresponding calculations 
to demonstrate explicitly that such an approach is much less efficient and, in fact,
fails to reproduce the experimental data for all analyzed energies, see Fig.~\ref{fig:plots}.
Therefore, in what follows, we focus entirely on the $\deltapart$-full scheme.

We also compare our results with the study~\cite{Pascalutsa:2007wb}
based on $\deltapart$-full {$\chi$PT} with $\delta$-counting.
The double-dashed lines in Fig~\ref{fig:plots} correspond to the 
central value of the $\deltapart^+$ magnetic moment 
$ \mu_{\deltapart^+}=3 (\nucleonmass/\deltamass)\mu_{\nucleon}\approx 2.3\mu_{\nucleon}$\footnote{The authors of 
Ref.~\cite{Pascalutsa:2007wb}
take into account loop corrections to $\mu_{\deltapart^+}$ generating also its imaginary part,
which are of higher order according to the power counting that we implement.}
suggested by the authors.
One observes a somewhat better agreement of our calculation with the data as compared to Ref.~\cite{Pascalutsa:2007wb},
which might be an indication that the power counting scheme based on
the modified small scale expansion that we adopt here is
more efficient for radiative pion photoproduction.
We recall that 
in the $\delta$-counting scheme, the pion-nucleon $q^3$ loops and nucleonic $q^3$ tree-level diagrams
are not included at NLO, see
Sec.~\ref{sec:PowerCountingInTheDeltaRegion} for discussion.
However, we find that their contributions are significant and help
to improve the description of the data, see the discussion of convergence below.
Another approximation used in Ref.~\cite{Pascalutsa:2007wb}, namely
the  expansion in the photon energy $\ego$,
makes the results less reliable when going to higher energies,
especially for $\sqrt{s}=$\SIlist{1313}{\MeV}.

Following Ref.~\cite{Chiang:2004pw}, we also analyze
the ratio $R$ of the differential cross sections for radiative and
ordinary pion photoproduction,  
see Sec:~\ref{sec:kinematics} for the definition.
As can be seen from Fig.~\ref{fig:plots}, our results for this ratio are also in 
reasonable agreement with the data. Moreover, the soft-photon limit
$R\overset{\ego\to 0}{\longrightarrow}1$ is reproduced exactly,
which serves as an additional crosscheck for our calculation.

\begin{figure}[tb]
\includegraphics[width=\textwidth]{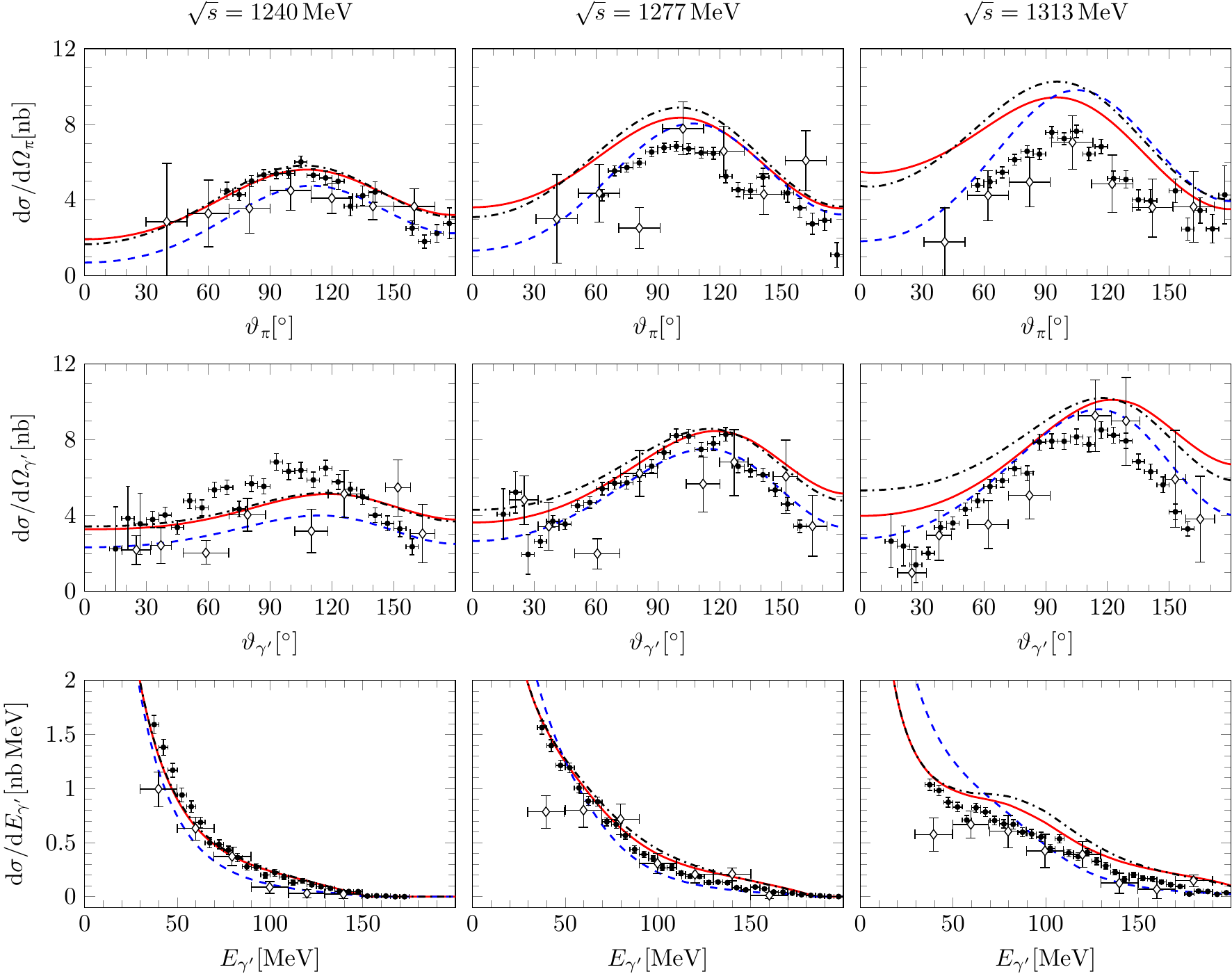}
\caption{Convergence of the small scale expansion for the reaction $\gamma p\to\gamma p\pion^0$
and sensitivity of observables to the value of $\deltamdmind{+}$.
The dashed (solid) lines correspond to the leading (next-to-leading) order result.
The dash-dotted lines denote the result of the calculation with $\deltamdmind{+}=0$.
The data are from \cite{Schumann:2010js} (filled circles) and \cite{Kotulla:2002cg} (diamonds).} \label{fig:convergence}
\end{figure}

Next, we look at the convergence properties of the (modified) small scale expansion
for radiative neutral-pion photoproduction. 
In Fig.~\ref{fig:convergence}, the dashed lines represent the results at leading ($\epsilon_\text{eff}^2$) order,
whereas the solid lines denote the results at next-to-leading ($\epsilon_\text{eff}^3$) order.
The NLO contributions are, in general, reasonably small compared to the LO
result, which indicates a good convergence.
Moreover, taking them into account improves the description of the data considerably.

We also show how sensitive the analyzed observables
are to the value of the $\deltapart^+$ magnetic moment by setting $\deltamdmind{+}=0$ 
($\bar{c}_6^{\deltapart}=3$, $\bar{c}_7^{\deltapart}=-1$),
see dash-dotted lines in Fig.~\ref{fig:convergence}.
The contribution of terms proportional to $\deltamdmind{+}$ is
generally rather small. In fact, it is  
almost negligible at $\sqrt{s}=$\SIlist{1240}{\MeV}  and rises with energy.
Nevertheless, statistically, it turns out to be sufficiently important for a reliable and 
accurate extraction of the $\deltapart^+$ magnetic moment as long as 
higher energies are taken into account.

We have considered three different fit configurations: 
apart from the already mentioned set of observables for three energies,
we also analyzed the cross section data at only the two lowest energies $\sqrt{s}=$\SIlist{1240;1277}{\MeV}
and also performed a fit to the lowest energy
$\sqrt{s}=$\SIlist{1240}{\MeV} only.
It is not obvious a priori that adding higher energies to the fit
would necessarily improve the statistical uncertainty of our extraction,
especially if the perturbative (small scale) expansion fails to converge in 
that higher-energy region.
For the fit to the lowest energy only, the $\chi^2$-function is not clearly peaked,
and a reliable extraction of $\deltamdmind{+}$ is impossible in this case.
This can be expected given a weak sensitivity of the considered observables to the magnetic moment
at this energy as discussed above.
The results of the two other fits are summarized in Table~\ref{tab:fitresCS}.
They are consistent with each
other for what concerns the resulting value of $ \mu_{\deltapart^+}$ (within the
error bars).
The fits to the sets of two and three energies 
yield values of $\chi^2/n_{\textrm{dof}}$ consistent with $1$ within the standard deviation 
$ \sigma_{\text{stat}}=\sqrt{2/n_{\textrm{dof}}} $.
However, the fit to the set of all three energies has the smallest $ \mu_{\deltapart^+}$ uncertainty.
Therefore, we choose this result as our best estimate:
\begin{align}
  \mu_{\deltapart^+}=(\num{1.5}\pm0.2)\,\mu_{\nucleon}\,.
\end{align}
This result agrees with the current {PDG} value within the errors, but
the accuracy is improved.
Notice that  
less then $5\%$  
of the error comes from the uncertainties in the
determination of the {LECs} from pion photoproduction.
We, however, emphasize that the quoted error does not
  take into account the uncertainty in the delta pole position
  employed in our analysis, which is probably sizable.
\begin{table}[t]
\caption{Results of the fit to various sets of energies for $ \mu_{\deltapart^+} $ and 
$ \tilde{c}^{\deltapart}_{67} $. } \label{tab:fitresCS}
\begin{tabular*}{\textwidth}{@{\extracolsep{\fill}}lSSSSS} 
 \hline
  &&&&& \\[-7pt]
{$\ws\; [\si{\MeV}]$} & {$ \tilde{c}^{\deltapart}_{67} $} & {$
                                                           \mu_{\deltapart^+}
                                                            \; [\mu_{\nucleon}]
                                                           $} &
                                                                {$n_{\textrm{dof}}$} & {$\chi^2/n_{\textrm{dof}} $}  & {$ \sigma_{\text{stat}} $}\\[2pt]
  \hline
  &&&&& \\[-7pt]   
1240 to 1313 &-1.5+-0.4&1.5+-0.2&283&1.00 & 0.08\\[2pt]
1240, 1277 &-1.1+-0.7&1.3+-0.4&182&0.93 & 0.10\\[2pt]
\hline
\end{tabular*}
\end{table}

\begin{figure}[tb]
\includegraphics[width=\textwidth]{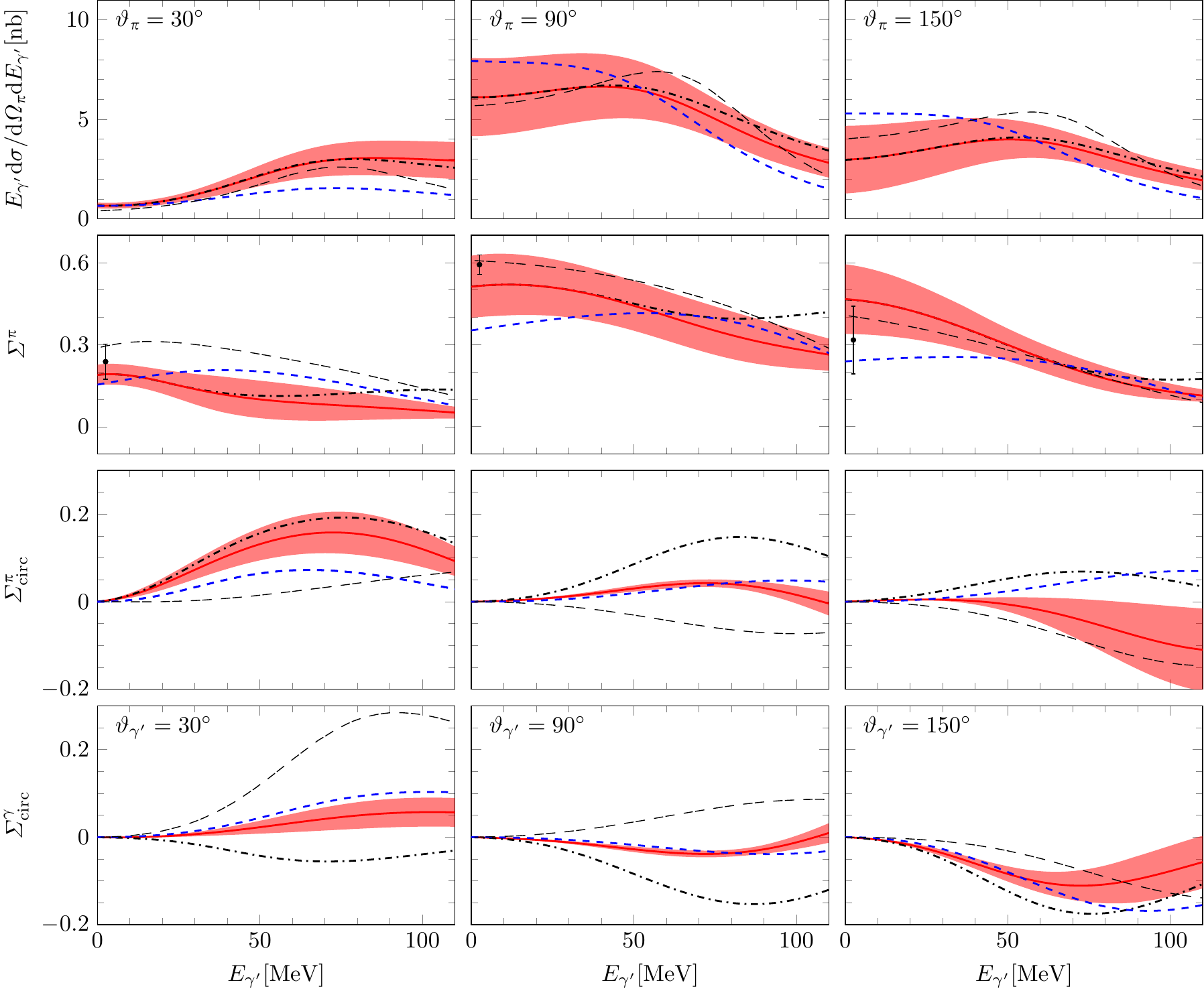}
\caption{Double differential cross section and polarization asymmetries 
for the reaction $\photon\proton \to \photon\proton\pion^0 $
as functions of the outgoing photon energy at $ \ws=\SI{1277}{\MeV} $. 
The dashed and solid lines correspond to the predictions at order $\epsilon_\text{eff}^2$ (LO)
and $\epsilon_\text{eff}^3$ (NLO), respectively.
The bands denote the truncation errors.
The double-dashed lines denote the results of
Ref.~\cite{Pascalutsa:2007wb}, while the dash-dotted lines denote the result of our calculation with $\deltamdmind{+}=0$.
The data points correspond to the linear photon polarization asymmetry
in the reaction $\photon \proton\to\pion^0 \proton$~\cite{Beck:1999ge,Leukel:2001},
as provided in Ref.~\cite{Pascalutsa:2007wb}.}\label{fig:addobs}
\end{figure}

It is interesting to see how sensitive some other observables are to the value of $\deltamdmind{+}$  
even though no experimental information on them is available yet.
We choose the same set of observables and the same  energy $ \ws=\SI{1277}{\MeV} $ 
as considered  in Ref.~\cite{Pascalutsa:2007wb}
for the ease of comparison.
In Fig.~\ref{fig:addobs}, we show the results for the double differential cross section $\ego\dd\sigma/\dd\ompi\dd\ego$,
the linear photon polarization asymmetry $\varSigma^{\pion}$ and the circular photon polarization asymmetries
$\varSigma^{\pion}_{\textrm{circ}}$ and $\varSigma^{\photon}_{\textrm{circ}}$ 
for specific angles of the outgoing pion or photon, see Sec.~\ref{sec:kinematics} for the definitions.
For the linear asymmetry, one can compare the results in the soft-photon limit $\ego\to 0$
with the corresponding asymmetry data for the reaction $\photon
\proton\to\pion^0 \proton$ from Refs.~\cite{Beck:1999ge,Leukel:2001}\footnote{We have extracted those data points from Ref.~\cite{Pascalutsa:2007wb}.}, 
and we observe an agreement of our calculation in this energy regime with the data within the errors. 
We show the leading-order ($\epsilon_\text{eff}^2 $) results (dashed lines)
and the next-to-leading ($\epsilon_\text{eff}^2 $) results (solid lines)
with the truncation error bands
as well as the results with 
the $\deltapart^+$ magnetic moment set to zero (dash-dotted lines).
The sensitivity of the double differential cross section $\ego\dd\sigma/\dd\ompi\dd\ego$
to the value of $\deltamdmind{+}$ is similar to the case of the unpolarized single differential cross sections.
Our results for this observable practically agree with Ref.~\cite{Pascalutsa:2007wb} within the error bands
for the pion angles $\vartheta_{\pion}=30^{\circ} $ and $\vartheta_{\pion}=90^{\circ} $.
For $\vartheta_{\pion}=150^{\circ} $, the agreement is slightly worse.
The convergence pattern follows essentially the one of $\dv{\sigma}{\ompi}$. 

The magnetic-moment contribution to the  polarization observables is in general more pronounced, see Fig.~\ref{fig:addobs}.
On the other hand, the convergence is rather poor in some cases,
which is no surprise since there are subtle cancellations
among various contributions typical for polarization asymmetries.
This can explain the disagreement with the results of Ref.~\cite{Pascalutsa:2007wb}.
In order to improve the description of these observables, one should
obviously include higher-order terms in the small scale expansion.
In this case, a more accurate treatment of pion photoproduction  
will be also necessary  including a more rigorous approach to uncertainties.
In particular, one might need to perform a combined
fit to the photoproduction and radiative-photoproduction observables.

\subsection{Radiative charged-pion photoproduction}
For the charged-pion channel, we repeat the calculations we have done for the neutral channel
and provide our predictions for the same set of observables and for the same set of energies.
Unfortunately, no experimental data are available for this channel. 
Therefore, it is instructive to analyze the sensitivity of various observables to the
$\deltapart$ magnetic moment for future experiments.

The $ \photon\proton\to\photon\neutron\pion^+ $ amplitude depends on the
magnetic moment of $\deltapart^+$ and $\deltapart^0$ (or, equivalently, on $\bar{c}_6^{\deltapart}$ and $\bar{c}_7^{\deltapart}$).
We fix the value of $\deltamdmind{+}$ from the fit to the neutral
channel,  see the previous subsection.
We adjust the remaining linear combination of  $\bar{c}_6^{\deltapart}$ and $\bar{c}_7^{\deltapart}$
to the value of $\deltamdmind{++}$
extracted  from the reaction $ \pion^+ p\to \pion^+\proton\photon $ \cite{Bosshard:1991zp,LopezCastro:2000cv}:
\begin{align}
\deltamdmind{++}=\num{6.14+-0.51}\mu_{\nucleon},
\end{align}
which yields 
\begin{align}
\bar{c}_6^{\deltapart} = \num{-14.9 +- 2.1}\quad\textrm{and}\quad \bar{c}_7^{\deltapart} = \num{9.0 +- 1.5}.
\end{align}

\begin{figure}[tb]
\includegraphics[width=\textwidth]{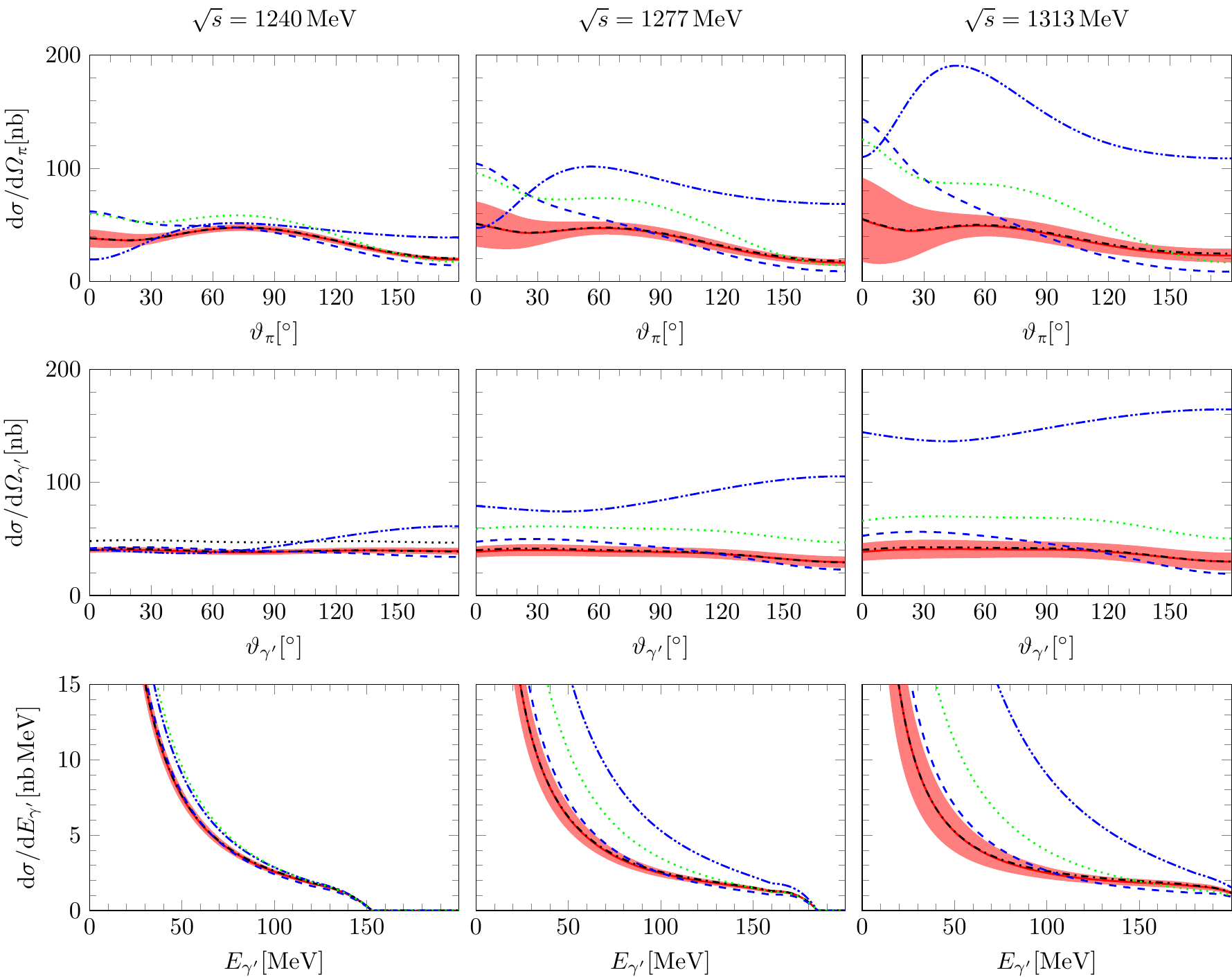}
\caption{Differential cross sections for the reaction 
$ \photon\proton \to \photon\neutron\pion^+ $. 
The dashed-double-dotted lines correspond to the $\deltapart$-less
$ q^3 $ calculation. The dotted, dashed and solid lines denote the
LO, NLO and N$^2$LO $\deltapart$-full results, respectively, with the
bands indicating the truncation errors.
The dash-dotted lines denote the result of the calculation with $\deltamdmind{+}=\deltamdmind{0}=0$.
} \label{fig:Ch2:plots}
\end{figure}

The LO, NLO and N$^2$LO results of our $\deltapart$-full calculation for the single differential unpolarized observables
for three energies  $\sqrt{s}=$\SIlist{1240;1277;1313}{\MeV}
are shown in Fig.~\ref{fig:Ch2:plots} with the bands indicating the
truncation errors.
Also shown are the N$^2$LO results obtained in the $\deltapart$-less scheme.
The difference between the $\deltapart$-less and $\deltapart$-full approaches is sizable
and increases with energy very rapidly.
The convergence of the EFT expansion as one goes from LO to N$^2$LO is satisfactory for the 
lowest energy, but it becomes less convincing for the energies $\sqrt{s}=$\SIlist{1277;1313}{\MeV}.

The unpolarized observables are practically insensitive
to the value of the $\deltapart$ magnetic moment, 
as can be seen by looking at the dotted curves in Fig.~\ref{fig:Ch2:plots}
corresponding to $\deltamdmind{+}=\deltamdmind{0}=0$, which almost coincide with the full results.
This is due to the fact that the leading-order amplitude for the charged channel 
is not $1/\nucleonmass$-suppressed in contrast to the neutral channel.
As a result, the absolute values of the cross sections in the charged channel are an order of magnitude larger.

\begin{figure}[tb]
\includegraphics[width=\textwidth]{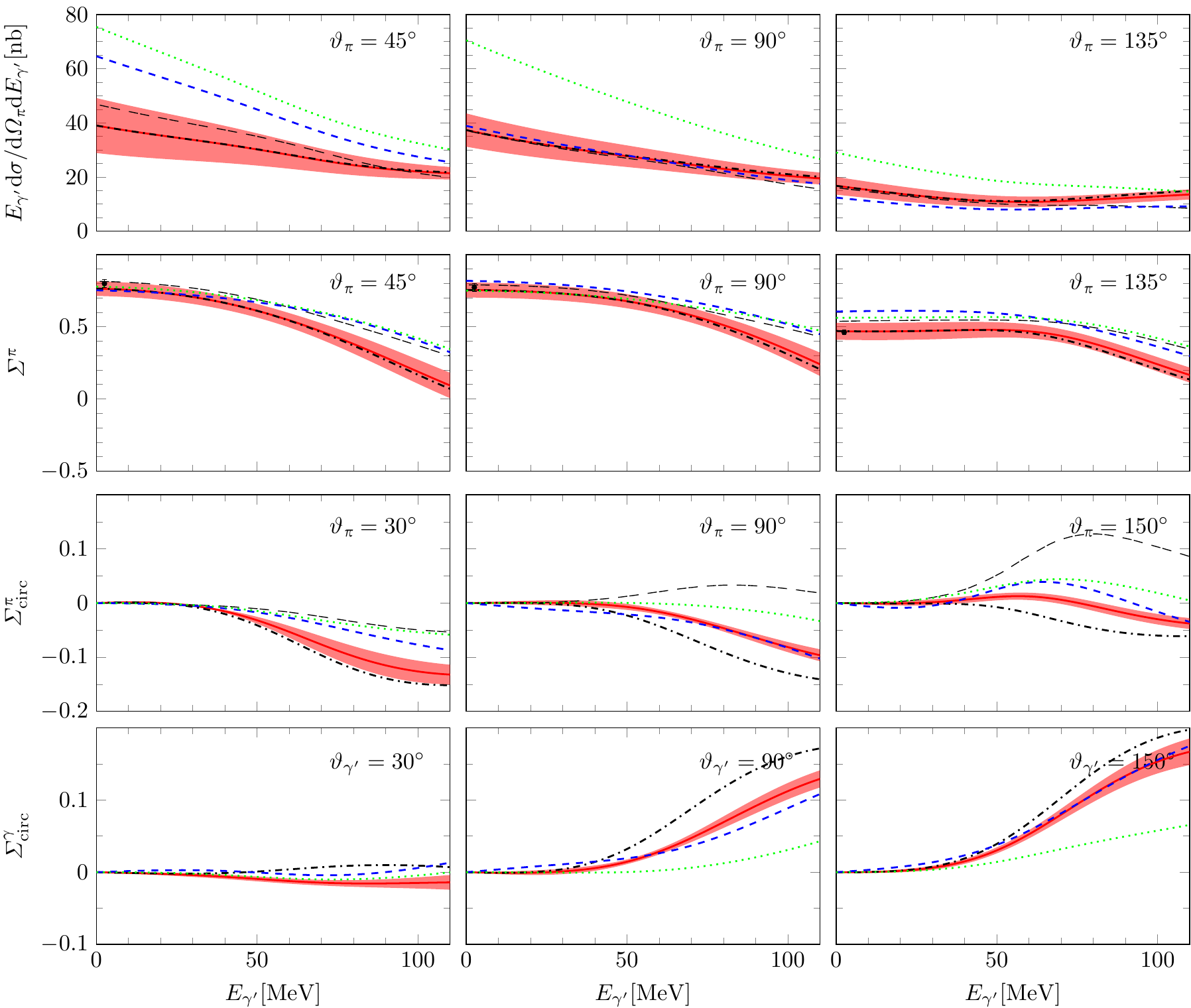}
\caption{Double differential cross section and polarization asymmetries 
for the reaction  $\photon\proton \to \photon\neutron\pion^+ $
as functions of the outgoing photon energy at $ \ws=\SI{1277}{\MeV} $. 
The dotted, dashed and solid lines denote the
LO, NLO and N$^2$LO $\deltapart$-full results, respectively, with the bands indicating the truncation errors.
The dash-dotted lines denote the results of the calculation with $\deltamdmind{+}=\deltamdmind{0}=0$.
The double-dashed lines denote the results of Ref.~\cite{Pascalutsa:2007wb}.
The data points correspond to the linear photon polarization asymmetry
in the reaction $\photon \proton\to\pion^+ \neutron$~\cite{Beck:1999ge},
as provided in Ref.~\cite{Pascalutsa:2007wb}.}\label{fig:Ch2:addobs}
\end{figure}

The results of the N$^2$LO  $\deltapart$-full calculation for the double differential cross section
$\ego\dd\sigma/\dd\ompi\dd\ego$
and the polarization asymmetries $\varSigma^{\pion}$, $\varSigma^{\pion}_{\textrm{circ}}$ and
$\varSigma^{\photon}_{\textrm{circ}}$ are depicted in Fig.~\ref{fig:Ch2:addobs}.
As in the case of the neutral channel, our calculation
of the linear asymmetry $\varSigma^{\pion}$ in the ultrasoft-photon limit agrees with 
the experimental data for the reaction $\photon \proton\to\pion^+ \neutron$.

The most sensitive to the $\deltapart$-magnetic-moment contribution are
the circular photon polarization asymmetries
$\varSigma^{\pion}_{\textrm{circ}}$ and
$\varSigma^{\photon}_{\textrm{circ}}$\footnote{Notice that the contribution of $\deltamdmind{0}$ is several times smaller compared to $\deltamdmind{+}$
in agreement with our power-counting analysis in Sec.~\ref{sec:PowerCountingInTheDeltaRegion}.}.
This confirms the findings of Ref.~\cite{Pascalutsa:2007wb}.
However, our results, in general, do not agree with the results of Ref.~\cite{Pascalutsa:2007wb}
(double-dashed lines in Fig.~\ref{fig:Ch2:addobs})
within the errors.
Analogously to the neutral channel, this is seemingly a consequence of the slow convergence
and subtle cancellations, especially for the
polarization asymmetries. 
Therefore, as in the case of the radiative $\pion^0$-photoproduction,
in order to be able to perform a reliable analysis of the polarization asymmetries,
one should presumably go to higher orders in the small scale expansion.

\newpage
\section{Summary and outlook} \label{sec:summary}
We have studied radiative pion photoproduction in the $\deltapart$ region within  covariant 
chiral perturbation theory including the $\deltapart$(1232) resonance as an explicit degree of freedom.
Specifically, we have analyzed the reactions 
$\photon\proton \to \photon\proton\pion^0 $ (neutral channel) and
$ \photon\proton\to\photon\neutron\pion^+ $ (charged channel).
The reaction amplitude has been calculated up to next-to-leading order for
the neutral channel and up to next-to-next-to-leading order for
the charged channel in the small scale expansion, modified for the
case of the $\deltapart$ region. These contributions include 
the full set of pion-nucleon order-$q^3$ loop diagrams as well as certain $\deltapart$-pole tree-level graphs
including the loop corrections to them. 

Several low-energy constants entering as input parameters for our calculation
have been obtained from a fit to the pion-photoproduction multipoles 
in the threshold and $\deltapart$ regions using the scheme consistent with our treatment of 
radiative pion photoproduction, but with a simplified treatment of experimental uncertainties.

The main goal of our study was an indirect determination of the dipole magnetic moment
of the  $\deltapart^+$ particle by fitting it to the available
experimental data for the unpolarized differential cross sections
in the reaction $\photon\proton \to \photon\proton\pion^0 $ for three values of initial energy.
The obtained fit is in good agreement with the data within errors.
Given the observed satisfactory convergence of the small scale
expansion for these observables,
this has allowed us to perform an accurate extraction of $\deltamdmind{+}$ with the resulting value
\begin{align}
  \mu_{\deltapart^+}=(\num{1.5}\pm0.2)\,\mu_{\nucleon}\,.
\end{align}
In comparison with previous extractions based on phenomenological models,
our result relies on a systematic EFT approach,
whereas in comparison with earlier EFT studies, our scheme
provides a more reliable estimate of theoretical errors 
by means of the Bayesian approach. Note that one should be cautious when interpreting
the truncation uncertainties that we provide as they are estimated based 
on the information on only two orders in the EFT expansion.

We also performed the calculations within the $\deltapart$-less scheme.
As expected for such an energy regime, the $\deltapart$-less approach
turns out to be much less efficient than the $\deltapart$-full
framework, and it fails to reproduce the experimental data at the considered order.

We also made predictions for several other observables, including
the linear and circular photon polarization asymmetries in order to
check their sensitivity to the $\deltapart$ magnetic moment.
Some of the polarization observables appear to be more sensitive to 
the value of $\deltamdmind{+}$ than the unpolarized differential cross sections.
However, the convergence of the small scale expansion in these cases is rather poor.
Therefore, a reliable analysis of these observables would require going to higher orders.

We also analyzed the same set of observables for the charged channel,
for which no experimental data are available at present.
We used the value of $\deltamdmind{+}$ from our fit to the neutral channel
and the value of $\deltamdmind{0}$ extracted from the reaction $ \pion^+ p\to \pion^+\proton\photon $.
We found that only the circular photon polarization asymmetries
possess sizable sensitivity
to the $\deltapart$ magnetic moment, however, with the same convergence issues as in the
case of the neutral channel.

Our results suggest that going to higher orders in the small scale expansion 
and using a more rigorous uncertainty-estimation procedure for pion photoproduction
may allow one to further improve the accuracy of the presented analysis.

\section*{Acknowledgments} 
We would like to thank Jambul Gegelia, Arseniy Filin and Patrick
Reinert for helpful discussions and Ulf-G.~Mei{\ss}ner 
and Astrid Nathalie Hiller Blin for useful
comments on the manuscript.
This work was supported in part by DFG  (Grant No. 426661267), 
by DFG and NSFC through funds provided to the
Sino-German CRC 110 ``Symmetries and the Emergence of Structure in QCD" (NSFC
Grant No.~11621131001, DFG Grant No.~TRR110) and by BMBF (Grant
No. 05P18PCFP1).

\newpage
\appendix
\section{Feynman diagrams for radiative pion photoproduction} \label{sec:Feynman-diagrams}
\renewcommand\thefigure{\thesection.\arabic{figure}}
\setcounter{figure}{0} 
In this section we present all considered Feynman diagrams for
radiative pion\hyp{}photoproduction.
\begin{itemize}
\item[--]The leading-order $O(q^1)$ tree diagrams are given in Fig.~\ref{fig:basictrees}.
\begin{figure}[p]
\includegraphics[width=\textwidth]{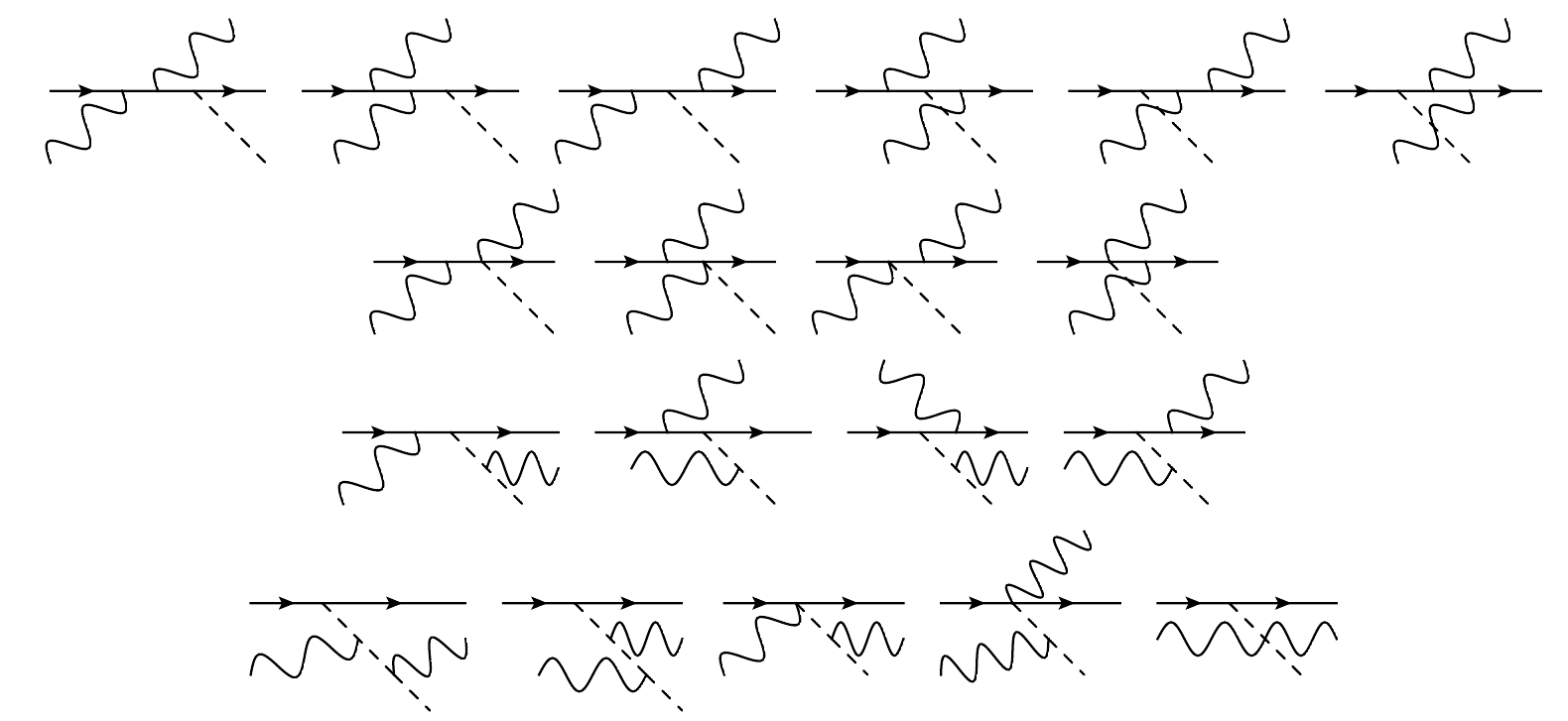}
\caption{Leading-order $O(q^1)$ tree-level radiative-pion-photoproduction diagrams. All vertices are of the lowest order, 
i.e. from $\lpin^{(1)}$ and $\lpipi^{(2)}$.} \label{fig:basictrees}
\end{figure}
\item[--]The order-$q^2$ tree-level diagrams are shown in Fig.~\ref{fig:qTo2Diagrams}.
\begin{figure}[htbp]
\includegraphics[width=\textwidth]{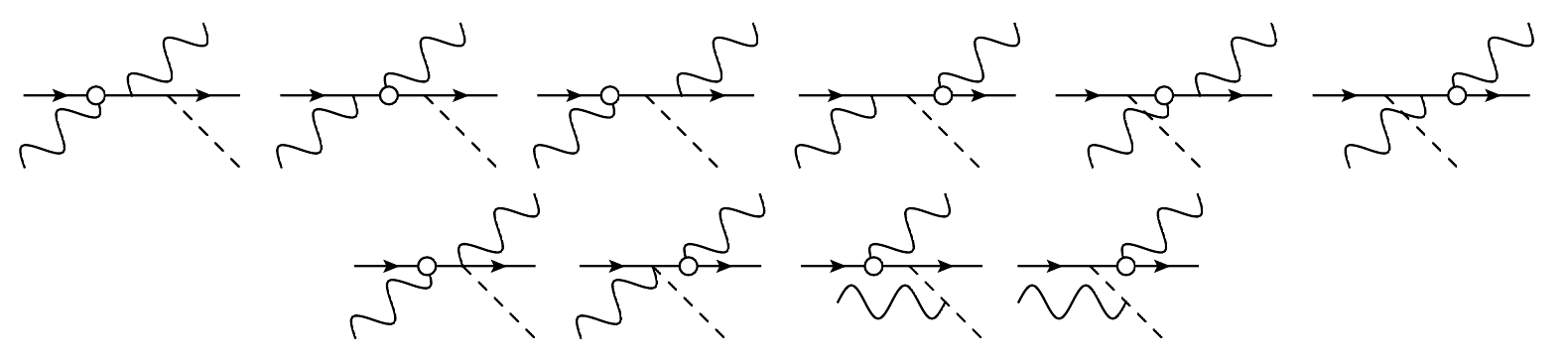}
  \caption{Tree-level radiative-pion-photoproduction diagrams of order $q^2$. The open circles represent second-order vertices from $\lpin^{(2)}$. 
Diagrams with crossed photon lines and 
the second-order corrections to the nucleon self-energy are not shown.} \label{fig:qTo2Diagrams}
\end{figure}
\item[--]The order-$q^3$ tree-level diagrams are shown in
Figs.~\ref{fig:qTo3Diagrams},~\ref{fig:highertrees}.
The diagrams in Fig.~\ref{fig:qTo3Diagrams} are obtained from the leading order tree diagrams of Fig.~\ref{fig:basictrees}
by either replacing a leading-order vertex with a subleading one or by inserting a nucleon or pion self-energy vertex. 
Fig.~\ref{fig:highertrees} contains additional $q^3$ tree-level topologies.
\begin{figure}[htbp]
\includegraphics[width=0.9\textwidth]{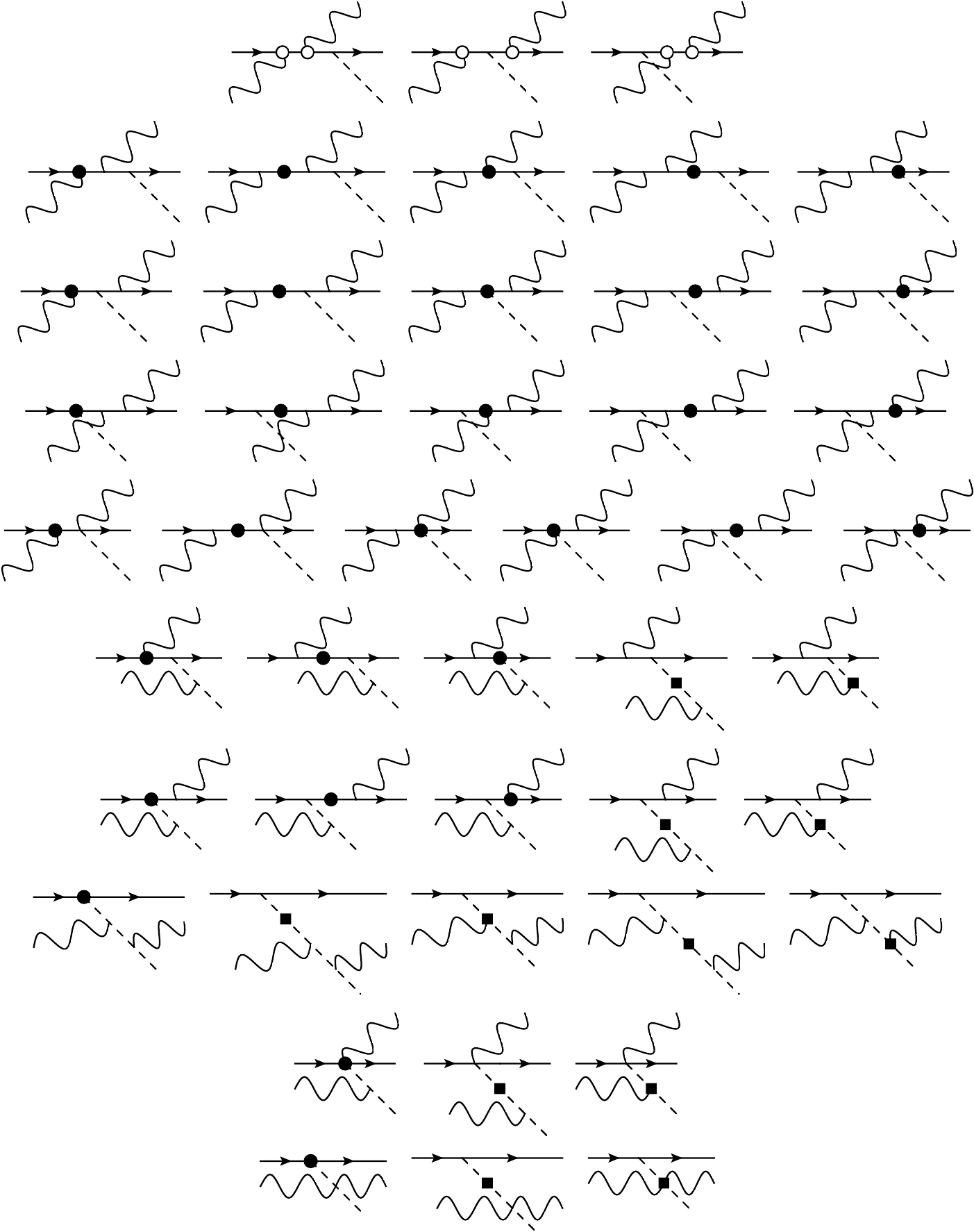}
\caption{Third-order tree-level radiative-pion-photoproduction topologies. The open circles, the filled circles and the squares represent vertices
from $\lpin^{(2)}$, $\lpin^{(3)}$ and $\lpipi^{(4)} $, respectively. 
Diagrams with crossed photon lines are not shown.} \label{fig:qTo3Diagrams}
\end{figure}
\begin{figure}[htbp]
\includegraphics[width=0.8\textwidth]{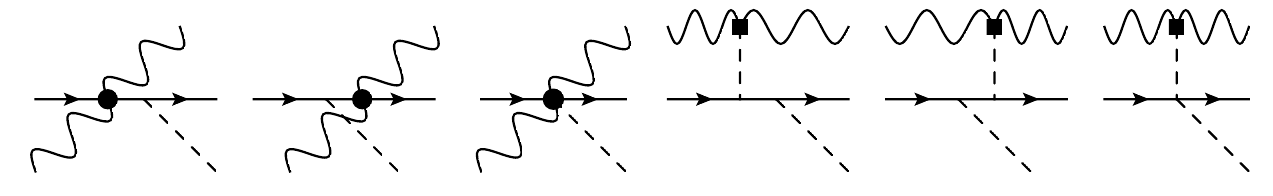}
\caption{Further tree-level radiative-pion-photoproduction topologies of order $q^3$. The filled circles are vertices from $\lpin^{(3)}$, the square represents the vertex from $\lgen_{\textrm{WZW}}^{(4)}$.} \label{fig:highertrees}
\end{figure}
\item[--]
Pion-nucleon loop diagrams appear at order $q^3$.
We split them into seven sets, which are shown in
Figs.~\ref{fig:Aloops}-\ref{fig:Ex2loops}.
The first four sets in Figs.~\ref{fig:Aloops}-\ref{fig:DEFloops}
are obtained from the four rows of tree-level diagrams in Fig.~\ref{fig:basictrees} by attaching a pion loop, wherever possible.
The diagrams in Fig.~\ref{fig:MAloops} are obtained from the leading-order nucleon pole graphs in the pion-photoproduction amplitude
(first two diagrams in Fig.~\ref{fig:basictreespiphp}) by attaching a pion loop, wherever possible, and attaching a photon 
to the pion inside the loop.
The diagrams in Fig.~\ref{fig:Ex1loops} are obtained in the same way from the leading-order pion pole graph and the contact graph in the pion-photoproduction amplitude
(last two diagrams in Fig.~\ref{fig:basictreespiphp}).
Fig.~\ref{fig:Ex2loops} contains all diagrams with two photons coupled to the pion inside a loop.
\begin{figure}[htbp]
\includegraphics[width=\textwidth]{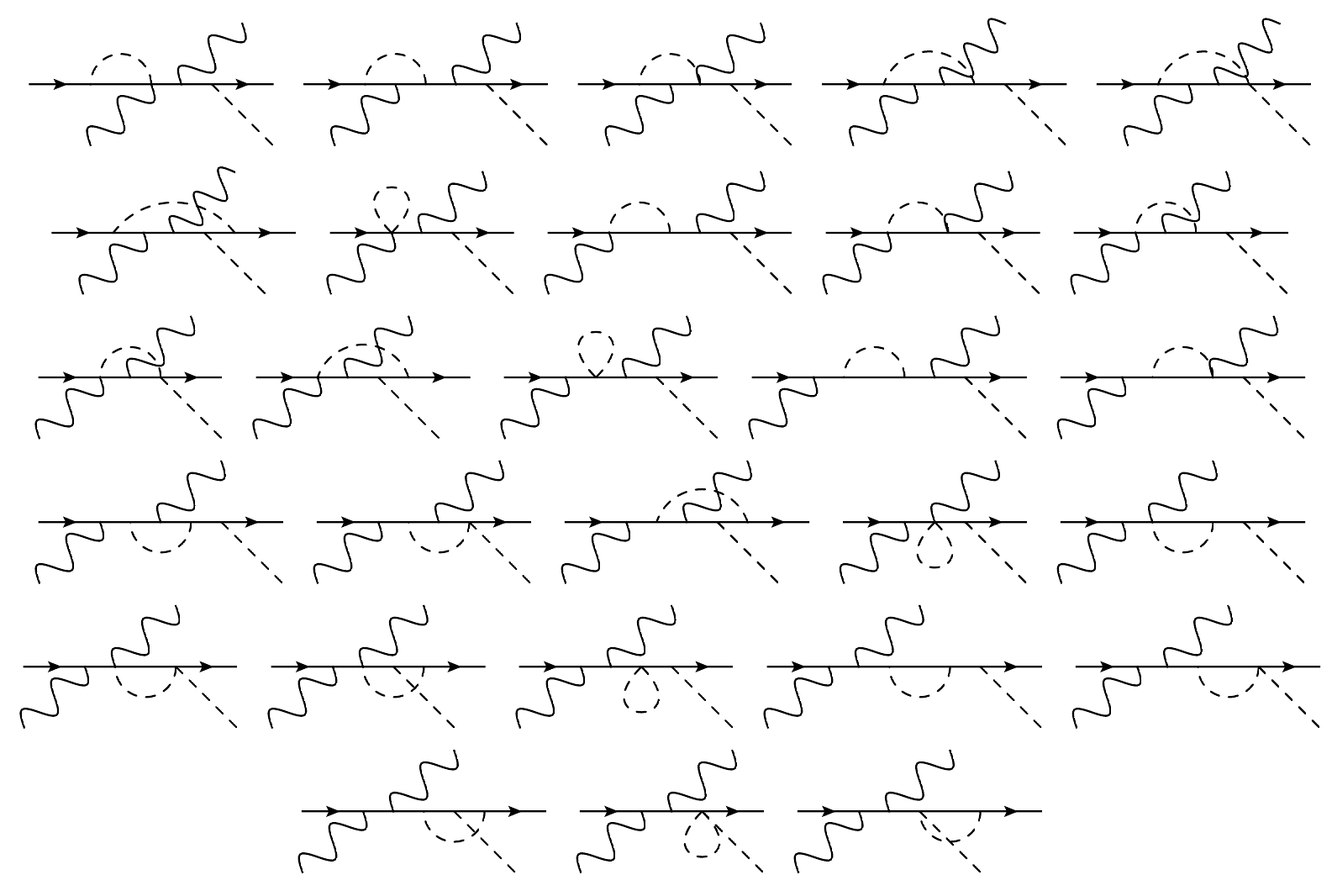}
\caption{First set of radiative-pion-photoproduction loop diagrams. Diagrams that can be obtained by permutations of the external boson (photon and/or pion) lines
are not shown.} \label{fig:Aloops}
\end{figure}
Note that every diagram depicted in figure~\ref{fig:Aloops} actually stands for six diagrams 
that can be obtained by permutations of the external boson (photon and/or pion) lines as in the first row of figure~\ref{fig:basictrees}. 
In other figures, when specified in a figure caption, crossed diagrams
(corresponding to crossing photon and nucleon lines) are not shown. 
\begin{figure}[htbp]
\includegraphics[width=0.9\textwidth]{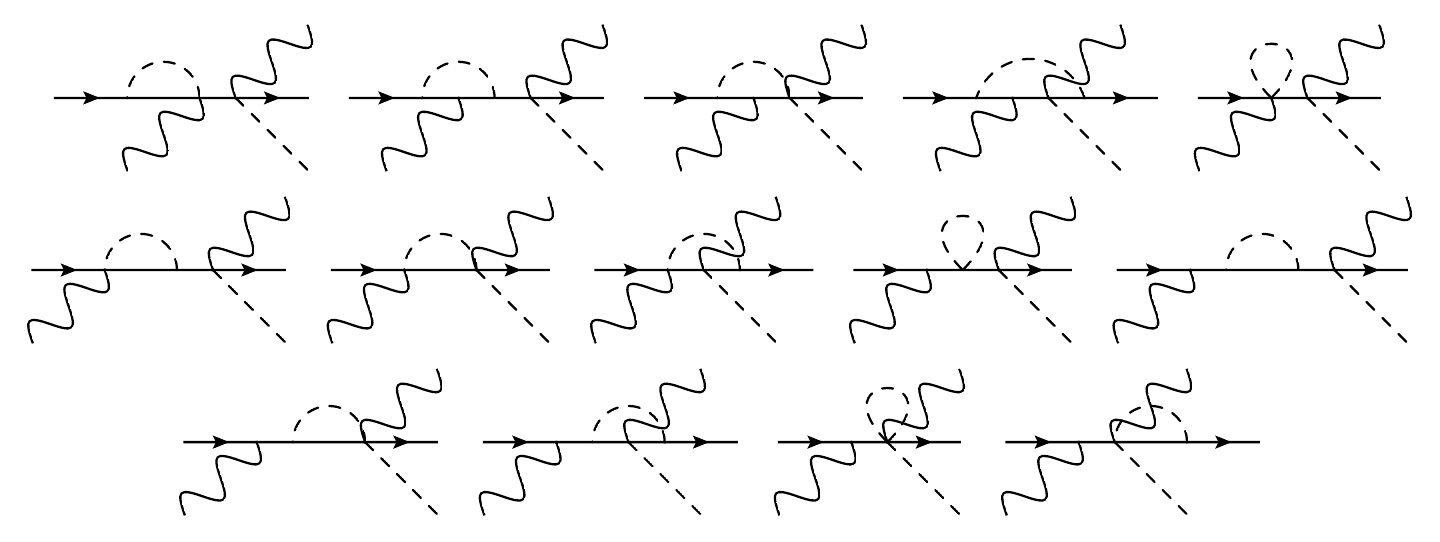}
\caption{Second set of radiative-pion-photoproduction loop diagrams. Crossed diagrams are not shown.} \label{fig:Bloops}
\end{figure}
\begin{figure}[htbp]
\includegraphics[width=0.8\textwidth]{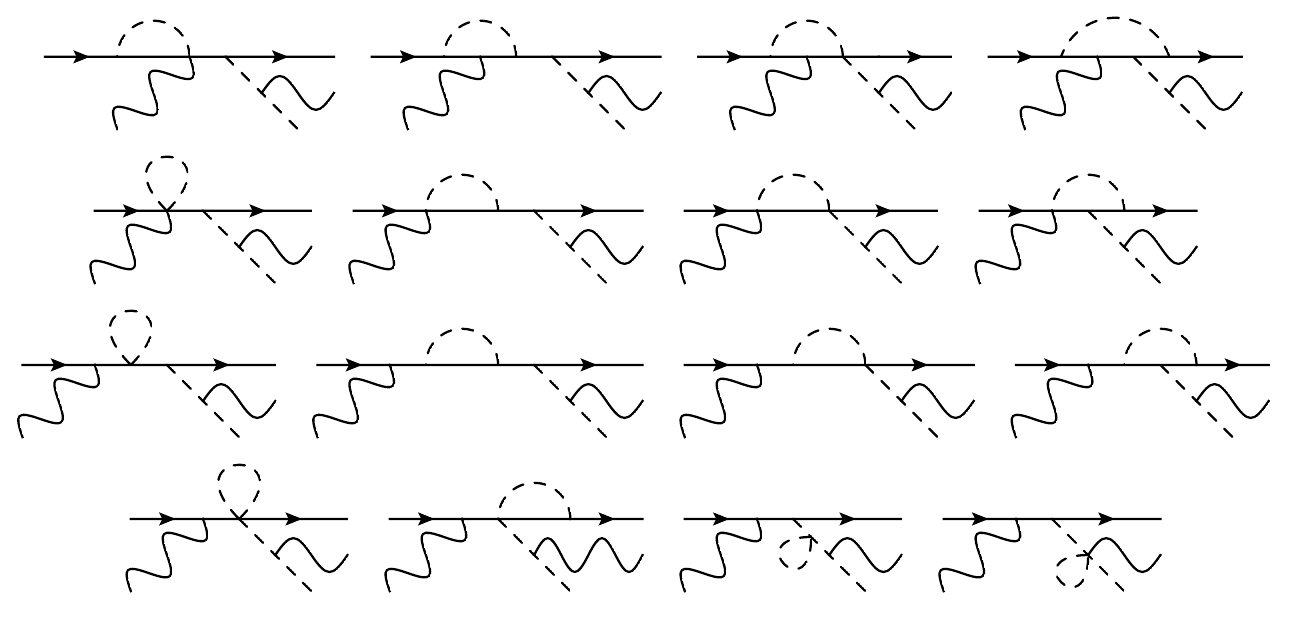}
\caption{Third set of radiative-pion-photoproduction loop diagrams. Crossed diagrams are not shown.} \label{fig:Cloops}
\end{figure}
\begin{figure}[htbp]
\includegraphics[width=\textwidth]{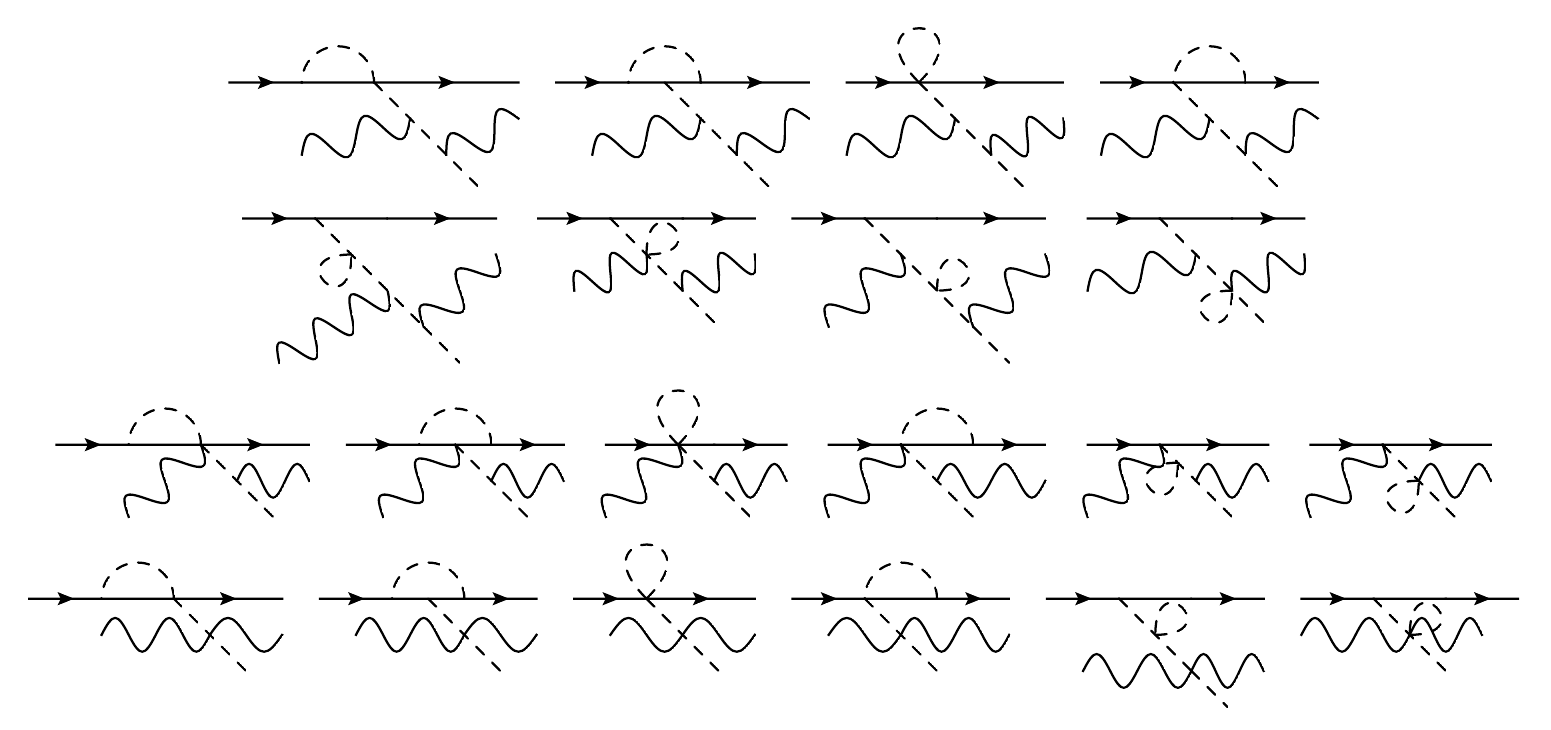}
\caption{Fourth set of radiative-pion-photoproduction loop diagrams. Diagrams with crossed photon lines are not shown.} \label{fig:DEFloops}
\end{figure}
\begin{figure}[htbp]
\includegraphics[width=\textwidth]{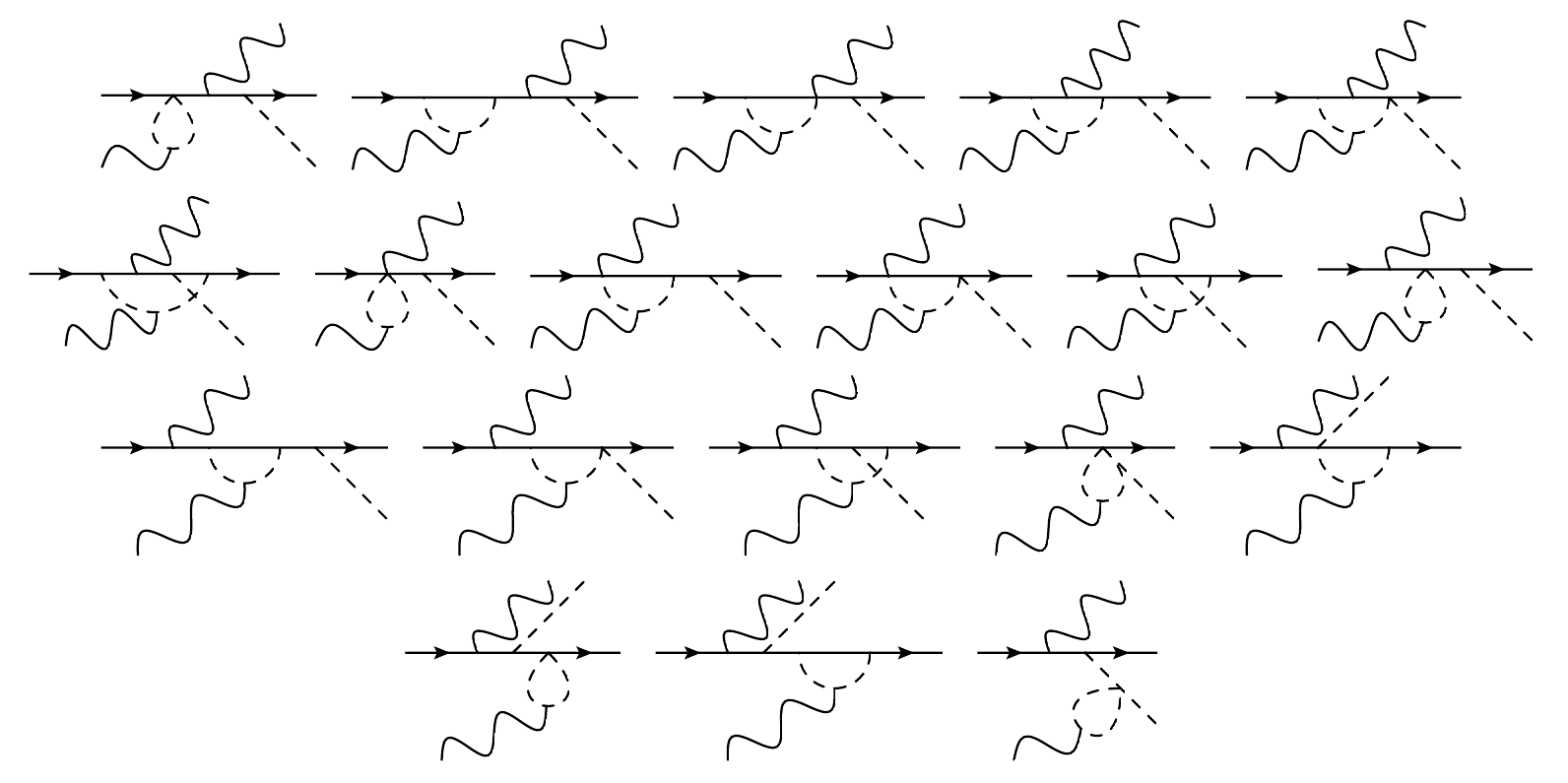}
\caption{Fifth set of radiative-pion-photoproduction loop diagrams. Crossed diagrams are not shown.} \label{fig:MAloops}
\end{figure}
\begin{figure}[htbp]
\includegraphics[width=\textwidth]{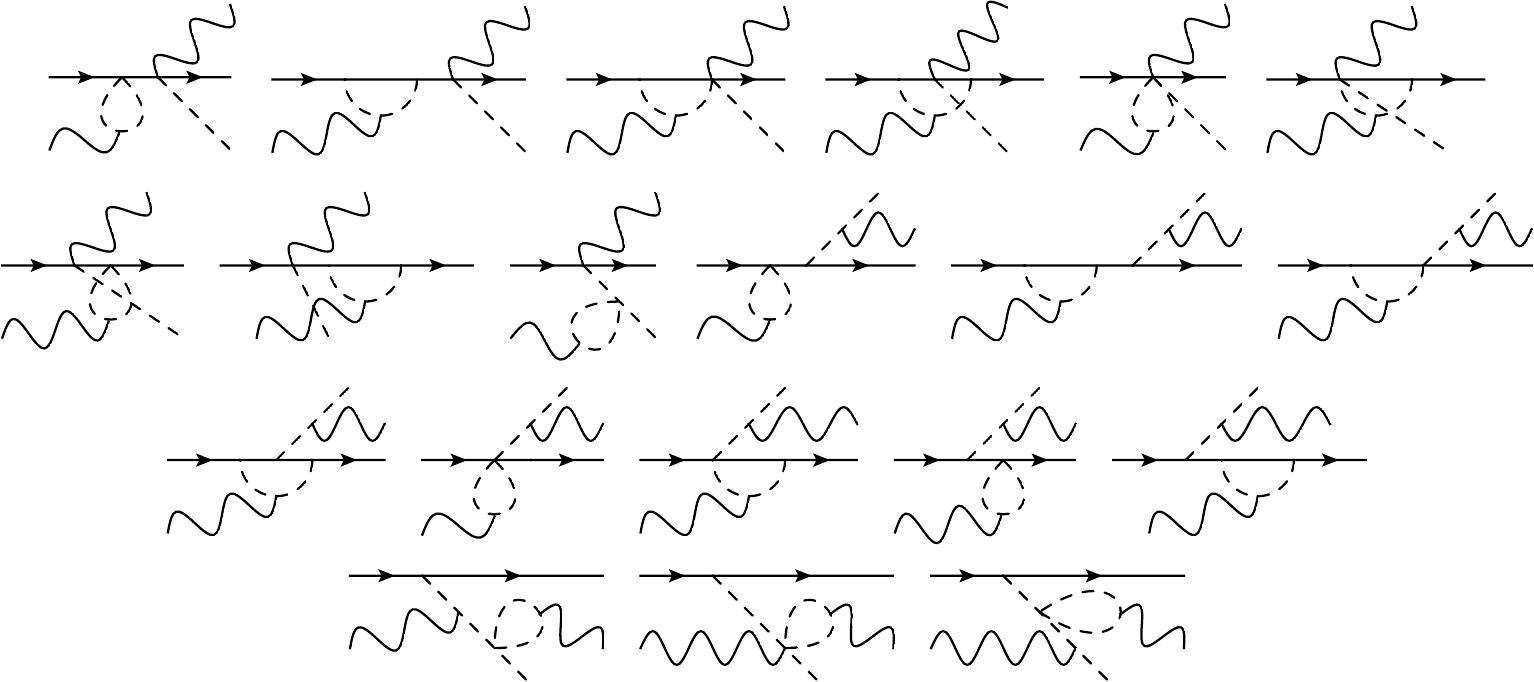}
\caption{Sixth set of radiative-pion-photoproduction loop diagrams. Diagrams with crossed photon lines are not shown.} \label{fig:Ex1loops}
\end{figure}
\begin{figure}[htbp]
\includegraphics[width=\textwidth]{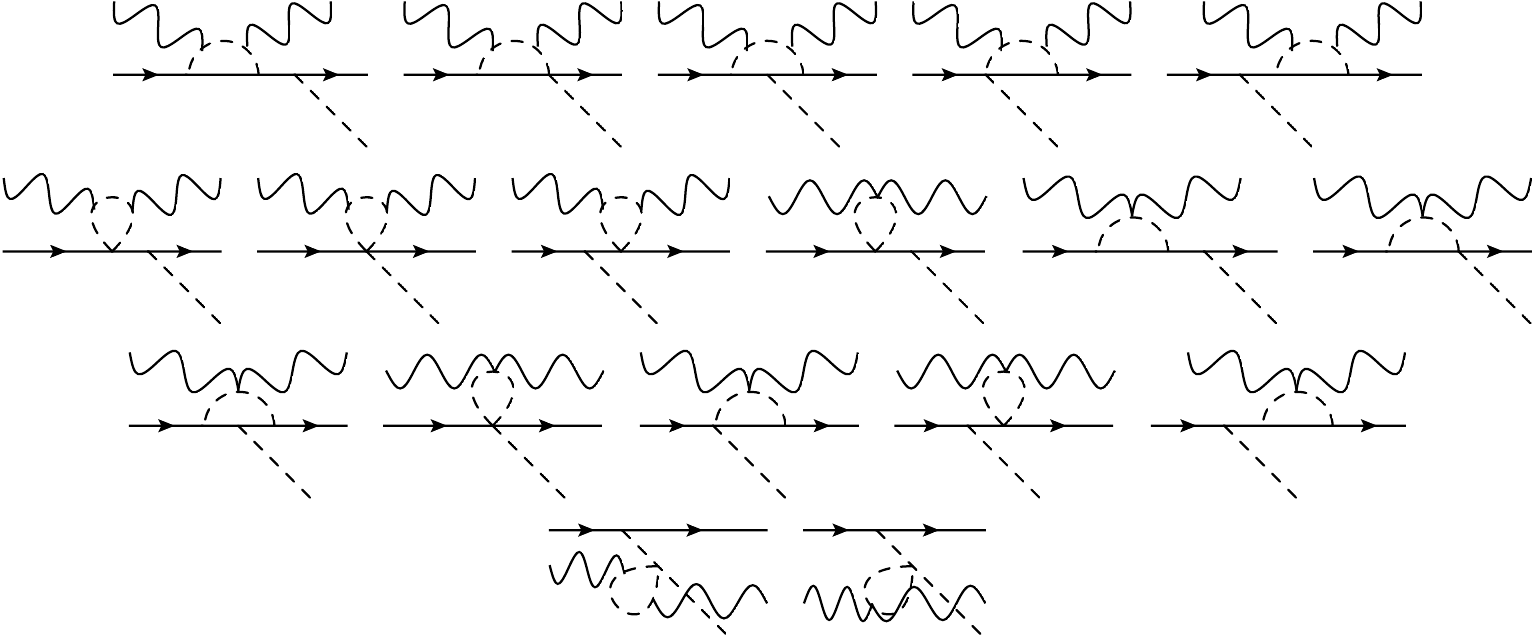}
\caption{Seventh set of radiative-pion-photoproduction loop diagrams. Diagrams with crossed photon lines are not shown.} \label{fig:Ex2loops}
\end{figure}
\item[--]
Tree-level diagrams with $\deltapart$ lines of order $\epsilon^2$ and $\epsilon^3$ are shown in Fig.~\ref{fig:deltadiagslo}
and Fig.~\ref{fig:deltaTreeE3}, respectively.
\begin{figure}[htbp]
\includegraphics[width=\textwidth]{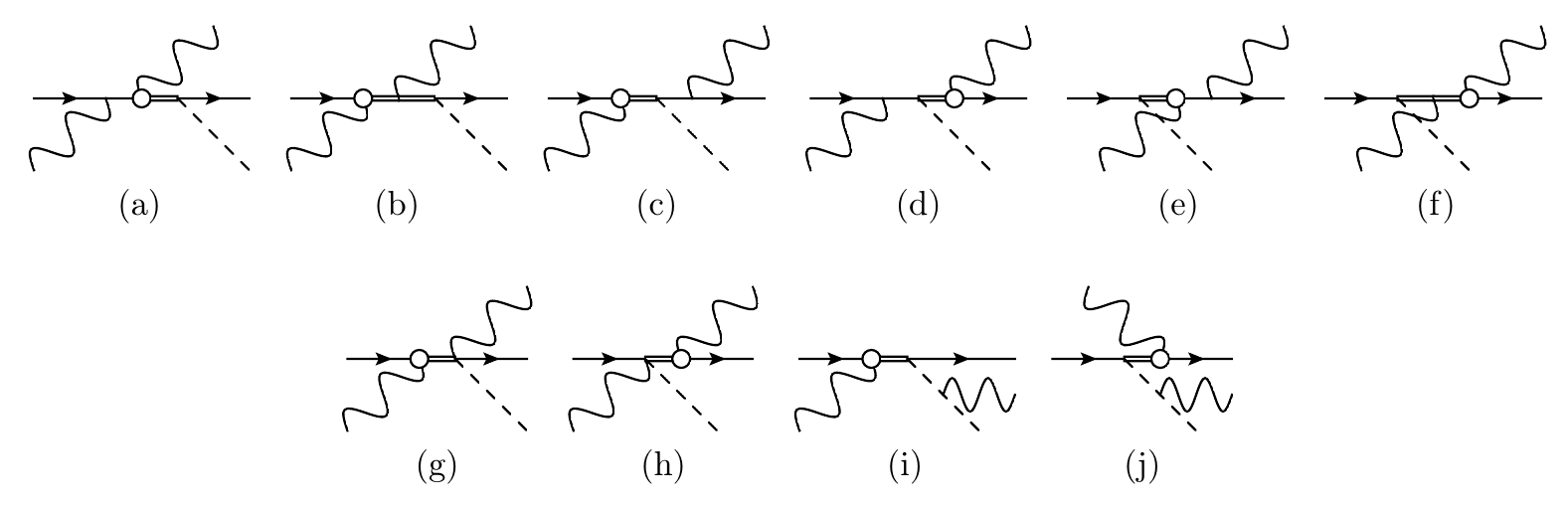}
\caption{Tree-level radiative-pion-photoproduction diagrams with
  $\deltapart$ lines of order $\epsilon^2$. 
The open circle denotes the second-order vertex from $\lpind^{(2)} $. Diagrams with crossed photon lines are not shown.} \label{fig:deltadiagslo}
\end{figure}
\begin{figure}[htbp]
\includegraphics[width=\textwidth]{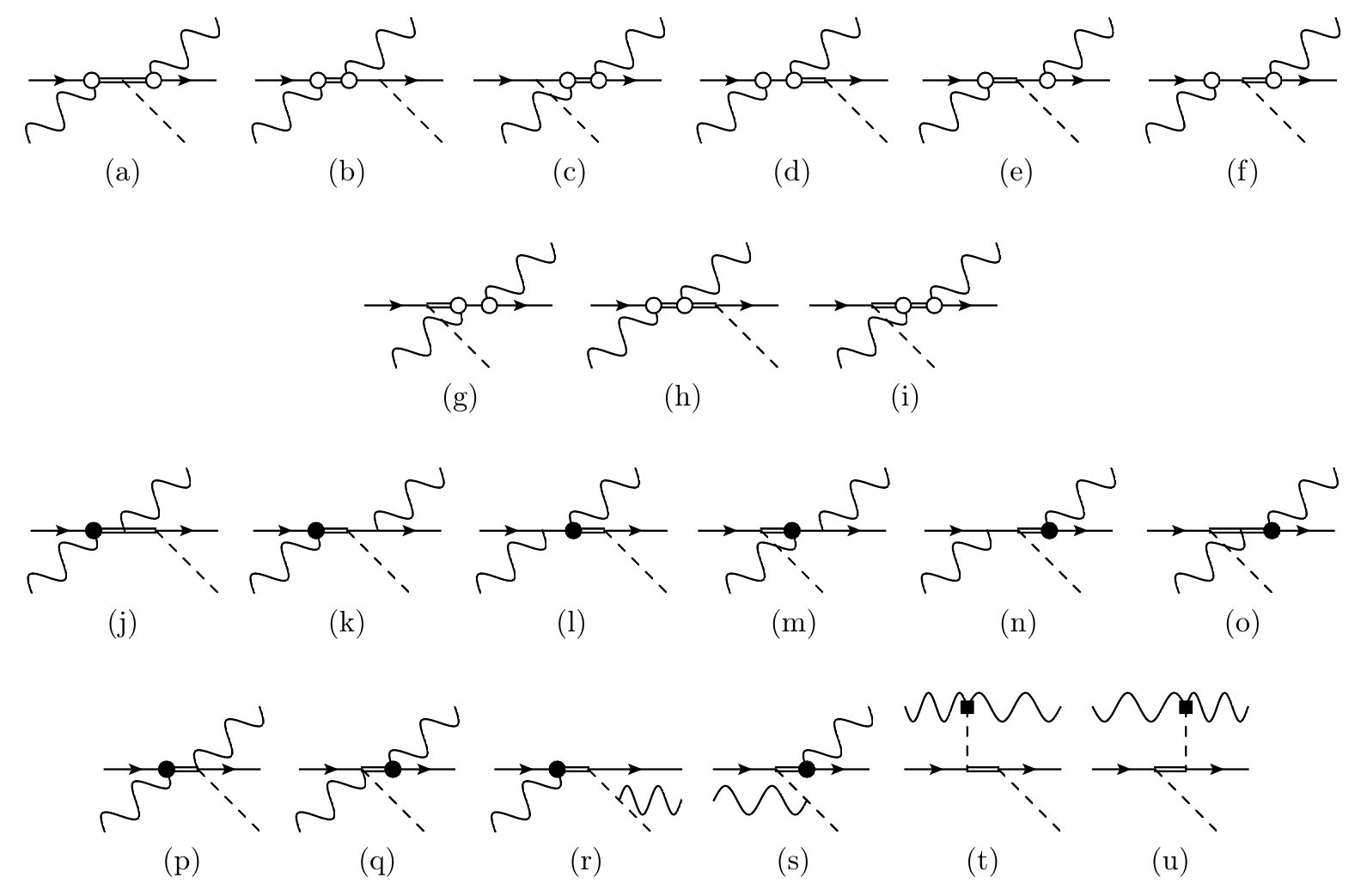}
\caption{Tree-level radiative-pion-photoproduction diagrams with
  $\deltapart$ lines of order $\epsilon^3$. 
The open circles denote second-order vertices from $\lpin^{(2)}$, $\lpind^{(2)}$ and $\lpid^{(2)}$.
The filled circle is the vertex from $\lpind^{(3)}$, the square represents the vertex from $\lgen_{\textrm{WZW}}^{(4)}$.
Diagrams with crossed photon lines are not shown.} \label{fig:deltaTreeE3}
\end{figure}
\item[--]
The loop corrections to the $\deltapart$-pole graphs of order $\epsilon^3$ considered in this work are presented in Figs.~\ref{fig:DeltaLoops1}-\ref{fig:DeltaLoops3},
where the last set (Fig.\ref{fig:DeltaLoops3}) contributes only to radiative charged-pion photoproduction.
The diagrams relevant for both reaction channels ($\photon\proton \to \photon\proton\pion^0$ and $ \photon\proton\to\photon\neutron\pion^+ $)
are split into the set that contains loop corrections to the electromagnetic $\nucleon\Delta$ transition form factor as a subgraph (Fig.~\ref{fig:DeltaLoops1})
and the remaining graphs (Fig.~\ref{fig:DeltaLoops2}).
\begin{figure}[htbp]
\includegraphics[width=0.8\textwidth]{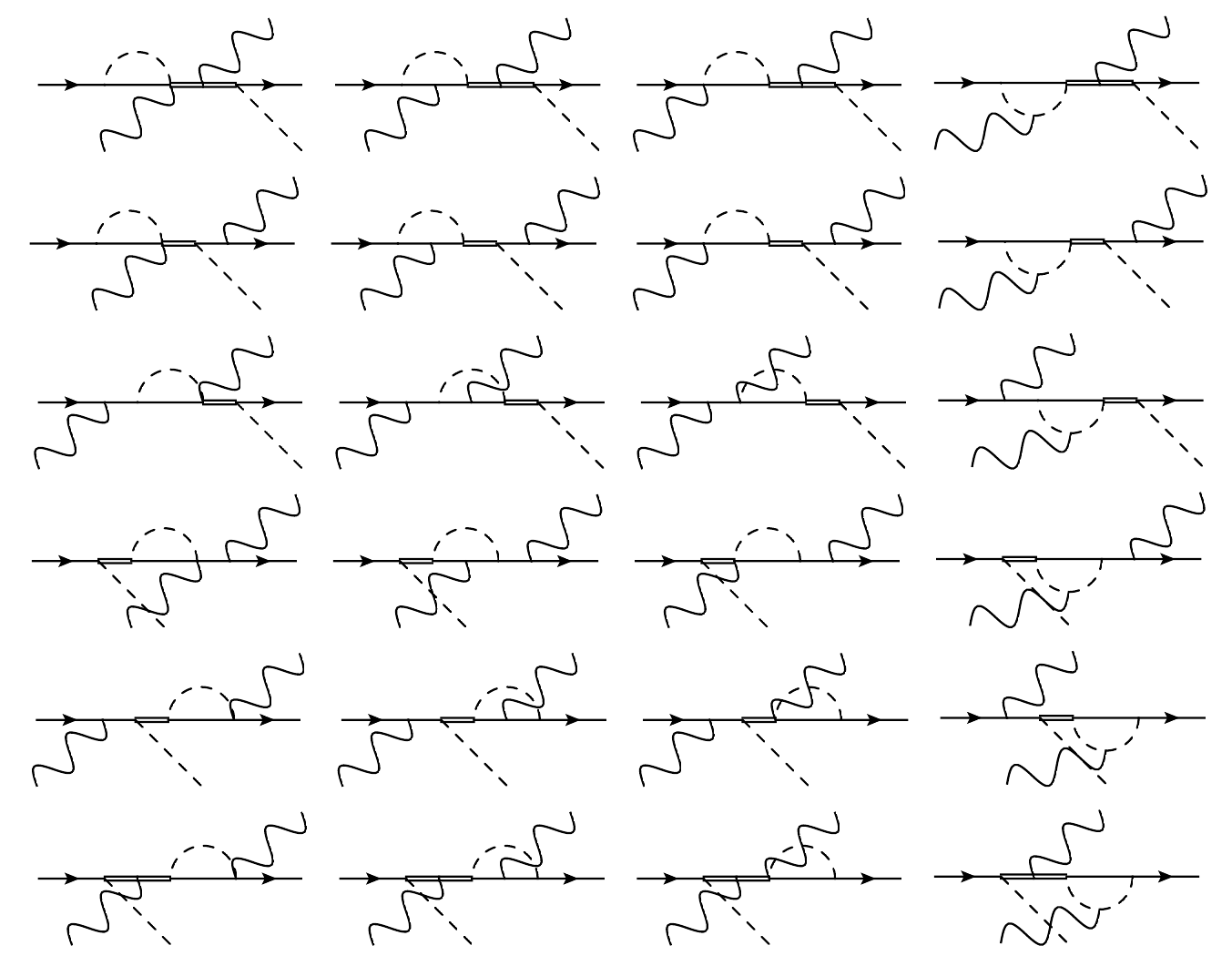}
\caption{First set of loop corrections to the $\deltapart$-pole radiative-pion-photoproduction graphs of order $\epsilon^3$. Diagrams with crossed photon lines are not shown.} \label{fig:DeltaLoops1}
\end{figure}
\begin{figure}[htbp]
\includegraphics[width=0.8\textwidth]{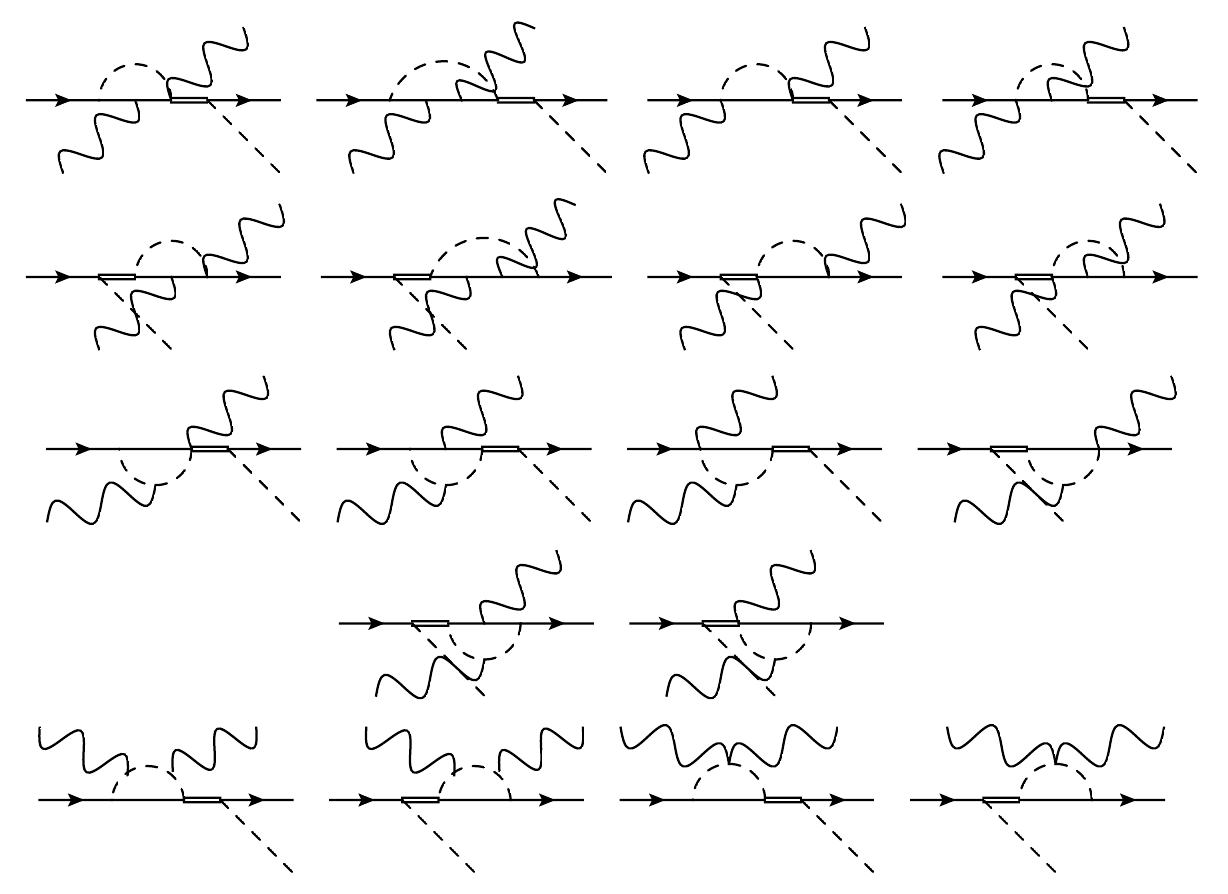}
\caption{Second set of loop corrections to the $\deltapart$-pole radiative-pion-photoproduction graphs of order $\epsilon^3$. Diagrams with crossed photon lines are not shown.}
\label{fig:DeltaLoops2}
\end{figure}
\begin{figure}[htbp]
\includegraphics[width=0.8\textwidth]{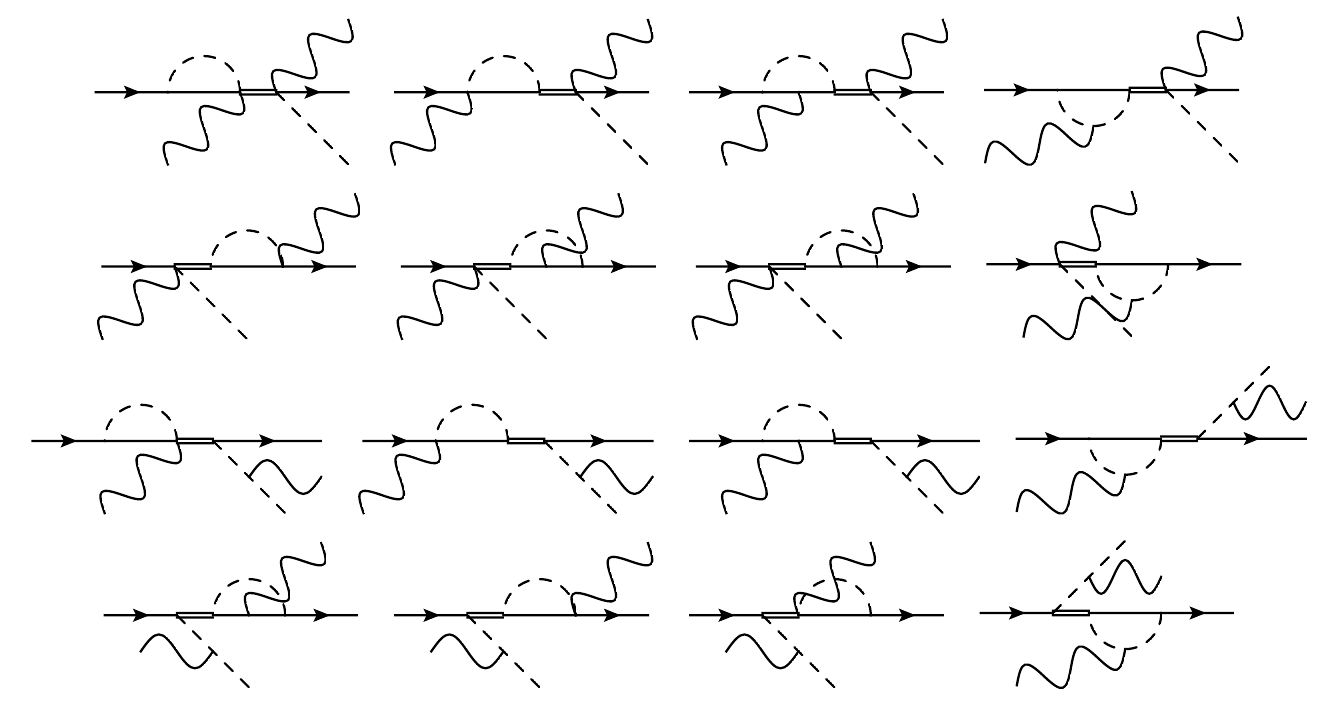}
\caption{Third set of loop corrections to the $\deltapart$-pole radiative-pion-photoproduction graphs of order $\epsilon^3$. 
Diagrams with crossed photon lines are not shown.}
\label{fig:DeltaLoops3}
\end{figure}
\end{itemize}

\newpage
\section{Feynman diagrams for pion photoproduction} \label{sec:Feynman-diagrams_photoproduction}
\setcounter{figure}{0} 
In this section we present all considered Feynman diagrams for pion photoproduction. 
\begin{itemize}
\item[--]
  The leading-order, i.e.~order-$q^1$, tree-level diagrams are shown in Fig.~\ref{fig:basictreespiphp}. 
\begin{figure}[htbp]
\includegraphics[width=0.6\textwidth]{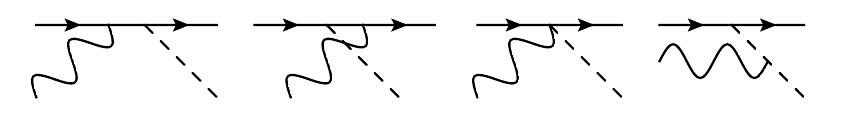}
\caption{Leading-order, i.e.~order-$q^1$, tree-level pion-photoproduction diagrams. All vertices are of the lowest order, 
i.e. from $\lpin^{(1)}$ and $\lpipi^{(2)}$.} \label{fig:basictreespiphp}
\end{figure}
\item[--]
The order-$q^2$ tree-level diagrams are shown in Fig.~\ref{fig:q2treespiphp}
and the order-$q^3$ tree-level diagrams in Fig.~\ref{fig:q3treespiphp}.
\begin{figure}[htbp]
\includegraphics[width=0.35\textwidth]{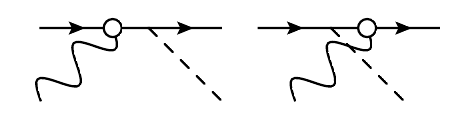}
\caption{Tree-level pion-photoproduction diagrams of order $q^2$. The open circles represent second-order vertices from $\lpin^{(2)}$. 
Diagrams with the second-order corrections to the nucleon self-energy are not shown.} \label{fig:q2treespiphp}
\end{figure}
\begin{figure}[htbp]
\includegraphics[width=0.9\textwidth]{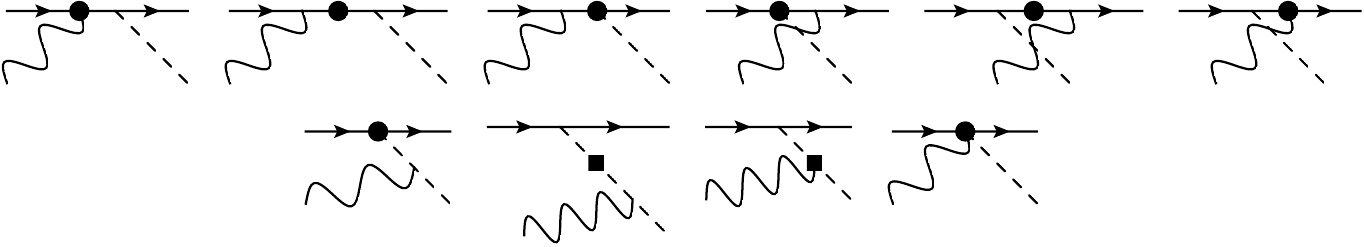}
\caption{Third-order tree-level pion-photoproduction diagrams. The  filled circles and the squares represent vertices
from $\lpin^{(3)}$ and $\lpipi^{(4)} $, respectively.} \label{fig:q3treespiphp}
\end{figure}
\item[--]
Pion-nucleon loop diagrams appear at order $q^3$.
We split them into four sets, which are shown in Figs.~\ref{fig:q3ALoopspiphp}-\ref{fig:q3restloopspiphp}.
The first two sets in Figs.~\ref{fig:q3ALoopspiphp},~\ref{fig:q3BLoopspiphp}
are obtained from the first two diagrams in Fig.~\ref{fig:basictreespiphp} by attaching a pion loop to the nucleon line, wherever possible.
The third set in Fig.~\ref{fig:q3CDLoopspiphp} is obtained in the same way from the last two diagrams in Fig.~\ref{fig:basictreespiphp}.
The remaining graphs are shown in Fig.~\ref{fig:q3restloopspiphp}.
 \begin{figure}[htbp]
\includegraphics[width=0.9\textwidth]{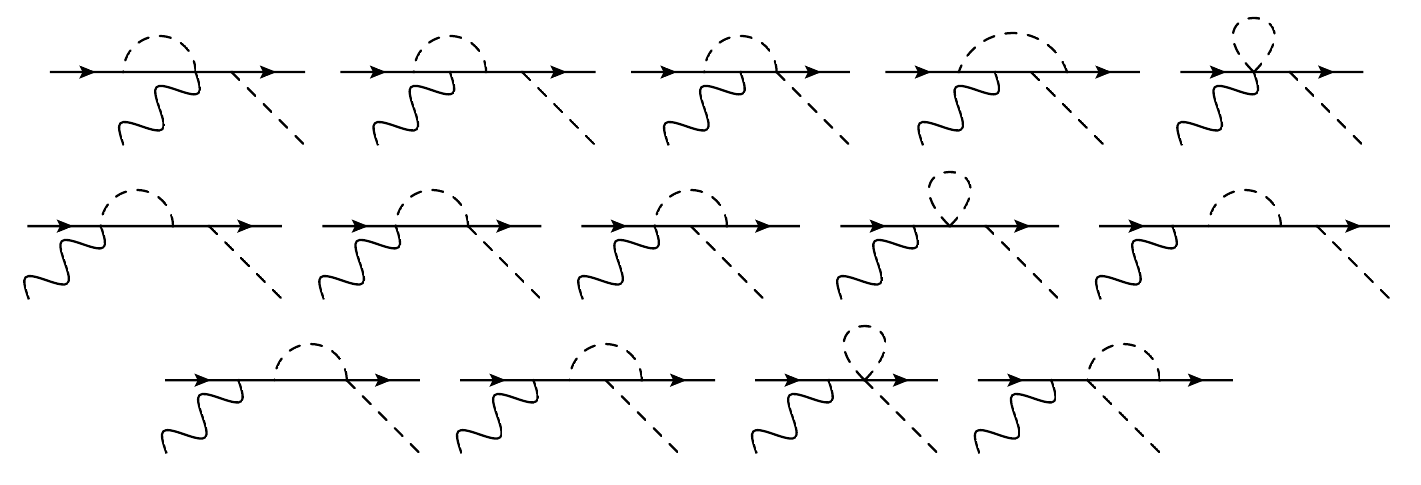}
\caption{First set of pion-photoproduction loop diagrams.} \label{fig:q3ALoopspiphp}
\end{figure}
\begin{figure}[htbp]
\includegraphics[width=0.9\textwidth]{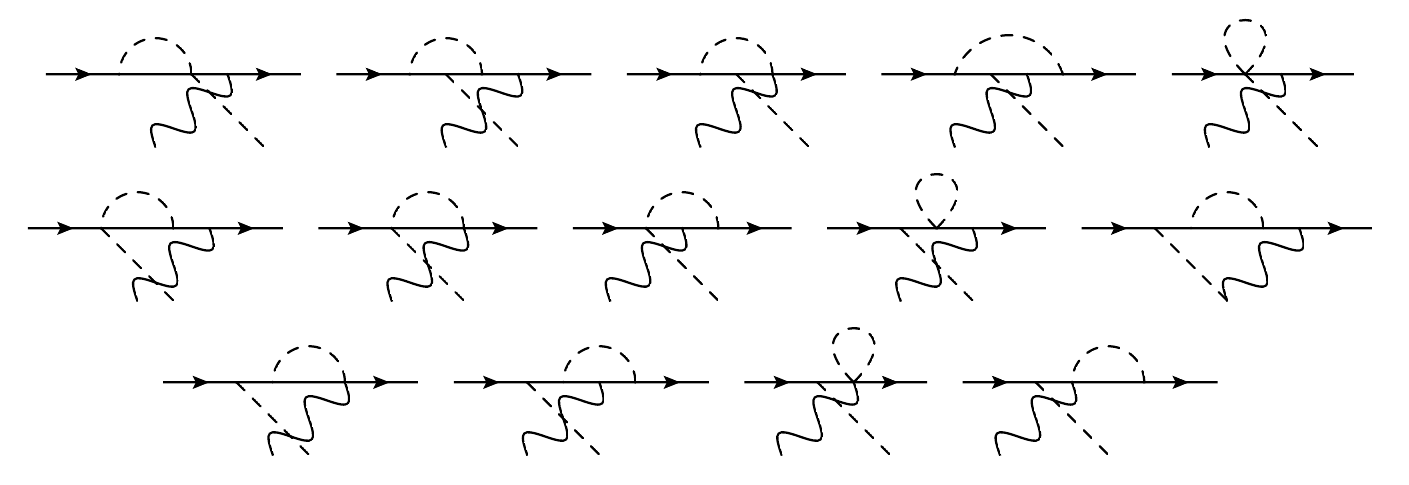}
\caption{Second set of pion-photoproduction loop diagrams.} \label{fig:q3BLoopspiphp}
\end{figure}
\begin{figure}[htbp]
\includegraphics[width=0.65\textwidth]{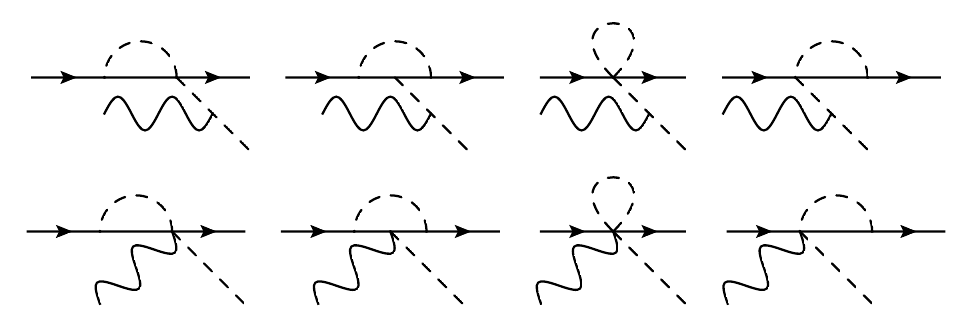}
\caption{Third set of pion-photoproduction loop diagrams.} \label{fig:q3CDLoopspiphp}
\end{figure}
\begin{figure}[htbp]
\includegraphics[width=0.7\textwidth]{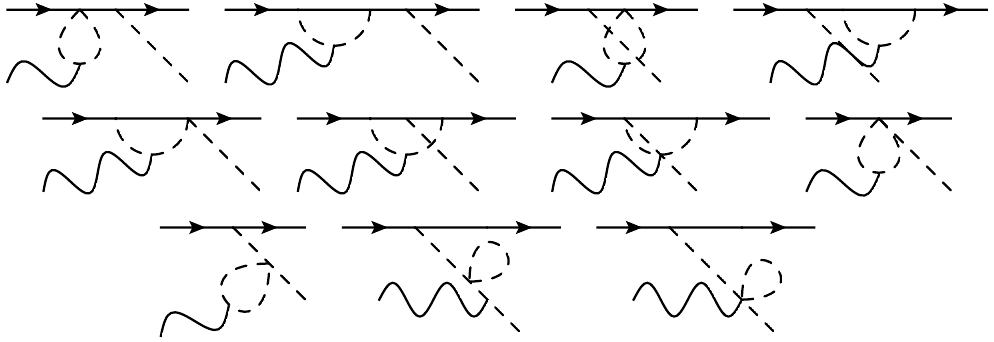}
\caption{Fourth set of pion-photoproduction loop diagrams.} \label{fig:q3restloopspiphp}
\end{figure}
\item[--]
The tree-level diagrams involving $\deltapart$ lines of order $\epsilon^2$ and $\epsilon^3$ are shown in Fig.~\ref{fig:eps2treespiphp}
and Fig.~\ref{fig:eps3treespiphp}, respectively.
\begin{figure}[htbp]
\includegraphics[width=0.35\textwidth]{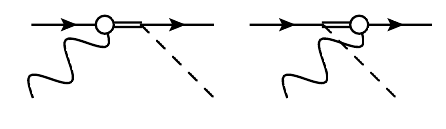}
\caption{Tree-level pion-photoproduction diagrams involving
  $\deltapart$ lines of order $\epsilon^2$. 
The open circle denotes the second-order vertex from $\lpind^{(2)} $.} \label{fig:eps2treespiphp}
\end{figure}
\begin{figure}[htbp]
\includegraphics[width=0.35\textwidth]{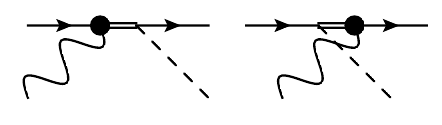}
\caption{Tree-level pion-photoproduction diagrams involving
  $\deltapart$ lines of order $\epsilon^3$. 
The filled circle denotes the vertex from $\lpind^{(3)}$.} \label{fig:eps3treespiphp}
\end{figure}
\item[--]
The loop corrections to the $\deltapart$-pole graphs of order $\epsilon^3$ considered in this work are presented in Figs.~\ref{fig:eps3loopspiphp}.
\begin{figure}[htbp]
\includegraphics[width=0.7\textwidth]{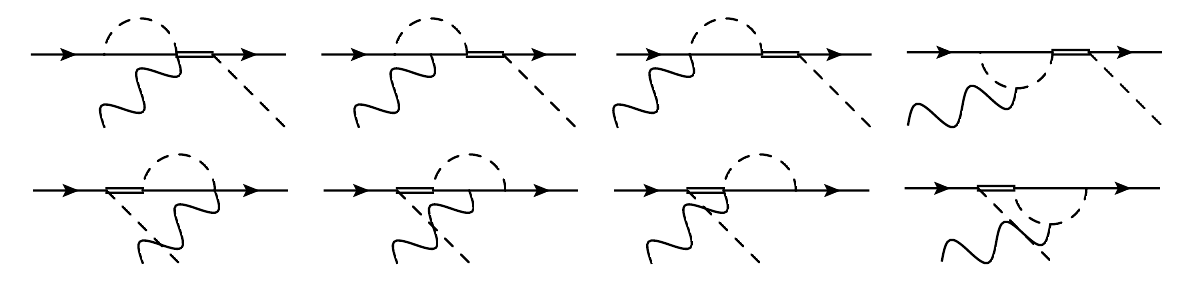}
\caption{Loop corrections to the $\deltapart$-pole pion-photoproduction graphs of order $\epsilon^3$.} \label{fig:eps3loopspiphp}
\end{figure}
\end{itemize}
%
\section{Counter terms} \label{sec:counterrerms}
In this Appendix, we present the expressions for the renormalized quantities and the counter terms. 
To keep the notation compact, we define the mass ratios $\alpha=\frac{\pionmass}{\nucleonmass} $ and $\beta=\frac{\deltamass}{\nucleonmass} $.
\paragraph{Pion mass and field renormalization.}
To the order we are working, the expression for the pion mass reads 
\begin{equation}
\barepionmass^2=\pionmass^2+\delmpi^{(4)}\,,\quad
\delmpi^{(4)}=\frac{\pionmass^2\left( A_0(\pionmass^2)-4 l_3 \pionmass^2\right)}{2 \piondecayconstant^2}\,.
\end{equation}
For the $Z$-factor and pion decay constant, we have  
\begin{equation}
Z_{\pion}=1+\delZpi^{(4)}\,,\quad
\delZpi^{(4)}=\frac{A_0(\pionmass^2)-2 l_4 \pionmass^2}{\piondecayconstant^2}\,,
\end{equation}
and
\begin{equation}
\bareFpi=\piondecayconstant+\delFpi^{(4)}\,,\quad
\delFpi^{(4)}=\frac{-A_0(\pionmass^2)-l_4 \pionmass^2}{\piondecayconstant}\,,
\end{equation}
respectively. The expressions for the loop integrals are provided in
appendix \ref{sec:loop_integrals}. 
\paragraph{Nucleon mass and field renormalization.}
For the nucleon mass, we obtain
\begin{align}
\barenucleonmass &= \nucleonmass + \delmn^{(2)} + \delmn^{(3)}\,,\nonumber\\
\delmn^{(2)} &= 4c_1\pionmass^2\,,\nonumber\\
\delmn^{(3)} &= -\frac{3\axialcoupling^2\nucleonmass}{2\piondecayconstant^2}(A_0(\nucleonmass^2)
+\pionmass^2B_0(\nucleonmass^2,\pionmass^2,\nucleonmass^2))\,,
\end{align}
while the expression for the nucleon $Z$-factor reads
\begin{eqnarray}
Z_{\nucleon}&=&1+\delZn^{(3)}\,,\nonumber\\
\delZn^{(3)}&=&\frac{3\axialcoupling^2}{\piondecayconstant^2\left(\alpha^2-4\right)}\Bigg(\frac{\pionmass^2}{16 \pimath^2}+\frac{ A_0(\pionmass^2) \left(5 \alpha^2-12\right)}{4}- \alpha^2 A_0(\nucleonmass^2)- \pionmass^2\left(\alpha^2-3\right) B_0(\nucleonmass^2,\pionmass^2,\nucleonmass^2)\Bigg)\,.
\end{eqnarray}
\paragraph{Pion-nucleon coupling constant.}
For the pion-nucleon coupling constant, we find the following result:
\begin{eqnarray}
\baregA 
&=& \axialcoupling + \delgA^{(3)}\,,\nonumber\\
\delgA^{(3)} &=&4 \left(d_{18}-2 d_{16}\right)
                 \pionmass^2+\frac{\axialcoupling}{\piondecayconstant^2}\Bigg(-\frac{3
                 \axialcoupling^2 \pionmass^2}{16 \pimath ^2
                 \left(\alpha^2-4\right)}+ \left(\frac{3
                 \axialcoupling^2
                 \alpha^2}{\alpha^2-4}+2\right)A_0\left(\nucleonmass^2\right)
                 +\left(\frac{\axialcoupling^2 \left(10 -4
                 \alpha^2\right)}{\alpha^2-4}-1\right)A_0\left(\pionmass^2\right) \nonumber\\
&-& \axialcoupling^2 \nucleonmass^2 B_0\left(\pionmass^2,\nucleonmass^2,\nucleonmass^2\right)  +\pionmass^2 \left(\frac{3 \axialcoupling^2 \left(\alpha^2-3 \right)}{\alpha^2-4}+2 \right)B_0\left(\nucleonmass^2,\pionmass^2,\nucleonmass^2\right)\nonumber\\
&-&\axialcoupling^2 \nucleonmass^2 \pionmass^2 C_0\left(\nucleonmass^2,\pionmass^2,\nucleonmass^2,\pionmass^2,\nucleonmass^2,\nucleonmass^2\right)\Bigg)\,.
\end{eqnarray}
\paragraph{Nucleon magnetic moments.}
The expressions for the LECs $c_6$, $c_7$ related to the nucleon
magnetic moments, see section \ref{sec:renormalization}, have the form:
\begin{eqnarray}
c_6 
&=& \bar c_6+\kronecker c_6^{(3)}\,,\nonumber\\
\kronecker c_6^{(3)}
  &=&\frac{\axialcoupling^2}{\piondecayconstant^2}\left(-\frac{5
      \nucleonmass^2}{32 \pimath ^2 }+\frac{3
      A_0\left(\nucleonmass^2\right)}{2}-\frac{3
      A_0\left(\pionmass^2\right)}{2}\right. +\frac{\nucleonmass^2\left(2 +3 \alpha^2\right) B_0\left(0,\nucleonmass^2,\nucleonmass^2\right)}{4}+\frac{\nucleonmass^2\left(15 \alpha^2-8\right) B_0\left(\nucleonmass^2,\pionmass^2,\nucleonmass^2\right)}{4}\nonumber\\
&-&3 \pionmass^2 B_0\left(0,\pionmass^2,\pionmass^2\right)+\nucleonmass^2 \pionmass^2\left(3 \alpha^2-4\right) C_0\left(0,\nucleonmass^2,\nucleonmass^2,\pionmass^2,\pionmass^2,\nucleonmass^2\right)\nonumber\\
&+&\left. \frac{\nucleonmass^2 \pionmass^2\left(3 \alpha^2-4\right)
    C_0\left(\nucleonmass^2,0,\nucleonmass^2,\pionmass^2,\nucleonmass^2,\nucleonmass^2\right)}{4
    }\right)\,,
\end{eqnarray}
and
\begin{eqnarray}
c_7 
&=& \bar c_7+\kronecker c_7^{(3)}\,,\nonumber\\
\kronecker c_7^{(3)}
  &=&\frac{\axialcoupling^2}{\piondecayconstant^2}\left(\frac{
    \nucleonmass^2}{8 \pimath ^2} - \frac{\nucleonmass^2\left(2 + 3
    \alpha^2\right)
    B_0\left(0,\nucleonmass^2,\nucleonmass^2\right)}{2}+\nucleonmass^2
    B_0\left(\nucleonmass^2,\pionmass^2,\nucleonmass^2\right)\right. +\frac{3\pionmass^2
    B_0\left(0,\pionmass^2,\pionmass^2\right)}{2} \nonumber \\
    &+&\frac{\nucleonmass^2 \pionmass^2\left(4 -3 \alpha^2\right)
        C_0\left(0,\nucleonmass^2,\nucleonmass^2,\pionmass^2,\pionmass^2,\nucleonmass^2\right)}{2}
        \left.+\frac{\nucleonmass^2 \pionmass^2\left(4 -3 \alpha^2\right) C_0\left(\nucleonmass^2,0,\nucleonmass^2,\pionmass^2,\nucleonmass^2,\nucleonmass^2\right)}{2}\right)\,.
\end{eqnarray}
\paragraph{Pion-nucleon-$\deltapart$ coupling constant:}
\begin{equation}
\barehA
= \pindcoupling + \kronecker \pindcoupling^{(2)}\,,\quad
\kronecker\pindcoupling^{(2)}
=b_3(\nucleonmass-\deltamass) + b_6\frac{\pionmass^2+\nucleonmass^2-\deltamass^2}{2\nucleonmass}\,.
\end{equation}
\paragraph{Electromagnetic $N\deltapart$ transition form factors.}
The auxiliary coefficients $ a_i $, $ b_i $ and $ c_i $ used below to shorten the notation 
are not related to {LECs} of the Lagrangian with similar names.
\allowdisplaybreaks
\begin{align}
b_1 &= \bar b_1 +\kronecker b_1^{(3)}\,,\nonumber\\
\kronecker b_1^{(3)} 
&= \Re\Big[a_0 +
a_1 A_0\left(\pionmass^2\right)
+ a_2 A_0\left(\nucleonmass^2\right) + b_3 B_0\left(\nucleonmass^2,\pionmass^2,\nucleonmass^2\right) + b_4 B_0\left(\deltamass^2,\pionmass^2,\nucleonmass^2\right)\nonumber\\
&\quad + c_5 C_0\left(0,\deltamass^2,\nucleonmass^2,\pionmass^2,\pionmass^2,\nucleonmass^2\right) 
+ c_{6} C_0\left(\nucleonmass^2,0,\deltamass^2,\pionmass^2,\nucleonmass^2,\nucleonmass^2\right)\Big]\,,\nonumber\\
a_0 
&= 4h_{15}\nucleonmass (\beta- 1) + h_{16}\nucleonmass \left(\beta^2 - 2\right)
+\frac{\axialcoupling \pindcoupling \nucleonmass \left(\beta (\beta - 3 )
+2 \alpha^2 \right)}{16 \pimath ^2 \piondecayconstant^2 \left(1 - \beta^2\right)}\,,\nonumber\\ 
a_1 
&=\frac{\axialcoupling \pindcoupling}{(\beta -1) \nucleonmass \piondecayconstant^2}\,,\nonumber\\
a_2 
&=\frac{\axialcoupling\pindcoupling (1 - 3 \beta )}{\left(\beta ^2-1\right) \nucleonmass \piondecayconstant^2}\,,\nonumber\\
b_3 
&=-\frac{\axialcoupling \pindcoupling \nucleonmass \left(\alpha ^2 
\left(-1 -2 \beta - 2 \beta ^2+ \beta ^3 \right)+2 \beta  
\left(2 - \beta + \beta ^2 \right)\right)}{\left(\beta ^2-1\right)^2 \piondecayconstant^2}\,,\nonumber\\
b_4 
&=-\frac{ \axialcoupling \pindcoupling \nucleonmass \beta
\left(-1 +\beta - 5\beta^2 + \beta^3 + \alpha ^2 (3 \beta +1) \right)}{\left(\beta ^2-1\right)^2 \piondecayconstant^2}\,,\nonumber\\
c_5 
&=-\frac{2 \axialcoupling \pindcoupling \nucleonmass^3 \alpha ^2
\left(-1 + \alpha ^2+\beta ^2\right)}{\left(\beta ^2-1\right) \piondecayconstant^2}\,,\nonumber\\
c_{6} 
&=\frac{2  \axialcoupling \pindcoupling \nucleonmass^3 \beta 
\left(2 -\alpha ^2 -\beta + \beta ^2 \right)}{\left(\beta ^2-1\right) \piondecayconstant^2}\,;
\end{align}
\begin{align}
h_1 &= \bar h_1 +\kronecker h_1^{(3)}\,,\nonumber\\
\kronecker h_1^{(3)} &= \Re\Big[a_0 +
a_1 A_0\left(\pionmass^2\right)
+ a_2 A_0\left(\nucleonmass^2\right) + b_3 B_0\left(\nucleonmass^2,\pionmass^2,\nucleonmass^2\right) + b_4 B_0\left(\deltamass^2,\pionmass^2,\nucleonmass^2\right)\nonumber\\
&\quad + c_5 C_0\left(0,\deltamass^2,\nucleonmass^2,\pionmass^2,\pionmass^2,\nucleonmass^2\right) 
+ c_{6} C_0\left(\nucleonmass^2,0,\deltamass^2,\pionmass^2,\nucleonmass^2,\nucleonmass^2\right)\Big]\,,\nonumber\\
a_0 
&= 4 h_{15} \nucleonmass +\frac{\axialcoupling \pindcoupling \nucleonmass 
\left(\beta -\alpha ^2\right)}{4 \pimath ^2 (\beta -1)^2 (\beta +1) \piondecayconstant^2}\,,\nonumber\\
a_1 
&= \frac{2 \axialcoupling \pindcoupling}{(\beta -1)^2 \nucleonmass \piondecayconstant^2}
+\frac{5 \gOne \pindcoupling}{324 (\beta -1)^2 \beta ^6 (\beta +1) \nucleonmass \piondecayconstant^2}\,,\nonumber\\
a_2 
&= -\frac{4   \axialcoupling \pindcoupling \beta}{(\beta -1)^2 (\beta +1) \nucleonmass \piondecayconstant^2}\,,\nonumber\\
b_3 
&= -\frac{2 \axialcoupling \pindcoupling \nucleonmass 
\left(\alpha ^2 \left(-1 - 2\beta - 2\beta^2 + \beta^3\right)+\beta  
\left(3 + \beta ^2\right)\right)}{(\beta -1)^3 (\beta +1)^2 \piondecayconstant^2}\,,\nonumber\\
b_4 
&= \frac{2   \axialcoupling \pindcoupling \nucleonmass \beta
\left(1+ 3\beta^2 - \alpha ^2 (1+ 3 \beta)\right)}{(\beta -1)^3 (\beta +1)^2 \piondecayconstant^2}\,,\nonumber\\
c_5 
&= -\frac{2  \axialcoupling \pindcoupling \nucleonmass^3 \alpha ^2
\left(-1 + 2 \alpha ^2+\beta ^2\right)}{(\beta -1)^2 (\beta +1) \piondecayconstant^2}\,,\nonumber\\
c_{6} 
&= -\frac{2 \axialcoupling \pindcoupling \nucleonmass^3 
\left(\alpha ^2 (1+ \beta)-\beta  \left(3+\beta ^2\right)\right)}{(\beta -1)^2 (\beta +1) \piondecayconstant^2}\,.
\end{align}

\section{Loop integrals}\label{sec:loop_integrals}
The loop integral functions are defined as
\begin{align}
A_0(m_0^2) &=\frac{1}{\ci}\int\frac{\dd^d l}{(2\pimath)^d}\frac{\mu^{4-d}}{l^2-m_0^2}\,,\nonumber\\
B_0(p_1^2,m_0^2,m_1^2) 
&=\frac{1}{\ci}\int\frac{\dd^d l}{(2\pimath)^d}\frac{\mu^{4-d}}{(l^2-m_0^2)((l+p_1)^2-m_1^2)}\,,\nonumber\\
C_0(p_1^2,p_2^2,(p_1-p_2)^2,m_0^2,m_1^2,m_2^2) 
&= \frac{1}{\ci}\int\frac{\dd^d l}{(2\pimath)^d}\frac{\mu^{4-d}}{(l^2-m_0^2)((l+p_1)^2-m_1^2)((l+p_2)^2-m_2^2)}\,.
\label{eq:loop_integrals}
\end{align}
The renormalization scale $\mu$ in all integrals is set to $\mu=\nucleonmass$.

\bibliographystyle{myunsrtnat}
\bibliography{gamma_n_gamma_pi_nlit}


\end{document}